\documentclass{article}

\usepackage{amsmath}
\usepackage{subfiles} 
\usepackage{xcolor}
\usepackage{blindtext}
\usepackage[margin=1in]{geometry}
\usepackage{caption}
\usepackage[normalem]{ulem}
\usepackage{float}

\numberwithin{equation}{section}
\numberwithin{figure}{section}   
\numberwithin{table}{section}   

\renewenvironment{quote}
  {\list{}{\rightmargin\leftmargin}%
   \item\relax\color{darkgray}\small}
  {\endlist}

\PassOptionsToPackage{unicode}{hyperref}
\PassOptionsToPackage{hyphens}{url}
\usepackage{amsmath,amssymb}
\usepackage{lmodern}
\usepackage{iftex}
\usepackage{textgreek}
\usepackage[mathletters]{ucs}

\usepackage{newunicodechar}
\newunicodechar{π}{\ifmmode\pi\else\textpi\fi}

\ifPDFTeX
  \usepackage[T1]{fontenc}
  \usepackage[utf8]{inputenc}
  \usepackage{textcomp} 
\else 
  \usepackage{unicode-math}
  \defaultfontfeatures{Scale=MatchLowercase}
  \defaultfontfeatures[\rmfamily]{Ligatures=TeX,Scale=2}
\fi

\IfFileExists{upquote.sty}{\usepackage{upquote}}{}
\IfFileExists{microtype.sty}{
  \usepackage[]{microtype}
  \UseMicrotypeSet[protrusion]{basicmath} 
}

\makeatletter
\@ifundefined{KOMAClassName}{
  \IfFileExists{parskip.sty}{%
    \usepackage{parskip}
  }{
    \setlength{\parindent}{0pt}
    \setlength{\parskip}{6pt plus 2pt minus 1pt}}
}{
  \KOMAoptions{parskip=half}}
  
\makeatother
\usepackage{xcolor}
\usepackage{graphicx}
\graphicspath{{./images/}}
\makeatletter
\def\maxwidth{\ifdim\Gin@nat@width>\linewidth\linewidth\else\Gin@nat@width\fi}
\def\maxheight{\ifdim\Gin@nat@height>\textheight\textheight\else\Gin@nat@height\fi}
\makeatother
\setkeys{Gin}{width=\maxwidth,height=\maxheight,keepaspectratio}
\makeatletter
\def\fps@figure{htbp}
\makeatother
\providecommand{\tightlist}{%
  \setlength{\itemsep}{0pt}\setlength{\parskip}{0pt}}
\ifLuaTeX
  \usepackage{selnolig}  
\fi
\IfFileExists{bookmark.sty}{\usepackage{bookmark}}{\usepackage{hyperref}}
\IfFileExists{xurl.sty}{\usepackage{xurl}}{} 
\hypersetup{
  pdftitle={Main Text},
  hidelinks,
      colorlinks=true,
    linkcolor=blue,
    filecolor=red,      
    urlcolor=blue,
  pdfcreator={LaTeX via pandoc}}

\let\oldhyperlink\hyperlink
\renewcommand{\hyperlink}[2]{\oldhyperlink{#1}{\uline{#2}}}
\let\oldhref\href
\renewcommand{\href}[2]{\oldhref{#1}{\uline{#2}}}

\usepackage{titlesec}

\titleformat{\section}
  {\normalfont\huge\bfseries}{\thesection}{1em}{}
\titlespacing*{\section}
  {0pt}{3.5ex plus 1ex minus .2ex}{4.6ex plus .2ex}

\titleformat{\subsection}
  {\normalfont\Large\bfseries}{\thesubsection}{1em}{}

\titleformat{\subsubsection}
  {\normalfont\large\bfseries}{\thesubsubsection}{1em}{}

\title{Variational Quantum Algorithms}
\author{Michał Stęchły}

\begin{document}


\begin{titlepage}
\renewcommand{\thefootnote}{\fnsymbol{footnote}}

\centering
\vspace*{\fill} 

\title{Variational Quantum Algorithms}
\author{Michał Stęchły}


{\Huge\textbf{Introduction to Variational Quantum Algorithms}}\\[3.5cm] 
{\Large Michał Stęchły{\footnotemark[1]}}  \\[1cm] 
{\large \today}\\ 

\vspace*{\fill} 

\footnotetext[1]{\href{mailto:michal@mustythoughts.com}{michal@mustythoughts.com}}

\end{titlepage}

\tableofcontents
\newpage


\hypertarget{section-1}{\section{Introduction}}

Dear Reader!

In the years 2019--2022, I wrote a series of blogposts aimed at helping people understand some of the intricacies of Variational Quantum Algorithms (VQAs) and demystifying how they work. This document is a PDF version of the blogposts.

I started writing them in 2019, when most of the community had big hopes for VQAs (and Noisy Intermediate-Scale Quantum (NISQ) devices in general) and there were still plenty of unknown unknowns. But by now (February 2024), there is an increasing consensus in the community that using VQAs on NISQ devices for any practical purposes faces fundamental issues and that fault-tolerant algorithms (like QPE) seem to be a still distant, but much safer bet.

Nevertheless---I enjoyed my time working on VQAs, I think they are a very interesting class of algorithms and I would be overwhelmed with joy if someone manages to use them for a practical application! Therefore even though I don't think this is the most promising research area right now, I don't think learning about them is a waste of time.

Over the years I have heard many times that my blogposts were very helpful for people to understand these algorithms. Therefore, I have decided that perhaps publishing them in the form of a single PDF might be useful for some people. This document is targeted at people with a basic understanding of quantum computing concepts and linear algebra.  I always strived to provide intuitive, but correct and not oversimplified explanations. I personally prefer that over mathematical rigor, so I hope those more mathematically inclined among you will forgive me.

If you like this document, you might want to visit my \href{https://mustythoughts.com/}{blog} to see if in the meantime I have written about some other topics that would be interesting for you.

\newpage

\hypertarget{section-2}{\section{Variational Quantum Eigensolver Explained}}

We will start with Variational Quantum Eigensolver (VQE).

\hypertarget{what-does-vqe-even-do}{%
\subsection{What does VQE even do, and why do we care?}\label{what-vqe-does}}
\subsubsection{What VQE does}

\textbf{VQE allows us to find an upper bound of the lowest eigenvalue of a given Hamiltonian.}

Without going too much into the details and math, I would like to explain what we can use VQE for and why it's important.

Let's break it down:

\begin{itemize}
\item Upper bound---let's say we have some quantity and don't know its value. For the sake of the example, let's say it's the water level in a well. We know that the water surface can't be above the edge of the well---if that were the case, the water would spill. So this is our \textbf{upper bound}. We also know that the level cannot fall below the bottom of the well---this is our \textbf{lower bound}. Sometimes, when we describe a physical system, finding the actual value is tough, but knowing the lower or upper bound can help us estimate the value.
\item Hamiltonian---the Hamiltonian is a matrix that describes the possible energies of a physical system. If we know the Hamiltonian, we can calculate the system's behavior, learn the system's physical states, etc. It's a central concept in quantum mechanics.
\item Eigenvalue---a given physical system can be in various states. Each state has a corresponding energy. The eigenvector describes these states; their energies are equal to the corresponding eigenvalues. In particular, the \textbf{lowest eigenvalue} corresponds to the \textbf{ground state energy}.
  \item Ground state---this is the state of the system with the lowest energy, which means it's the ``most natural'' state---i.e.~a given system always tends to get there, and if it is in the ground state and is left alone, it will stay there forever.
\end{itemize}

These ``physical systems'' and ``states'' may sound abstract, so let's try an example that's a little bit more down-to-earth. I cannot think of anything more straightforward than a hydrogen atom.

So, we have a two-particle system with a proton in the center and one electron. We can describe it with a Hamiltonian. The ground state of the electron is its first orbital. In this configuration, it has some energy corresponding to the lowest eigenvalue of the Hamiltonian. If we hit it with a photon with a specific energy, it can get to the excited state (any state that's not the ground state)---described by another eigenvector and eigenvalue.

With all that in mind, we can reformulate our statement about VQE:

\textbf{VQE can help us to estimate* the energy of the ground state of a
given quantum mechanical system.**}

With the following caveats:

\begin{itemize}
    \tightlist
    \item *estimate by providing us an upper bound
    \item **if we know the Hamiltonian of this system \footnote{To be mathematically exact, upper/lower bounds are defined for sets, not single values. So whenever I mention ``upper bound of a value,'' I mean ``upper bound of the one-element set containing given value.'' Also, I know that stating ``Hamiltonian is a matrix'' might be viewed as oversimplification, but I find it a very helpful oversimplification.}
\end{itemize}

Now it's time to get to the second, arguably more important part:

\hypertarget{why-do-we-care}{%
\subsubsection{Why do we care?}\label{why-do-we-care}}

There are at least two reasons why we care about using a quantum computer to get the energy of a ground state:

\begin{enumerate}
\tightlist
\item Knowing a ground state is useful.
\item  It's complicated to obtain it using a classical computer.
\end{enumerate}

Ground state energy is fundamental in quantum chemistry. We use this value to calculate a number of other, more interesting properties of molecules, like their reaction rates, binding strengths or molecular pathways (the different ways in which a chemical reaction can occur). Many of these calculations involve several steps, so an error in the first step will introduce an error to all the subsequent steps. Therefore if one of the values we use in the first step (e.g.~energy of the ground state) is closer to the reality, so will be our final result.

\hypertarget{the-variational-principle}{%
\subsection{The variational principle}}
\label{subsection:the-variational-principle}

In this and the following sections, I will go into more technical details of why and how VQE works.

The first piece is the variational principle.

Let's say we have a Hamiltonian $H$ with eigenstates and associated eigenvalues. Then the following relation holds:

\begin{equation}
    H | \psi_{\lambda} \rangle = \lambda | \psi_{\lambda} \rangle
\end{equation}

which is basically a characteristic equation (eigenequation) for $H$.

The problem is that we usually don't know what the eigenstate
$| \psi_{\lambda} \rangle$ is and what's the value of the $\lambda$. So how can we get some estimate of that?

Well, to get the energy value for the given state $ |\psi \rangle$, we can use the following expression:

\begin{equation}
    \langle \psi | H | \psi \rangle = E( \psi )
    \label{eq:expectation_value}
\end{equation}

(We usually refer to this as calculating the ``expectation value'' of $H$.)

If we instead of any $| \psi \rangle$ we use an eigenstate we get:

\begin{equation}
    \langle \psi_{\lambda} | H | \psi_{\lambda} \rangle = E_{\lambda}
\end{equation}
    
For the eigenstate associated with the smallest eigenvalue we would get
$E_0$ (ground state energy). It's by definition the lowest value we
can get, so if we take an arbitrary state $| \psi \rangle$ with an
associated energy $E_\psi$, we know that $E_\psi \ge
E_0$. We get equality if we're lucky enough that our state is equal
to the ground state.

The method described here is known as variational method and the following relation is known as the variational principle:

\begin{equation}
    \langle \psi | H | \psi \rangle \ge E_0
\end{equation}

Our $E_\psi$ is an \textbf{upper bound} for the ground state energy, so exactly what we wanted. The problem is that knowing any upper bound doesn't yield much benefit. The closer the upper bound is to the actual value, the better. In principle, we could simply try all the possible states and pick the one with the lowest energy. However, this approach isn't very practical---there are way too many states to accomplish that in a reasonable amount of time. Also, most of the states will give you really bad answers, so there is no point in trying them
out.\footnote{I know I've done a couple of shortcuts here to keep it shorter and easier to digest. If you want to get a formal proof for the variational principle, it's described pretty well
\href{https://en.wikipedia.org/wiki/Variational_method_(quantum_mechanics)}{on wiki}.}

This is where the \textbf{ansatz} takes the stage!

\hypertarget{what-the-hell-is-an-ansatz}{%
\subsection{What the hell is an
ansatz?}\label{what-the-hell-is-an-ansatz}}

Once you start reading papers about VQE or QAOA, you'll start noticing that people care a lot about ansatzes. I missed the definition of ansatz when I started learning, which meant that for a long time, it was something rather abstract to me, so here I will try to give you some intuition regarding this topic.

The problem (as mentioned in the previous section) is that we would like to explore the space of all the possible states in a reasonable manner.\footnote{Actually, this is an oversimplification, as often you don't
  want to explore the whole state space, as you know some states are garbage. I discuss it a bit more in the subsection about ansatz design in \hyperlink{section-05}{\emph{chapter 5}.}}
  We do that using parameterizable circuits---i.e.~circuits which have gates with some parameters. So let's take a simple circuit consisting of a single RY gate as an example. It has a single parameter---rotation angle. With this simple circuit, we can represent a whole spectrum of quantum states---namely $\sqrt{\alpha}|0 \rangle + \sqrt{1-\alpha}|1\rangle $.

\begin{figure}
\centering
\includegraphics[width=6cm]{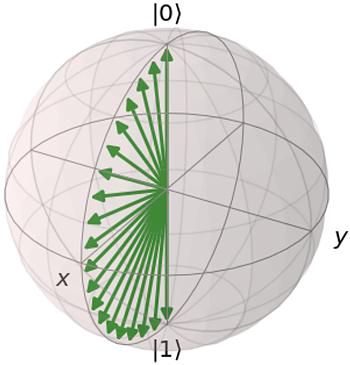}
\caption{Bloch sphere depicting the family of states that is reachable through an RY gate.}
\label{fig:bloch_RY}
\end{figure}

In Fig. \ref{fig:bloch_RY}, you can find a visualization of this family of states using a Bloch sphere. I'm not going to explain what the Bloch sphere is---if you're not familiar with it, you can read about it \href{http://akyrillidis.github.io/notes/quant_post_7}{here} or elsewhere.

However, there are many states that we can't represent---e.g.
$\frac1{\sqrt{2}}  ( |0\rangle +i |1\rangle)$.
In order to cover this state, we need to introduce another gate, e.g.~RX. And of course, find the corresponding angles.

So RX or RX RY are examples of simple ansatzes. The situation gets more complicated when we have to deal with multiple qubits or entanglement. We will get there in next sections, there is also \href{https://arxiv.org/abs/1905.10876}{this great paper} that covers this topic in depth. But in general, we want to have an ansatz which:

\begin{itemize}
\tightlist
\item Covers many states---ideally, we want it to cover the whole space of possible states.
\item  Is shallow---the more gates you use, the higher the chance of something going wrong. We're talking about NISQ after all.
\item Doesn't have too many parameters---the more parameters, the harder it is to optimize.
\end{itemize}

If the description above doesn't give you enough intuition of what it's about, let's try with the following metaphor.

\begin{figure}
\centering
\includegraphics[width=10cm]{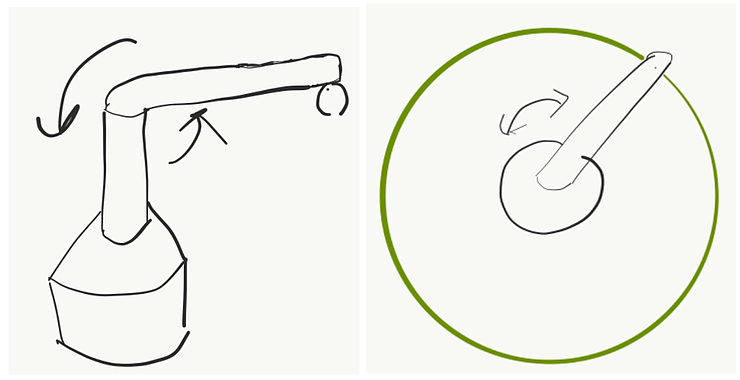}
\caption{Robot with one fixed rotating arm.}
\label{fig:robot_1}
\end{figure}

Imagine you have a robot. This a simple robot with one rigid arm that can rotate around the central point, depicted in Fig \ref{fig:robot_1}.

Let's say the robot has a pen and it can draw. In this case, no matter how hard you try, the most you can get out of it is a circle. How to improve it?

\begin{figure}
\centering
\includegraphics[width=10cm]{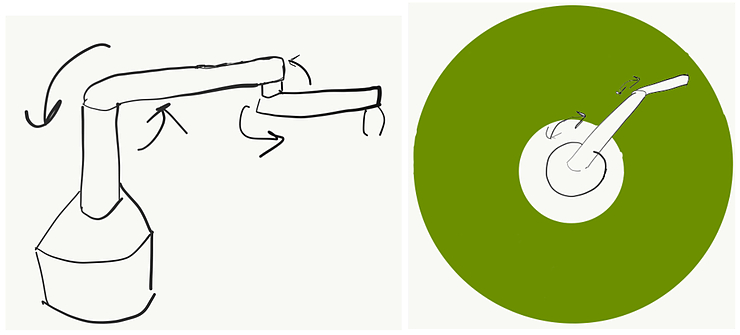}
\caption{Robot with two rotating arms.}
\label{fig:robot_2}
\end{figure}

Well, you can add another rotating arm at the end of the first one. Now you can cover much more---a big circle empty in the middle. See Fig. \ref{fig:robot_2}.

Ok, and what if I told you that actually the arms can only move in a discrete way---i.e.~they always rotate by a fixed angle. If it's small, then you can still cover most of the space. If not---well, you will get a nice pattern. Anyway, it might be beneficial to add additional arms in this imperfect scenario to cover the space better.

Why don't you add another 20 arms? It sounds impractical for several reasons. The more arms you add, the harder it is to construct and control them, and it's more error-prone. By the way, you can check out \href{https://youtu.be/-qgreAUpPwM}{this video} to see how crazy you could get with a similar setup in theory. Imagine what these drawings would look like if you got even a 0.1\% error in the angles for each arm.

So what does it have to do with our ansatzes? Having a good robot setup will allow you to cover a lot of space and at the same time it would be relatively simple. The same is true for ansatzes. A good ansatz is one that allows you to cover many states and which has a reasonable number of parameters---so it's easier to control and optimize.

\hypertarget{what-does-the-circuit-look-like-vqe}{%
\subsection{What does the circuit look
like?}\label{what-does-the-circuit-look-like-vqe}}

\emph{While writing this subsection I realized it requires explaining some additional concepts. If it seems difficult, don't worry---you don't need to understand it fully to get the big picture. If this is too much at once, you can always come back later.}

As we all know, to do quantum computing we use quantum circuits. So what does one for VQE look like?

It consists of three parts:

\begin{itemize}
\tightlist
\item Ansatz
\item Hamiltonian
\item Measurement
\end{itemize}

We know what the ansatz part is about---it prepares the state we need in order to apply the variational principle.

Hamiltonian---there is a limitation on what Hamiltonian we can use with VQE. We need Hamiltonian constructed as a sum of Pauli operators ($X$, $Y$, $Z$) and their tensor products. So for example we can have the following Hamiltonian: $XY + YZ + Z$. Or for the multi-qubit case:

\begin{equation}
    X_1 \otimes X_2 \otimes Y_1 - Z_1 \otimes X_2 + Z_1
\end{equation}

What might be surprising is the fact that we don't actually put any of the corresponding $X$, $Y$, or $Z$ gates into the circuit. We use them to choose on which basis we want to do a measurement.

Most (if not all) contemporary quantum computers perform measurements on the so-called ``computational basis.'' What it means is that they will tell you whether your state turned out to be 0 or 1. However, that's not the only way to go. What if we wanted to distinguish between states
$|+\rangle = |0\rangle + |1\rangle$ and $|-\rangle =|0\rangle - |1\rangle$ ?

\begin{figure}
\centering
\includegraphics[width=6cm]{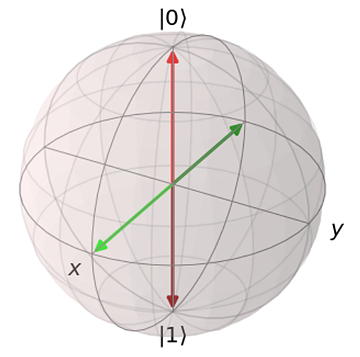}
\caption{Bloch sphere showing the states $|0\rangle$ and $|1\rangle$ (red) and $|+\rangle$ and $|-\rangle$ (green)}
\label{fig:bloch_bases}
\end{figure}

Well, measuring in the computational basis, you'll get 0 and with the same probabilities for both of them, so they're indistinguishable. To see what might help we can use the Bloch sphere to visualize both pairs of states (Fig. \ref{fig:bloch_bases}).

As you can see, in both cases they are on the opposite ends of the sphere. It means that we can switch between them by performing a rotation around the Y axis by 90 degrees.

This gives us a tool to distinguish between $|+\rangle$ and $|-\rangle$ states even if we can measure only whether the state is $|0\rangle$ or $|1\rangle$. If we start from $|+\rangle$ and perform $RY(π/2)$ rotation (rotation around the Y axis by 90 degrees), we will end up in the $|1\rangle$ state. If we now measure it, we will always get 1. Similar for the other state.

\begin{figure}
\centering
\includegraphics[width=6cm]{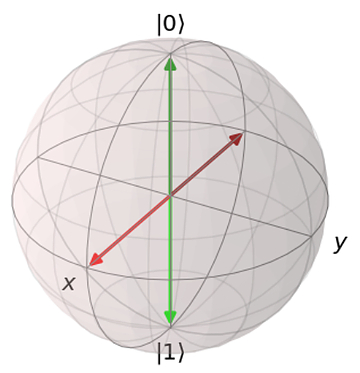}
\caption{The Bloch sphere in Fig. \ref{fig:bloch_bases} after applying $RY(π/2)$. Notice the colors switched.}
\label{fig:bloch_bases_rotated}
\end{figure}

After rotating applying $RY(π/2)$ to all the states, we will get the Bloch sphere in Fig. \ref{fig:bloch_bases_rotated}.

Okay, so coming back to our Hamiltonians. The first thing we need to do is divide the sum into its elements, so-called ``terms.'' Now, for each term we look at every Pauli operator and append appropriate rotations to the circuit:

\begin{itemize}
  \item $RY(-π/2)$ if it's X
  \item $RX(π/2)$ if it's Y
  \item We do nothing if it's Z
\end{itemize}

There's one more caveat here: we don't do it all at once---we create a separate circuit for each term.

And once we've done this, we are ready to perform the measurements. But how exactly do these measurements relate to the variational principle mentioned earlier? Remember equation \ref{eq:expectation_value}?

$| \psi \rangle$ is the state created by the ansatz. Hamiltonian is composed of sums of Pauli terms, so we can separate it into individual terms and deal with them one at the time: 

\begin{equation}
H = H_1 + H_2 + \dots + H_n
\end{equation}

The question at hand is how to get from here to
$\langle \psi | H_i | \psi \rangle$. We can utilise the fact that
$\langle \psi | H_i | \psi \rangle$ represents the expected value of
$H_i$, which can be approximated by observing (measuring) the state
$| \psi \rangle$ using the right basis (one corresponding to the eigenvalues of $H_i$) and averaging the results. The more times we do
that, the better precision we'll get.

One more thing about the averaging---when we do a measurement we get either 0 or 1. Taking all the 0s and 1s and averaging them would give us an expectation value of the measurements, not an expectation value of the Z operator. If we look at the Z operator, we see that it has two eigenstates $| 0 \rangle$ and $|1 \rangle$ associated with eigenvalues $1$ and $-1$ accordingly. To calculate
$\langle \psi | Z | \psi \rangle$, we need to measure our circuit many times and then substitute every ``$0$'' with ``$1$'' and every ``$1$'' with ``$-1$.''

For more details and some helpful derivations, I recommend reading chapter 4 of \href{https://www.goodreads.com/book/show/5299445-quantum-computing-for-computer-scientists?from_search=true\&from_srp=true\&qid=AOrNwKA4DZ\&rank=1}{``Quantum Computing for Computer Scientists'' by N. Yanofsky and M. Mannucci}.

Ok, so let's take a look at the following example.

Let's say the Hamiltonian is:

\begin{equation}
    H = 2*Z + X + I
\end{equation}

(There's only one qubit, so I'm omitting indices.) It is represented by the following matrix:

\begin{equation}
    \begin{bmatrix}
        3       & 1 \\
       1       & -1\\
    \end{bmatrix}
\end{equation}

We would like to learn its ground state using VQE. To begin with---we need to choose our ansatz. We don't want to overcomplicate it, let's say our ansatz will be $RY(\theta)$.
This will allow us to explore the states
$\cos(\theta) | 0 \rangle + \sin(\theta) | 1 \rangle$.

Our Hamiltonian has three parts:

\begin{equation}
\begin{aligned}
    H_1 &= 2*Z  \\
    H_2 &= X    \\
    H_3 &= I   
\label{eq:hamiltonian-components}
\end{aligned}
\end{equation}

The first part is fairly simple. As you recall, if we have a Z gate, we don't need to add any rotation. So our circuit for the first part of the Hamiltonian will consist just of the ansatz.

You might wonder what should we do with the ``2'' in front of Z? Well, we will treat it just as weight---after we do measurements and we will be calculating the energy, the results coming from $H_1$ will count
twice as much as the results coming from $H_2$ or $H_3$.
In case of $H_2$ we have one gate: X. So we simply need to add a
$RY(-\pi/2)$ gate to our circuit.
$H_3$ is the easiest one of all of them---it's just a constant factor we need to add to our calculations and we don't need to do any measurements.

So in this example, we need two circuits to use our VQE. You can see them below in Fig. \ref{fig:1_qubit_vqe}:

\begin{figure}[H]
\centering
\includegraphics{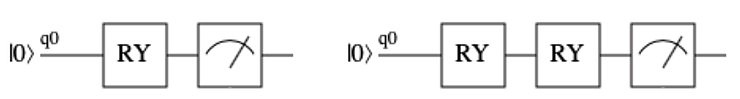}
\caption{Circuits needed for our simple VQE. First RYs are ansatzes, and they are parametrized by theta. Second RY in the right picture is actually $RY(-\pi/2)$.
The last symbol is how we depict measurement in quantum circuits.}
\label{fig:1_qubit_vqe}
\end{figure}

Recall our Hamiltonian's parts from equation \ref{eq:hamiltonian-components}.

So if $\theta = 0$, then we will get

\begin{equation}
\begin{aligned}
    \langle \psi(0) | H_1 | \psi(0) \rangle &= 2 \cdot 1 \\
    \langle \psi(0) | H_2 | \psi(0) \rangle &= 0 \\
    \langle \psi(0) | H_3 | \psi(0) \rangle &= 1
\end{aligned}
\end{equation}

So our total is

\begin{equation}
    \langle \psi(0) | H | \psi(0) \rangle = 2 + 0 + 1 = 3
\end{equation}

If $\theta=\pi$, we get $2\cdot(-1)$ for $H_1$, 0 for $H_2$ and 1 for
$H_3$, so

\begin{equation}
    \langle \psi(\pi) | H | \psi(\pi) \rangle = -1
\end{equation}

Going through all the possible values of $\theta$ would be cumbersome, so I will tell you how to approach the problem of finding a good value of theta in \hyperlink{hybrid-approach}{the next section}.

This example was pretty simple and you might wonder: ``Ok, but how do I deal with the case when there is more than one Pauli matrices in one term?'' It turns out that whenever you start multiplying Pauli matrices, in the end, you always get another Pauli matrix multiplied by some constant. So, you can always simplify your term to end up with a single Pauli matrix.

And what if I have more qubits? Again, we treat each term separately, perform all the rotations that we need at the end and measure all the qubits that we need.

If you would like to calculate these by hand, you might find \href{https://drive.google.com/file/d/1yrH4_5AeioXbi03AmP9TBspyWlEx2ydO/view?usp=sharing}{the notes} by \href{https://christophtrybek.github.io/}{Christoph Trybek} helpful.

\hypertarget{hybrid-approach}{%
\subsection{Hybrid approach}}
\label{hybrid-approach}

As mentioned earlier, VQE is a hybrid, quantum-classical algorithm. It means that we don't perform all the computations on a quantum computer, we use a classical one too. So what does that look like?

You remember that ansatz has a lot of parameters, right? The problem is that we don't know which parameters are good and which ones are crappy. So, how do we learn that? Well, we treat it as a regular optimization problem.

Usually, we start from a random set of parameters---which corresponds to a random state (keep in mind it's created using the ansatz that we use---so it's not a ``random state chosen from the set of all possible states,'' but a ``random state chosen from the set of all possible states we can get with given ansatz''). We perform measurements so we know the corresponding energy (we also use the term ``expectation value'').

Then, we slightly change the parameters, corresponding to a slightly changed state. After we perform measurements, we should get a different energy value.

If the new energy value is lower, it means that we're moving in a good direction. So, in the next iteration, we will further increase those parameters that have increased and decrease those that have decreased. If the new energy value is higher, we must do the opposite. Hopefully, after repeating this procedure many times, we will get to a place where no matter what changes we make to the parameters, we can't get any better---this is our minimum.

The optimization strategy I described is pretty trivial and it's good only as an example. Choosing the right optimization method for your problem might decide whether you'll succeed and is pretty difficult---there are no silver bullets here. You can choose from a range of methods, like SGD (Stochastic Gradient Descent), BFGS or Nelder-Mead method.

Ok, but this subsection was supposed to be about explaining how the hybrid quantum-classical thing works, right?

So the idea is that we use a quantum computer (sometimes called QPU---quantum processing unit) for one thing only---to get the energy value for a given set of parameters. Everything else---so the whole optimization part---happens on a regular computer.

To make it more tangible here's a (probably incomplete) list.

What a QPU does:

\begin{itemize}
\item Given a set of parameters, returns us a set of measurements.
\end{itemize}

What a classical computer does:

\begin{itemize}
\tightlist
\item Given a set of measurements calculate the energy value (it's usually some kind of averaging)
\item Performs optimization procedure
\item  Calculates what a new set of parameters should be
\item Check whether we have reached the minimum
\item Makes sure we don't make more iterations than we should have done
\end{itemize}

\hypertarget{why-is-this-appropriate-for-nisq-devices}{%
\subsection{Why is this appropriate for NISQ
devices?}\label{why-is-this-appropriate-for-nisq-devices}}

As already mentioned, NISQ stands for ``Noisy Intermediate Scale Quantum'' (John Preskill explains what it exactly means in this \href{https://www.youtube.com/watch?v=h4nUyF9cSaw}{video} or \href{https://arxiv.org/abs/1801.00862}{paper})---so not perfect million-qubit devices that are needed to run the famous Shor's or Grover's algorithms, but rather several-qubit prototypes.

Some of the main disadvantages of NISQ devices are:

\begin{itemize}
\tightlist
\item There's a lot of noise---e.g.~every time you apply a gate, there's some substantial chance (e.g.~0.1\%) that you'll actually apply a different gate. And this is just one of many flavors of noise you can get.
\item The qubits don't live for too long---after some time you can be pretty confident that whatever information was stored in your system, it's lost. So the number of operations you can perform in this time is limited.
\item You don't have too many qubits---right now (late 2019) the biggest chips have around 50. Also, there might be some restrictions on which qubits can interact with each other.
\end{itemize}

This translates to a pretty clear set of restrictions on your quantum
circuit:

\begin{itemize}
\tightlist
\item You can use only N operations per qubit (so called ``circuit depth'')
\item You can't assume everything will go according to plan and there will  be no errors along the way. Even if there are errors, you should still get a reasonable (though not perfect) answer. We call this ``robustness.''
\item Obviously, you can't use more qubits than you have.
\end{itemize}

So why VQE is a sensible proposition for NISQ devices?

\begin{enumerate}
\item It's pretty flexible when it comes to the circuit depth, so you can make some tradeoffs. You can use a pretty simple ansatz if you want to keep your circuit very shallow. But once you have a machine that allows you for a deeper circuit---you can use more sophisticated ansatzes and your performance might increase as a result.
\item VQE has been shown to be somewhat resistant to noise---i.e.~even in the presence of noise (up to some levels), it gives you the correct results. Some even claim that a pinch of noise might help with the optimization process because it helps to escape local minima.
\item There are some practical problems that do not require a lot of qubits. One example is nitrogen fixation---an important industrial process (Honestly, it's not directly about VQE, but you can read the details in \href{http://arxiv.org/abs/1605.03590}{this paper}).
\end{enumerate}

\hypertarget{example}{%
\subsection{Code examples}\label{example}}

If you would like to know how the VQE works,
\href{https://github.com/mstechly/mustythoughts_plus/blob/master/VQE_QAOA/VQE_explained_example.ipynb}{here is a link to a jupyter notebook with an example in pyQuil}.

It's not sophisticated, but it follows exactly the steps you have seen here already. I encourage you to play with it, see how it works, try to improve it. Write your own, proper version of VQE and check if it gives you the same results as different open source implementations.

And if you do---make sure you let me know!

\href{https://twitter.com/davit_khach}{Davit Khachatryan} has written an excellent tutorial for VQE---it's much more detailed than my example and also gives an example of a two-qubit
circuit. You can find it {here}. I strongly recommend it!

\begin{figure}[H] 
\centering
\includegraphics{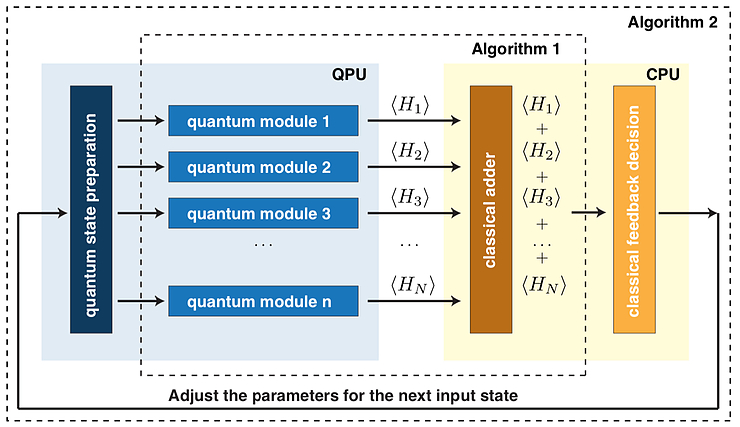}
\caption{Diagram depicting the VQE algorithm, from \href{https://arxiv.org/pdf/1304.3061.pdf}{the original VQE paper}.}
\label{fig:original_vqe}
\end{figure}

Also, I think now you have finally all the pieces to appreciate the following picture  from \href{https://arxiv.org/pdf/1304.3061.pdf}{the original VQE paper} (Fig. \ref{fig:original_vqe}). It shows the overview of the whole algorithm and I think it might be helpful in understanding how all different elements come together.

You can read ``Quantum state preparation'' as ``ansatz'' and ``quantum module'' as ``adding rotations to enable measurements on the right basis.''

\newpage

\hypertarget{section-3}{\section{Quantum Approximate Optimization Algorithm Explained}}

In this section, I will describe QAOA (Quantum Approximate Optimization Algorithm)---what's the motivation behind it, how it works, and what it's good for.

One of the reasons why QAOA looks so promising is that it can help us solve combinatorial optimization problems better. Why should anyone care? Let's see!

\hypertarget{combinatorial-optimization-problems}{%
\subsection{Combinatorial optimization
problems}\label{combinatorial-optimization-problems}}

\hypertarget{evil-imps-sending-notes-in-the-classroom}{%
\subsubsection{Evil imps sending notes in the
classroom}\label{evil-imps-sending-notes-in-the-classroom}}

Imagine a classroom where students are sending notes between each other. The teacher is annoyed by this and thanks to a meticulous invigilation knows exactly how many notes pass between each pair of students. The class is moving to a new room, where there is a huge gap in the middle, preventing students from sending notes to the other side. How should the teacher divided the class into two groups in order to minimize the number of notes being sent?

Or another example---in a far away kingdom, there were multiple merchants who all traded with each other. However, an evil imp wanted to impair the trade in this kingdom, so he decided to create an intrigue which would separate the merchants into two opposite factions, not trading between each other. The question was---who should be in which faction to maximize the losses?

What do both these situations have in common? They both can be described as a MaxCut problem---given a graph, find a way to divide it into two groups, such that the edges going between the two groups have the biggest possible weight.

In the first example, the nodes of the graph represent the students and the connections represent how many notes a given pair of students exchange. In the second example, the nodes are the merchants and the connections are the number of ducats, drachmas or talents their mutual trade is worth.

Let's see what this graph might look like:

\begin{figure}[H]
\centering
\includegraphics[width=6cm]{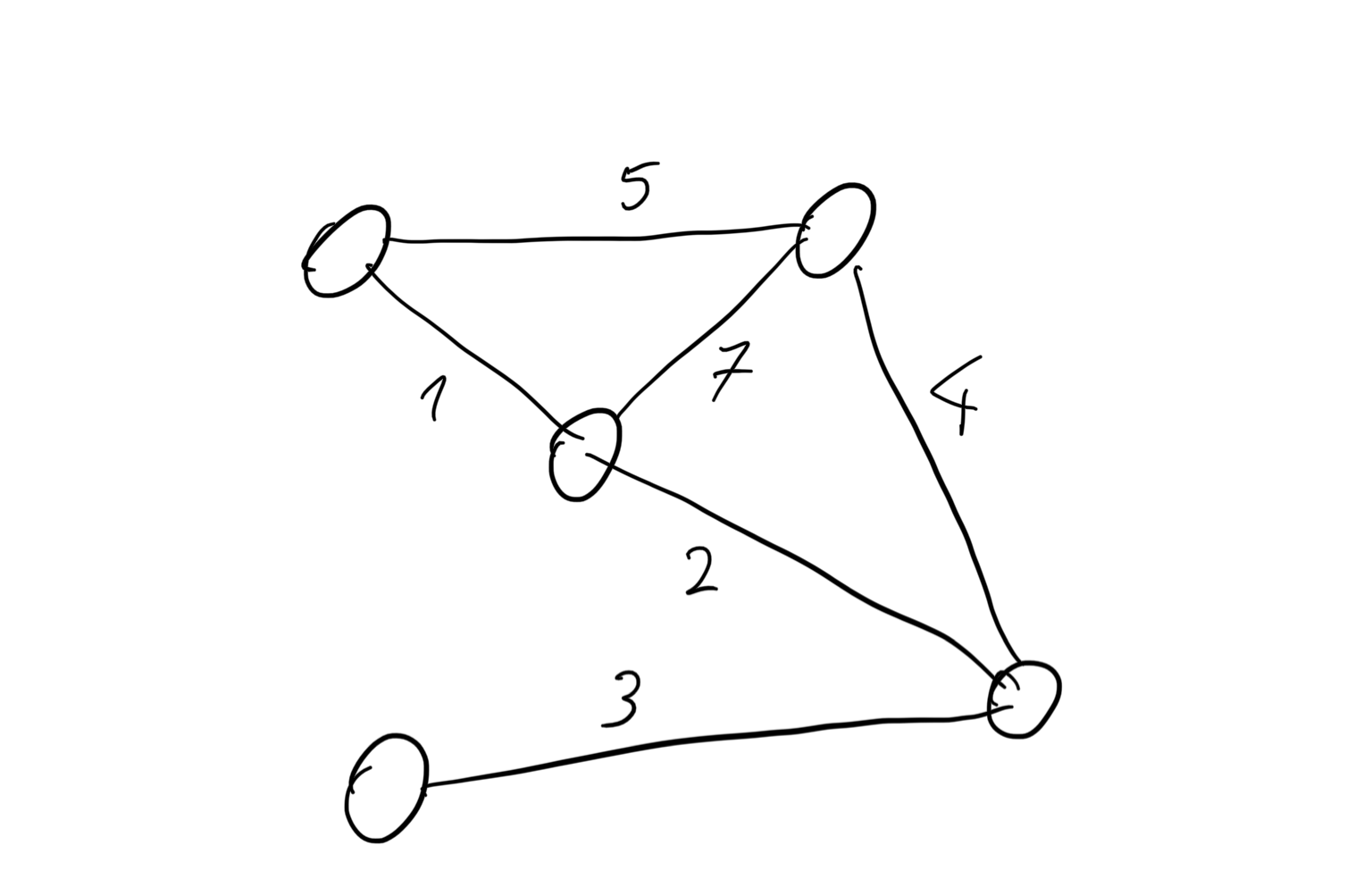}
\caption{An example weighted graph depicting a maxcut problem}
\label{fig:students_graph}
\end{figure}

and what's the solution of this problem:

\begin{figure}[H]
\centering
\includegraphics[width=6cm]{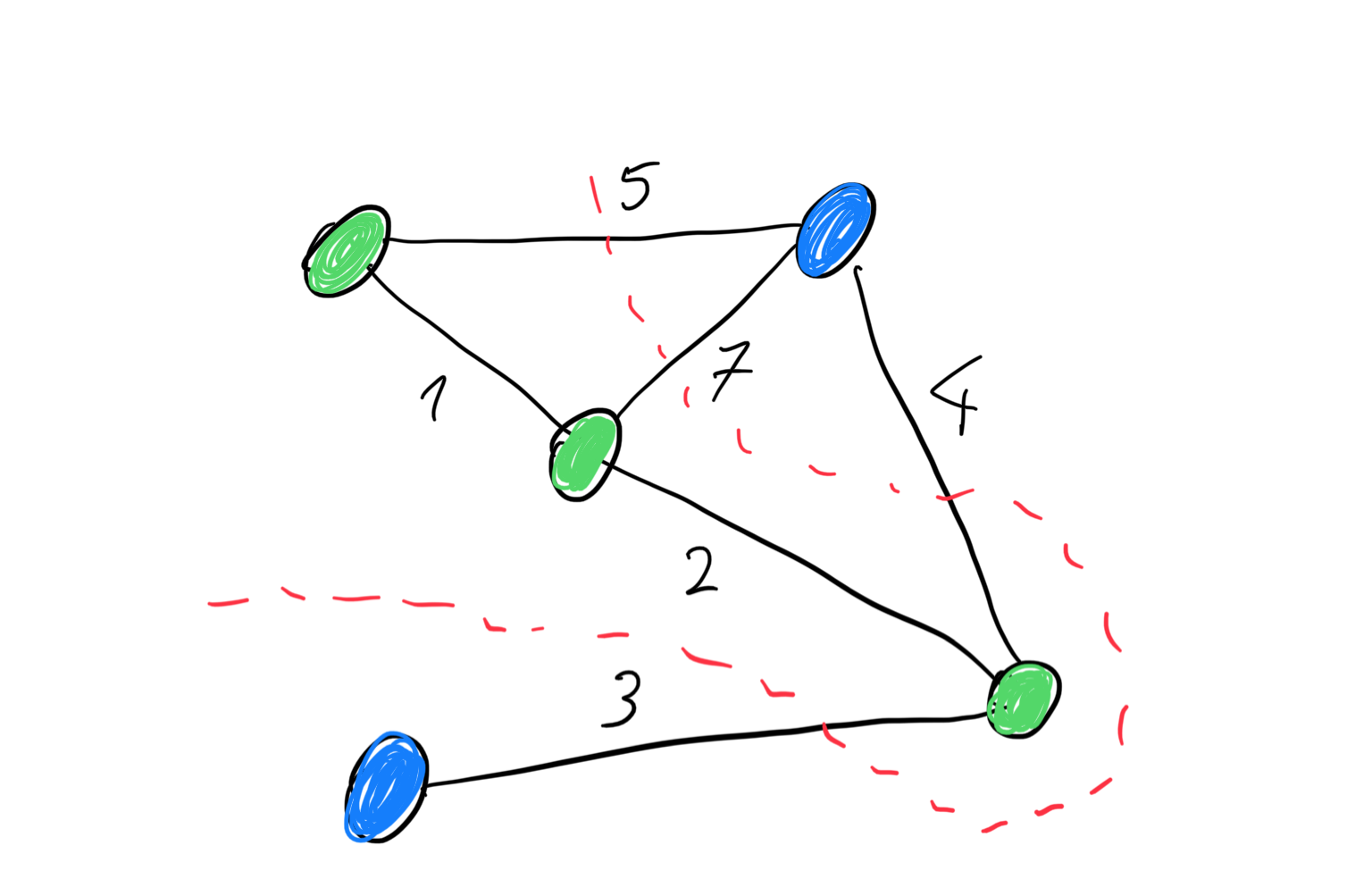}
\caption{The red line depicts the ``maximum cut,'' and vertices are colored according to what side of the cut they are on.}
\label{fig:merchants_graph}
\end{figure}

\hypertarget{why-maxcut}{%
\subsubsection{Why MaxCut?}\label{why-maxcut}}

A small digression here: if you've seen your portion of graph and optimization problems, MaxCut might not be the most obvious example. However, we use it here for a few reasons:
\begin{enumerate}
    \item MaxCut is the problem that the authors of \href{https://arxiv.org/abs/1411.4028}{the original QAOA paper} solved with this algorithm
    \item It is the most well-studied example of QAOA application in the literature (probably because of 1).
    \item Solution to MaxCut is a binary string (so just a string of 0s and 1s), and there is no problem with encoding the results.
    \item As far as I can tell, it has some neat mathematical properties (especially for k-regular graphs) which make it easier to analyze. If you're interested, you can find more answers in \href{https://arxiv.org/abs/1411.4028}{the original paper}, as well as other papers by the same and these two talks: \href{https://youtu.be/J8y0VhnISi8}{talk 1} and \href{https://youtu.be/wJwsLkHWYMA}{talk 2}.
\end{enumerate}

\hypertarget{why-combinatorial-optimization}{%
\subsubsection{Why combinatorial
optimization?}\label{why-combinatorial-optimization}}

As you might have already guessed, MaxCut is an example of a
combinatorial optimization problem (I'll abbreviate it to COP). This is a class of problems where we want to find the best solution from a finite set of solutions---it usually involves graphs or ordering some objects. Some well-known examples are the Traveling Salesman Problem or the Knapsack Problem.

The first problem with combinatorial optimization is that it is, well, combinatorial. This usually means that the number of all possible solutions grows extremely fast with the size of the problem---so called ``exponential growth'' or ``combinatorial explosion.'' Exponential means that the number of possible solutions typically depends on $a^n$ ($a$ is a constant depending on a problem and $n$ is the problem size). Combinatorial means that $n!$ is involved---which is even worse than exponential.

That's not a problem in and of itself---in the case of continuous optimization we have to deal with an infinite number of possible solutions (since we use real numbers) and somehow we have efficient methods of doing that. The fact that we need to deal with discrete values actually makes things harder---or at least different. Many continuous functions we have to deal with in optimization are smooth---i.e. we change $x$ just a little, and the $f(x)$ also changes just a little, which is really helpful. For discrete functions, if we change only one bit of the solution from 0 to 1, we might get dramatically different results.

It's also worth noting that many such problems are NP-hard---for those unfamiliar with this term it basically means that there are no efficient methods for solving exactly big instances of a given problem. If you're interested in details you can read on \href{https://en.wikipedia.org/wiki/NP-hardness}{Wikipedia}, watch \href{https://www.youtube.com/watch?v=YX40hbAHx3s}{this video}, or read \href{https://www.goodreads.com/book/show/400716.Introduction_to_the_Theory_of_Computation?from_search=true\&from_srp=true\&qid=riKxp0V42Y\&rank=1}{this book}.

Since these problems are so difficult to solve, why do we bother? Because they are applicable in so many places that I won't even try to list them here---apparently there is even
\href{https://www.goodreads.com/book/show/5248928-industrial-applications-of-combinatorial-optimization?from_search=true\&from_srp=true\&qid=xxJYpGVqpr\&rank=1}{a book} just listing out where you can apply COPs.

\hypertarget{qaoa-introduction}{%
\subsection{QAOA---introduction}\label{qaoa-introduction}}

QAOA is an algorithm which draws from various concepts. We'll start with a description of QAOA and then we'll go one by one through all the pieces needed to actually understand why it works.

Let's say we have a cost Hamiltonian $H_C$ representing our problem (we will see how to create one in the section about Ising models). We can also construct an operator:

\begin{equation}U(H_C, \gamma) = e^{-i \gamma H_C}\end{equation}

Then, we can take a Pauli X matrix: $H_B = \sum_{j=0}^n \sigma_j^x \sigma_j^x$ and construct an operator from it in the same way:

\begin{equation}U(H_B, \beta) = e^{-i \beta H_B}\end{equation}

In QAOA we construct the state

\begin{equation} | \gamma, \beta \rangle = U(H_B, \beta_p) U(H_C, \gamma_p) … U(H_B, \beta_1) U(H_C, \gamma_1) | s \rangle \end{equation}

where $p$ is usually called ``number of steps'' and denotes just how many times do we repeat applying $U(H_B, \beta) U(H_C, \gamma)$.
$|s\rangle$ is the initial state, usually $|0...0 \rangle$ or
$H |0...0 \rangle = |+...+ \rangle$. We often call $\beta$ and
$\gamma$ ``angles.''

In order to find what are the values of angles that produce a reasonable state, we do exactly the same thing as we did with VQE in \hyperlink{section-02}{chapter 1}. We start from some initial parameters, measure the state and update the angles to get a state closer to our solution in the next iteration.

The first time I looked at these equations I was like: ``Ok, let's implement it and see how it works!'' But the more I've been learning, the more I started to ask questions such as: ``What do we need $H_B$
for?'', ``Why does $H_B$ have this exact form?'', ``Why do we use these funny $e^{-iH}$ operators?'' and, predominantly, ``Why should it even work at all?''

Let's find out the answers in the following paragraphs!

\hypertarget{background-aqc}{%
\subsection{Background: Adiabatic Quantum Computation
(AQC)}\label{background-adiabatic-quantum-computation-aqc}}

\hypertarget{whats-aqc}{%
\subsubsection{What's AQC?}\label{whats-aqc}}

Let's say we have a quantum system described with a simple Hamiltonian
$H_S$ with a known ground state. We also have another Hamiltonian
$H_C$, whose ground states we want to learn. We can now construct a new Hamiltonian in the form of:

\begin{equation}
    H(\alpha) = (1 - \alpha) H_S + \alpha H_C
    \label{eq:aqc}
\end{equation}

For $\alpha=0$ our system is described with $H_S$, for $\alpha=1$ by $H_C$.

The adiabatic theorem states that if we start in the ground state of
$H_S$ and very slowly start to increase $\alpha$ up to 1, then throughout the process the system will always stay in the ground state of $H(\alpha)$---which means that at the end it will be in the ground state of $H_C$
It's also worth mentioning that $\alpha$
could be an actual physical quantity.

Of course, that's not an easy thing to do and certain conditions apply, but in principle, if the evolution is slow enough, we can find a solution to any problem if we can encode its solution in the ground
state of $H_C$. We use the word ``evolution'' for the process of
changing the quantum system by changing $\alpha$. Also, we often use
$\alpha=\frac{t}{T}$, where $t$ is current time and $T$ is total or end time.

\hypertarget{ising-models}{%
\subsubsection{Ising models}\label{ising-models}}

There is one big assumption in the previous paragraph, which could easily ruin any potential usefulness of this method: ``if we can encode its solution in the ground state of $H_C$.'' Well, if we don't know how to encode the problem, then AQC won't help us. Fortunately, we know such encodings for many combinatorial optimization problems and we use ``Ising models'' to do that.

Imagine a chain of particles. Each particle has a spin which is either ``up'' or ``down'' (or $+1$ and $-1$) and it wants to have a different spin than its neighbours. We can write down the Hamiltonian for such system:

\begin{equation}
    H(\sigma) = - \sum_{<i, j>} J_{ij} \sigma_i \sigma_j
\end{equation}

Here $\sigma_i$ is the spin of the i-th particle and by $\sigma$ we denote a string of all the spins---e.g. $[-, -, +, + -, \ldots{} ]$---and $J_{ij}$ is the strength of the interaction between the particles. This mathematical description actually works equally well with lattices instead of chains. Also, we can change the behaviour of the system by changing the sign of $J_{ij}$---if it's negative it behaves as described, but if it's positive, then the particles will try to have the same spin direction.

This model maps very well on quantum computers---we just need to change $
\sigma_i$ to $\sigma_i^z$ (pauli Z operator acting on $i$-th qubit).

It turns out that we can encode many different COPs using Ising models---I won't describe how to do that here, but you can find many examples in \href{https://arxiv.org/pdf/1302.5843.pdf}{this paper}. For our discussion, it suffices to know that it is doable and that we can do that using only Pauli Z operators.

\hypertarget{problems-with-aqc}{%
\subsubsection{Problems with AQC}\label{problems-with-aqc}}

Ok, so we can encode most problems of interest into Hamiltonians and we have a method that basically guarantees that we find a proper solution.
So what's the catch?

Well, there are a couple of them:

\begin{itemize}
    \tightlist
    \item In order to work AQC requires to be very well isolated from the outside world---which is really hard to do in practice.
    \item It might require a very long time to run.
    \item You won't be able to run it on a gate-based quantum computer. Well, while it is possible in principle, it would require much better hardware than we have today and doesn't seem to be practical in any foreseeable future
\end{itemize}

Basically, the harder your problem is, the more isolated your quantum system must be, and the longer it has to run.

\hypertarget{relation-to-quantum-annealing}{%
\subsubsection{Relation to quantum
annealing}\label{relation-to-quantum-annealing}}

This seems like a good place to mention quantum annealing (QA)---the paradigm that Canadian company D-Wave is following. The main idea behind QA is the same as for AQC---start from a ground state of an easy Hamiltonian and slowly evolve your state towards the ground state of the interesting Hamiltonian. What's the difference? QA is an imperfect implementation of AQC which trades some of AQC's power for easier implementation. Although not everybody might agree on this since the line between what the community calls AQC versus QA is very blurry, basically we can say that QA is a heuristic algorithm that:

\begin{enumerate}
    \tightlist
    \item Can be more flexible when it comes to the rate of change between two Hamiltonians (finite practical time which is on the order of microseconds in current devices) and
    \item Operates at finite temperature.
\end{enumerate}

This comparison leaves a lot of subtleties out, but I think it's a good first approximation to start thinking about it. If someone would like to learn more, these two publications might shed some light: \href{https://arxiv.org/abs/1901.01903}{Comparison of QAOA with QA and SA} and \href{https://www.goodreads.com/book/show/23092024-adiabatic-quantum-computation-and-quantum-annealing?from_search=true\&from_srp=true\&qid=buAmNDqPcP\&rank=1}{AQC and QA: Theory and Practice}.

\hypertarget{background-quantum-mechanics}{%
\subsection{Background: Quantum
Mechanics}\label{background-quantum-mechanics}}

In this section, I'd like to review two ideas that come directly from basic quantum mechanics---time evolution of a quantum system and trotterization.

\hypertarget{time-evolution}{%
\subsubsection{Time evolution}\label{time-evolution}}

There is an operation which is quite common in quantum computing---taking an operator $A$ and creating another of the form
$U_A = e^{-itA}$
But why would you do that? Well, it comes from the Schrödinger equation:

\begin{equation} i \hbar \frac{d}{dt} | \Psi(t) \rangle = H | \Psi(t) \rangle \end{equation}

What it says is that the change (in time) of the state of a quantum system described by a time-independent Hamiltonian $H$ depends solely on the form of this Hamiltonian.

And if we solve the equation we see that this dependency is:

\begin{equation}| \Psi(t) \rangle = e^{-iHt} | \Psi(0) \rangle \end{equation}

What does this tell us? That if we take some state, $ | \Psi(0) \rangle$ and act on it (evolve it) with a Hamiltonian $H$ for a period of time $T$, we will get a state $ | \Psi(T)\rangle = e^{-iHT} | \Psi(0) \rangle$.

If this sounds abstract, imagine an electron orbiting the proton (hydrogen atom again!). This system is defined solely by the interaction between these two particles and their initial states and we know how to describe the interaction in the form of a Hamiltonian. So by solving the Schrödinger equation we are able to get a formula describing the state of the system after time T. And if we do the math, we know it will look like this:

\begin{equation} | \Psi(T) \rangle = e^{-iHT} | \Psi(0) \rangle\end{equation}

The key point to remember? Whenever you see an operator which looks like this $e^{-iHt}$
you can understand it as an evolution of the quantum system described by the Hamiltonian $H$.

Until this point we've made one silent assumption: we've assumed that the Hamiltonian we wanted to solve was time-independent. But when we take a look at the Hamiltonian we used for AQC (equation \ref{eq:aqc}) and change $\alpha$ to $t$, we cannot pretend it's time-independent. Depending on how complicated this time-dependency is, this might complicate the solution we've presented a little bit or a lot. In the latter case, the problem might be impossible to solve, so it's reasonable to make some approximations.

\hypertarget{trotterization}{%
\subsubsection{Trotterization}\label{trotterization}}

Another concept important for understanding QAOA is ``Trotterization.'' Most of the time we want to find the ground state of a Hamiltonian which is too complicated for us to deal with directly---then it might be useful to approximate it. An example can be (surprise), a time-dependent Hamiltonian.

\begin{figure}
\centering
\includegraphics[width=6cm]{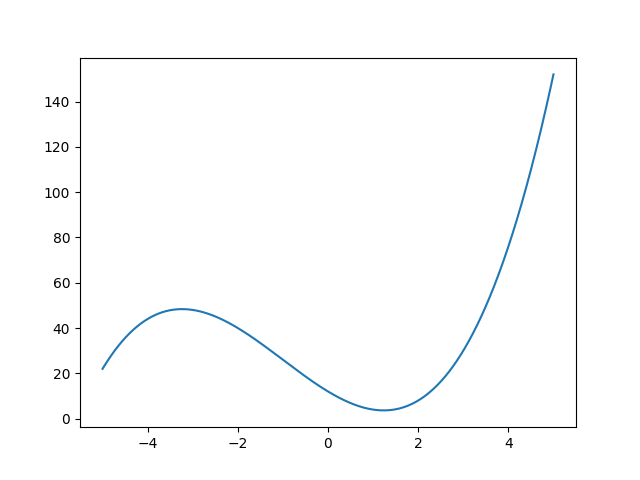}
\caption{A curve described by a function $x^3 + 3(x-2)^2$}
\label{fig:curve}
\end{figure}

To help us grasp the idea we'll get back to the ``classical world'': imagine a curve. Let's say this curve is described by some function. (See Fig. \ref{fig:curve})

\begin{figure}
\centering
\includegraphics[width=12cm]{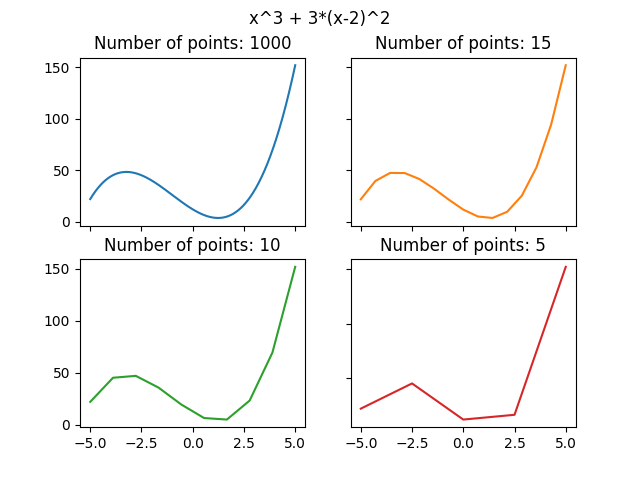}
\caption{Piecewise linear approximation of the curve in Fig \ref{fig:curve}.}
\label{fig:curve_piecewise}
\end{figure}

We can approximate it by a \href{https://en.wikipedia.org/wiki/Piecewise_linear_function}{piecewise linear function}. The more ``pieces'' we use, the better approximation we get. (See Fig. \ref{fig:curve_piecewise})

We can do the same with the time evolution of a quantum system for an operator $U(t)$. $U(t)$ is actually a shorthand of $U(t, t_0)$,which means that we act in time $t$ starting from $t_0$
We can write: (\href{https://ocw.mit.edu/courses/nuclear-engineering/22-51-quantum-theory-of-radiation-interactions-fall-2012/lecture-notes/MIT22_51F12_Ch5.pdf}{source})

\begin{equation}U(t, t_0) = U(t_n, t_{n-1}) U(t_{n-1}, t_{n-2}) … U(t_2, t_1), U(t_1, t_0)\end{equation}

The bigger $n$ we use, the better approximation we can get. Keep in mind that the time differences $\Delta t_i = t_{i+1} - t_i$ don't need to be equal.

Another approximation, particularly useful in quantum mechanics, is called the ``Suzuki-Trotter expansion'' or ``Trotterization'' (it is somewhat similar to other expansions we use in mathematics, like a Taylor series expansion. If you're not familiar with this concept, I recommend watching \href{https://www.youtube.com/watch?v=3d6DsjIBzJ4}{this video by 3Blue1Brown}).

If our Hamiltonian has the following form $e^{A+B}$
we can use the following relation:

\begin{equation}
e^{A+B} = lim_{n \rightarrow \infty} (e^{\frac{A}{n}} e^{\frac{B}{n}} )^n
\end{equation}

We can look at it as approximating time evolution by $A+B$ by applying alternatively $A$ and $B$ for time intervals $\frac{1}{n}$.

Now we have all the knowledge we need to understand what's the motivation behind QAOA.

\hypertarget{qaoa-revisited}{%
\subsection{QAOA revisited}\label{qaoa-revisited}}

Now that we have some more context, let's look at QAOA again. QAOA creates the state:

\begin{equation}
    | \gamma, \beta \rangle = U(H_B, \beta_p) U(H_C, \gamma_p) … U(H_B, \beta_1) U(H_C, \gamma_1) | s \rangle
    \label{eq:qaoa-state}
\end{equation}

where
\begin{enumerate}
    \tightlist
    \item $U(H_B, \beta) = e^{-i \beta H_B}$,
    \item $U(H_C, \gamma) = e^{-i \gamma H_C}$, and
    \item $|s\rangle$ is the starting state (we defined it before). 
\end{enumerate}

First, we can now see that the operators
\begin{equation}
\begin{aligned}
U_B &= U(H_B, \beta), \quad \text{and} \\
U_C &= U(H_C, \gamma)
\end{aligned}
\end{equation}
correspond to evolving the state with the Hamiltonian $H_B$ and $H_C$ for time $\beta$ and $\gamma$.

Second, we can also see that having several layers of $U_B U_C$ looks a lot like a trotterization of $e^{(H_B+H_C)}$.
It's not ``pure'' trotterization, since we have different $\beta$ and $\gamma$ for each step, so it resembles the ``piecewise approximation.'' But approximation of what? 

Third, by combining the two previous points we can see that QAOA (somewhat) mimics adiabatic evolution from the ground state of $H_B$
to the ground state of $H_C$
which encodes our solution (remember Ising models?). So answering the question in the previous paragraph---QAOA could be (given right choice of parameters) viewed as the
approximation of AQC.

This still leaves a couple of important questions open:
\begin{itemize}
    \tightlist
    \item What's the relation between QAOA and AQC?
    \item Why should QAOA work?
    \item Why do we actually need $H_B$?
\end{itemize}

\hypertarget{relation-to-aqc}{%
\subsection{Relation to AQC}\label{relation-to-aqc}}

QAOA is not a perfect analogy to AQC. While there are a lot of similarities, there are also significant differences.

First of all, the goal is different. In AQC our goal is to get the ground state of the Hamiltonian. In QAOA (Quantum \textbf{Approximate}
Optimization Algorithm) our goal is not really to get the exact ground state. It is to get a state which has good enough overlap with the ground state, so that we get an approximate solution.

Second of all, we could choose our angles in such ways that the evolution of a state using QAOA would mimic the evolution using AQC. But we don't do that---in QAOA we can choose angles arbitrarily (usually using some classical optimization strategy).

And lastly, there's something that I struggled a lot to understand (honestly, I still struggle with it ;) ). In AQC we start from $H_B$
and slowly move towards $H_C$. In QAOA we actually frantically alternate between $H_B$ and $H_C$---more on that later.

\hypertarget{why-should-it-work}{%
\subsubsection{Why should it work?}\label{why-should-it-work}}

In principle, if you can get an infinite number of steps and choose angles that exactly mimic the adiabatic path, you will get the right results. It reminds me of the \href{https://en.wikipedia.org/wiki/Universal_approximation_theorem}{universal approximation theorem} for neural networks. UAT says that if you have a one-layer neural network that is big enough, it can approximate any function, so basically solves any problem. The problem is with the ``big enough'' part---the requirements are astronomical for relatively simple problems. That's why we usually don't use gargantuan one-layer networks, but rather just stack smaller layers on top of each other. With QAOA the case is similar---in principle, we could just keep increasing the number of steps and stay close to the parameters for the adiabatic path, but in practice, we choose not to. We try to solve the problem with a smaller number of steps by finding the right set of angles.

So we have some intuitions as to why it should work, but to be honest, we don't know if it will work (at least without infinite resources)---it's an important open question. Some of the researchers (me ;) ) are optimistic, because as it was with neural networks, they had very limited practical applications until some improvements made them an extremely useful and widely used tool. But we still don't know whether this is the case with QAOA---solving a toy problem with an algorithm is different from solving a real problem with it.

Actually, there's a quote illustrating it from a recent paper by two of the three authors of the \href{https://arxiv.org/abs/1411.4028}{original QAOA paper}:

\begin{quote}
No one knows if a quantum computer running a quantum algorithm will be able to outperform a classical computer on a combinatorial search problem.\footnote{This feels even more true in 2024!}
\end{quote}

\hypertarget{why-is-h_b-important}{%
\subsubsection{\texorpdfstring{Why is $H_B$
important?}{Why is H\_B important?}}\label{why-is-h_b-important}}

$H_C$ is easy---it corresponds to the problem that we want to solve and hence we know why it looks the way it looks. But what about $H_B$?
Why do we choose it to be just $\sum_i^N \sigma_i^x$ and not something else? Why do we even need it at all?

First, it doesn't need to be just $\sum_i^N \sigma_i^x$. We want it to be something that does not commute with $H_C$ and this choice of
$H_B$ meets this requirement and is super easy to implement (after the exponentiation we get a layer of $R_x$ gates). \href{https://physics.stackexchange.com/questions/9194/what-is-the-physical-meaning-of-commutation-of-two-operators}{This discussion} might help you better understand why the commutation of operators is significant. Another way to look at it is that the choice of
$H_B$changes the expressibility of our ansatz---it allows us to reach states that we would otherwise never reach (remember the robot arm from \hyperlink{what-the-hell-is-an-ansatz}{the VQE section earlier}?).

Second, why do we need it at all? One part of the answer is that it is just a part of the setup that came from AQC and the way we apply it comes from Trotterization. Unfortunately, it doesn't give us a lot of intuition or reason why not to drop it altogether. For me, it's easier to understand this problem if we look at it from the perspective of not having $H_B$ at all. So what would happen in such a case?

We would be just repetitively applying $U_C$. But once we got into a state which is the eigenstate of $H_C$ we wouldn't get any further. This is basic linear algebra---if we apply an operator to its eigenvector, it can change its length, but not direction. The same applies if we had a $H_B$ which commutes with $H_C$. So we need this intermediate step of applying $H_B$ which allows us to escape from the local minimum. How do we make sure we escape it? Well, that's where the classical optimization loop is useful---we try to find the right values of the parameters $\beta$ and $\gamma$ which make it happen.

I came up with a metaphor which might be helpful to get some more intuition. Imagine you're in a forest full of traps and you want to find a treasure. $H_C$ allows you to feel the treasure and move towards it,
while $H_B$ provides you with a set of rules on how to move to avoid
the traps. Without $H_C$ you have no idea which direction to go and
without $H_B$ you cannot make any moves. This metaphor definitely has
some limitations, but should work for starters :)

\hypertarget{what-does-the-circuit-look-like-qaoa}{%
\subsection{What does the circuit look
like?}\label{what-does-the-circuit-look-like-qaoa}}

\begin{figure}
\centering
\includegraphics[width=6cm]{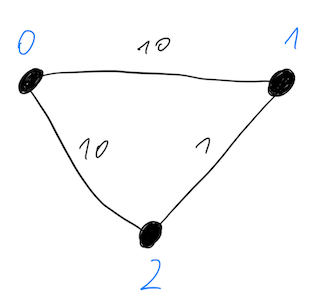}
\caption{A weighted triangular graph, with two edges of weight 10 and one edge of weight 1.}
\label{fig:triangular-graph}
\end{figure}

Ok, so let's take a peek at how it looks in practice. The simple problem that we want to solve is Maxcut on a triangular graph (see Fig. \ref{fig:triangular-graph}). Two edges have a weight of 10 and one weight has a weight of 1.

The correct solution to this problem is a cut which divides this graph into two groups: \{0\} and \{1, 2\}. We can encode solution into a binary string where 0 and 1 indicates in which group given node is, for example:

\begin{itemize}
    \tightlist
    \item 001: nodes 1 \& 2 are in group 0, node 0 is in group 1
    \item 110: nodes 1 \& 2 are in group 1, node 0 is in group 0 
\end{itemize}

As you can see, we use representation where the rightmost bit is the least significant.

The cost function for MaxCut is the sum of costs of all the individual edges. Each individual cost is equal to:

\begin{equation}
    C_{ij} = \frac{1}{2} w_{ij} (1 - z_i z_j)
\end{equation}

where $z_i=1$ if
$i-th$ node is in group 0 and $z_i=-1$ if it's in group 1. So if they're both in the same group $C_{ij}$ is equal to 0 and it's 1 otherwise. I know such representation might not make a lot of sense at first glance, but once you start solving this problem you will see it's reasonable. Now we can translate this into Hamiltonian form:

\begin{equation}
    \frac{1}{2} w_{ij} (1 - \sigma_i^z \sigma_j^z)
\end{equation}

So in case of our graph, the Hamiltonian of interest is equal to:

\begin{equation} H_C = \frac{1}{2} \cdot 10 \cdot (1 - \sigma_0^z \sigma_1^z) +  \frac{1}{2} \cdot 10 \cdot (1 - \sigma_0^z \sigma_2^z) +  \frac{1}{2} (1 - \sigma_1^z \sigma_2^z) \end{equation}

Then we can take care of $H_B$---this is easy, it's just
$\sigma^x_0 + \sigma^x_1 + \sigma^x_2$. Now we need to build $U_C$ and $U_B$ from these and then create a circuit out of that. How?

\begin{figure}
\centering
\includegraphics{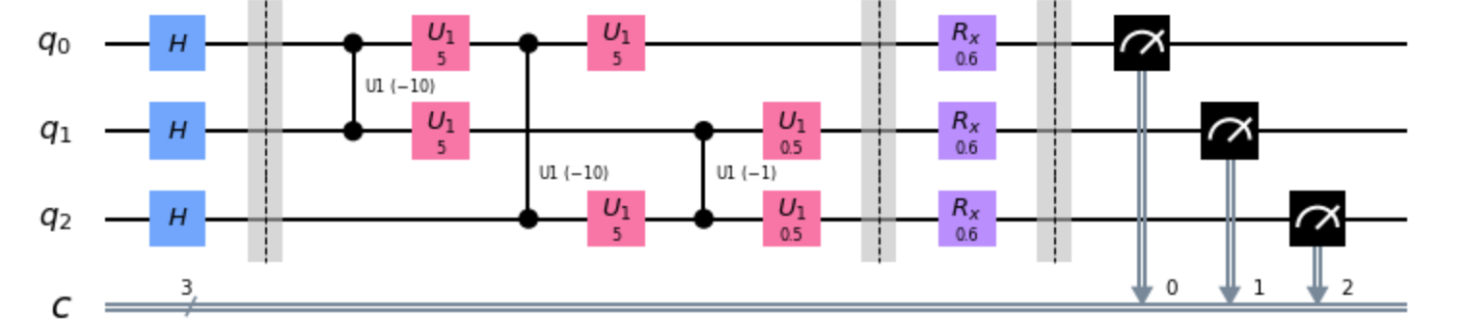}
\caption{Example QAOA circuit for 3 qubits. $U_1 = e^{i \frac{\lambda}{2}}  R_Z(\lambda)$. This is for
$\gamma = 0.5$ and $\beta=0.3$.}
\label{fig:qaoa-circuit}
\end{figure}

Well, describing this procedure is beyond the scope of this blogpost. For those of you who are curious, you can find it in \href{https://arxiv.org/pdf/1001.3855.pdf}{this paper}, section 4. I have also not done this manually, I just used code from \href{https://qiskit.org/textbook/ch-applications/qaoa.html}{a qiskit tutorial}. You can see the result in Fig. \ref{fig:qaoa-circuit}.

This circuit consists of 4 parts: 
\begin{enumerate}
    \item Hadamard gates---this is just for preparing the initial state---i.e.~equal superposition of all states.
    \item $U_C$---in our graph we had 3 edges and we can see three blocks
of gates here. Each consists of one two-qubit controlled $U_1$ and two single qubit $U_1$ gates. As you can see first two are parametrized with $10 \gamma$, since the weights of the corresponding edges were equal to 10.
    \item  $U_B$ ---This is just a layer of $R_X$ gates parametrized with $2 \beta$.
    \item Measurements.
\end{enumerate} 

\hypertarget{why-is-it-appropriate-for-nisq}{%
\subsection{Why is it appropriate for
NISQ?}\label{why-is-it-appropriate-for-nisq}}

If you don't know what the issues with NISQ devices are, I described it already in the \hyperlink{section-02}{chapter 1 about VQE}. So here I'll just point out why QAOA seems like a good algorithm for NISQ:

\begin{enumerate}
\def\labelenumi{\arabic{enumi}.}
\tightlist
\item
 The depth of the algorithm is scalable---the more steps we want to
  have, the more power QAOA has. With most current devices we're limited
  to 1-2 steps (which are of little practical use), but this number will
  get higher with better hardware.
  \href{https://arxiv.org/abs/2004.04197}{Recently Google published
  results} for $p=3$.\footnote{Recently at the time of writing of course.}
\item
  Similarly to VQE, QAOA is somewhat resistant to noise (which is also
  shown in \href{https://arxiv.org/abs/2004.04197}{the same paper from
  Google}).
\item
  Even though I've described only one particular way of creating the
  ansatz, there's actually some flexibility and therefore you might be
  able to tune the algorithm toward specific architecture.
\end{enumerate}

\hypertarget{relation-to-vqe}{%
\subsection{Relation to VQE}\label{relation-to-vqe}}

VQE and QAOA are often mentioned together. VQE is often described as an algorithm good for chemistry calculations and QAOA as good for combinatorial optimization. It's worthwhile to understand the relation between the two.

QAOA can be viewed as a special case of VQE. In the end, the whole circuit we've described above is just an ansatz, parameterizable by the angles. And since the Hamiltonian contains only Z terms, we do not need to change the basis for measurements.

There are, however, a couple of things that make QAOA different: 

\begin{itemize}
    \tightlist
    \item The form of the ansatz is limited to the alternating form we've already seen.
    \item We are restricted in terms of which types of Hamiltonians we can use (i.e.~Ising Hamiltonians). This also means that we don't need to do additional rotations before measurements as we did in VQE.
    \item The last (and I think the most fundamental) difference is the goal of the algorithm. In VQE we want to find the ground state energy and in order to do that we need to reproduce the ground state. In QAOA our goal is to find the solution to the problem. To do that we don't need to find the ground state---we just need to find a state which has a high enough probability of finding the right solution.
\end{itemize}

\hypertarget{closing-notes}{%
\subsection{Closing notes}\label{closing-notes}}

Here are a couple of QAOA tutorials that you might find useful. I can't say I've done them step by step, but I think they will be a good intro into how to use QAOA in practice: 
\begin{itemize}
    \item \href{https://colab.research.google.com/drive/1yFIzqXWDwWe1l1c5ATFhuFcmDQiQUk25}{MaxCut tutorial by Guillame Verdon}
    \item \href{https://lucaman99.github.io/new_blog/2020/mar16.html}{Various graphs tutorial by Jack Ceroni}
    \item \href{https://pennylane.ai/qml/demos/tutorial_qaoa_intro/}{QAOA tutorial in PennyLane, also by Jack Ceroni}
    \item  \href{https://github.com/mstechly/quantum_tsp_tutorials}{My tutorial for Traveling Salesman Problem}
\end{itemize}

\newpage

\hypertarget{section-4}{\section{Variational Quantum Algorithms --- How Do They Work?}}

The first two parts of this document were about specific algorithms---\hyperlink{section-2}{VQE} and \hyperlink{section-3}{QAOA} and in this part, I'd like to explain the general structure as well as the different components of such algorithms.

Before we begin, let's explain what a VQA is: VQA is an algorithm that tries to find the minimum value of some function using: 
\begin{itemize}
    \tightlist
    \item A parameterized quantum circuit as one of the steps for calculating a function's value.
    \item Classical optimization to find the optimal parameters of the circuit.
\end{itemize}

If you want to read a more detailed definition, I recommend \href{https://dash.harvard.edu/bitstream/handle/1/42029810/ROMEROFONTALVO-DISSERTATION-2019.pdf?sequence=1\&isAllowed=y}{Jhonathan Romero Fontalvo's dissertation, section 1.2}.

Also, there are two more things just to set the expectations right: 
\begin{enumerate}
    \item  I've written this assuming running the programs on an actual QPU (Quantum Processing Unit), not a simulator. Accounting for both cases made some of the diagrams harder to read, and it's pretty straightforward to imagine what changes are needed if we run it on a simulator.
    \item This is a popular article explaining the complexity of VQAs. It's not a scientific paper, and some of the choices I made might turn out to not be the most universal or 100\% correct. In such cases, please let me know; I'd love to discuss it.
\end{enumerate}

With that being said, let's get started with the

\hypertarget{basic-setup}{%
\subsection{Basic setup}\label{basic-setup}}

When I first started learning about Variational Quantum Algorithms
(VQAs), I wasn't able to appreciate how complex they are.

In my mind, a quantum algorithm looked like this:

\begin{figure}[H]
\centering
\includegraphics{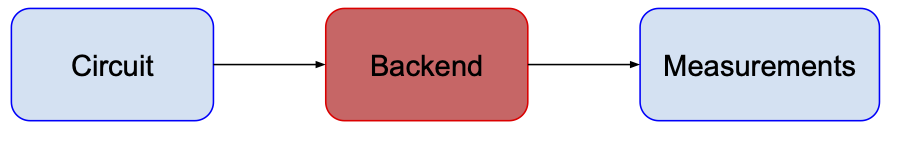}
\caption{Basic diagram of a VQA}
\end{figure}

\emph{Note: ``backend'' is a jargon term people in QC use for different
hardware devices and their interfaces.}

However, when I started learning and thinking more about running such
algorithms on near-term quantum computers, I quickly realized this view
is oversimplified; there are several crucial intermediate steps in a VQA
that must be taken to make the best use of today's devices.

One of the important questions one needs to ask is ``How do I know what
circuit I need to run?'' The choice of the circuit is mainly motivated
by the problem we want to solve and in the case of VQAs, we use
so-called ``Ansatz circuits'' (see
\hyperlink{section-02}{chapter 1} for more details). You can think about it as a circuit template---it has some parameters which correspond (directly or indirectly) to
parameters of gates as well as some ``hyperparameters''---e.g.~the
number of layers, the types of gates for a given layer, etc.

\begin{figure}[H]
\centering
\includegraphics{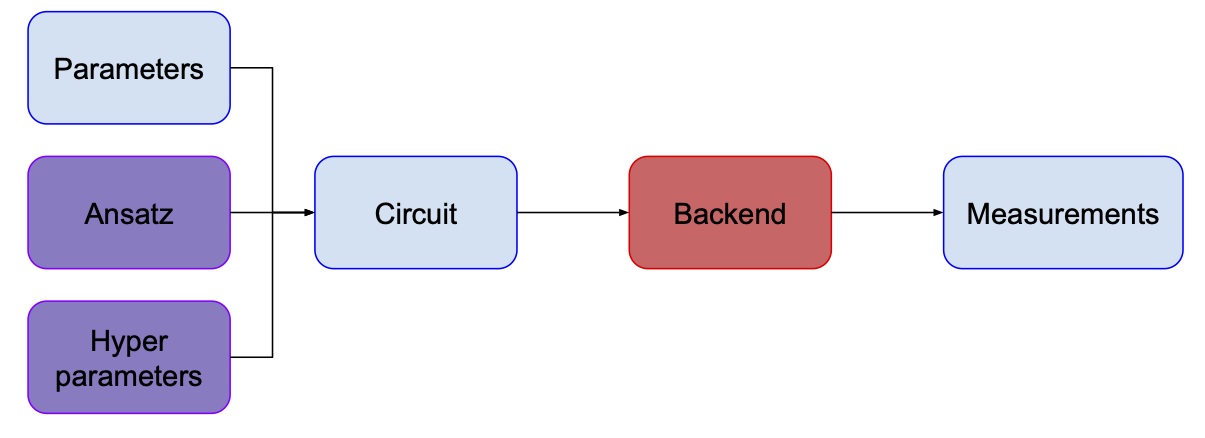}
\caption{Adding an ansatz to our previous diagram}
\end{figure}

As you can see on the diagram, I'm using different colors for different
blocks. Here's what they mean: 
\begin{enumerate}
    \item Blue boxes are objects created/changed when we run the algorithm.
\item Purple boxes are objects which define the problem we're solving and algorithm's hyperparameters.
\item Green boxes are objects which can perform some actions. They change the content of the blue boxes and the way they do it is informed by the purple ones.
\item Backend is red, because it's the actual quantum computer and I wanted it to stand out :)
\end{enumerate}

In addition to the specific problem, the choice or design of the quantum
circuit is determined by the backend/device. Devices we have available
today cannot execute any quantum circuit we can imagine. The two main
reasons for that are:

\begin{itemize}
\tightlist
\item
  Implemented (or ``native'') gate sets---in the classical Boolean
  logic, people sometimes say that the NAND gate is a ``universal''
  gate. This means that you can construct any circuit using only NAND
  gates. In quantum computing, it's slightly more complicated. From what
  I've seen, usually a set of three gates is used to construct a
  universal set, but there are some schemes which involve 2 gates (or as
  it turns out even 1 gate!) (see
  \href{https://en.wikipedia.org/wiki/Quantum_logic_gate\#Universal_quantum_gates}{wiki}
  and \href{https://quantumcomputing.stackexchange.com/a/2015}{here's
  some more discussion} on this topic). There are
  many different gate sets that you use for that purpose that are (at
  least on paper) equally good. But when you start implementing them on
  the real hardware, it turns out that some of these gates are easier to
  realize than others. So usually, when it comes to running a program on
  an actual quantum computer, you have only a handful of gate types
  available.
\item
  Connectivity---in the ideal case, you assume that all the qubits are
  connected to each other. But in reality, this is often not the case---for example that would be infeasible for superconducting chips.
  Imagine having a chip where qubits are arranged in a line. Directly
  performing a CNOT gate on the first and last qubits is not possible,
  because these two qubits are not connected.
\end{itemize}

We can overcome these device-imposed restrictions by introducing an
additional step that translates an ``abstract circuit'' into an
``executable circuit,'' which is called compilation. In the diagram
below, I've hidden the connectivity and available gate set under ``QPU
specs'' box (since it depends on the specific backend we're using I
added a dashed line to indicate causation).

I've also added a new box ``Execution,'' to make the diagram more
consistent and clear.

\begin{figure}[H]
\centering
\includegraphics{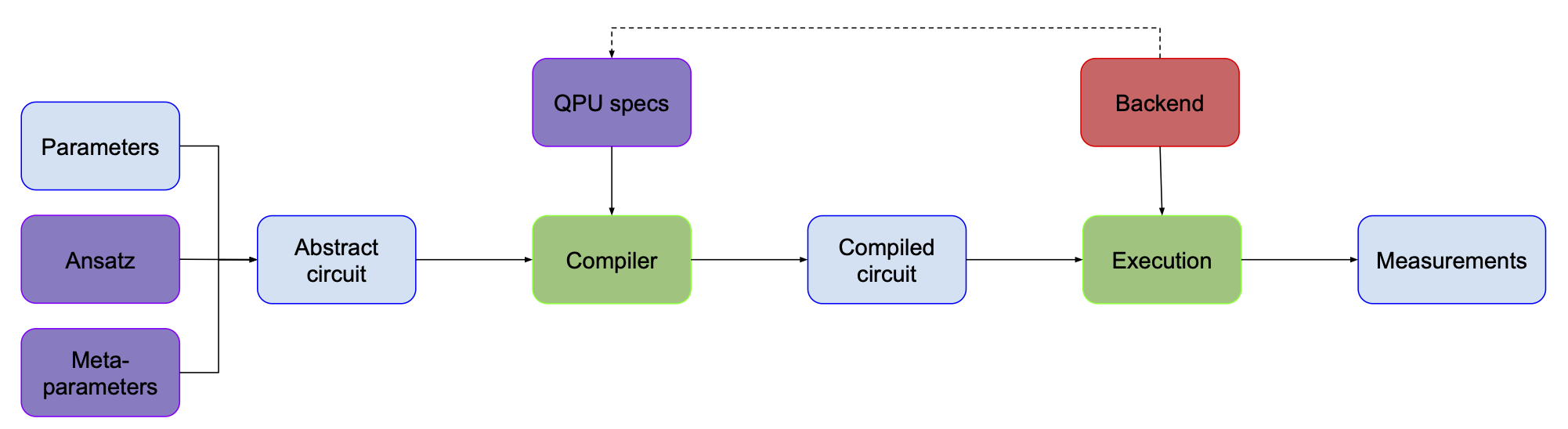}
\caption{Adding compilation to our previous diagram}
\end{figure}

There's one last step from the first picture that we haven't talked
about yet---getting measurements.

This one's trivial, isn't it? You just take a circuit, run it on your
backend and measure the output, right?

Well, not necessarily. There are a couple of things to consider.

First, readout correction. As we mentioned before, quantum computers
nowadays are imperfect machines. There's no reason to think that these
imperfections have spared the measurement process. When you measure your
qubit, various errors can happen. Sometimes it's impossible to say
whether an error occurred, but sometimes we can detect its occurrence
based on how the physical system has behaved. Then we can either discard
such a sample or correct it. It is worth mentioning that measurement
errors can be an order of magnitude higher compared with other types of
errors (see \href{https://youtu.be/FklMpRiTeTA?t=1024}{this
talk}).

It is a good place to mention other types of error mitigation or error
correction. In the most basic form it doesn't need to be used for VQAs,
as they are designed to be somewhat resistant to noise. While they can
benefit from it (see
e.g.~\href{https://arxiv.org/abs/1804.06969}{here}),
I'm leaving it out from our discussion. Adding error correction is what
I would consider an advanced modification of basic VQA, which is outside
of the scope of this article. An additional complication is that
depending on implementation it could live in different places in our
diagram---e.g.~ansatz or compilation.

Second, as you might remember from the VQE article, for a given
Hamiltonian we need to measure each Pauli term separately, i.e.~we need
to run a separate circuit. However, this is the most naïve approach---in reality, there are certain grouping strategies that make it possible
to measure multiple terms simultaneously. This also means that choosing
a certain grouping strategy might influence how the circuit (in
particular its last part) will look like.

Third, there's a very fundamental issue of deciding how many samples
(sometimes also called ``shots'') you want to get. For a given problem
one sample might be too little, but one million might be way more than
necessary. In most cases we just decide on a particular number,
e.g.~10000. But to make things even more complicated, you might want to
use various numbers of shots for different groups or even change it in
different iterations of the optimization loop.

So here's what our picture looks like now:

\begin{figure}[H]
\centering
\includegraphics{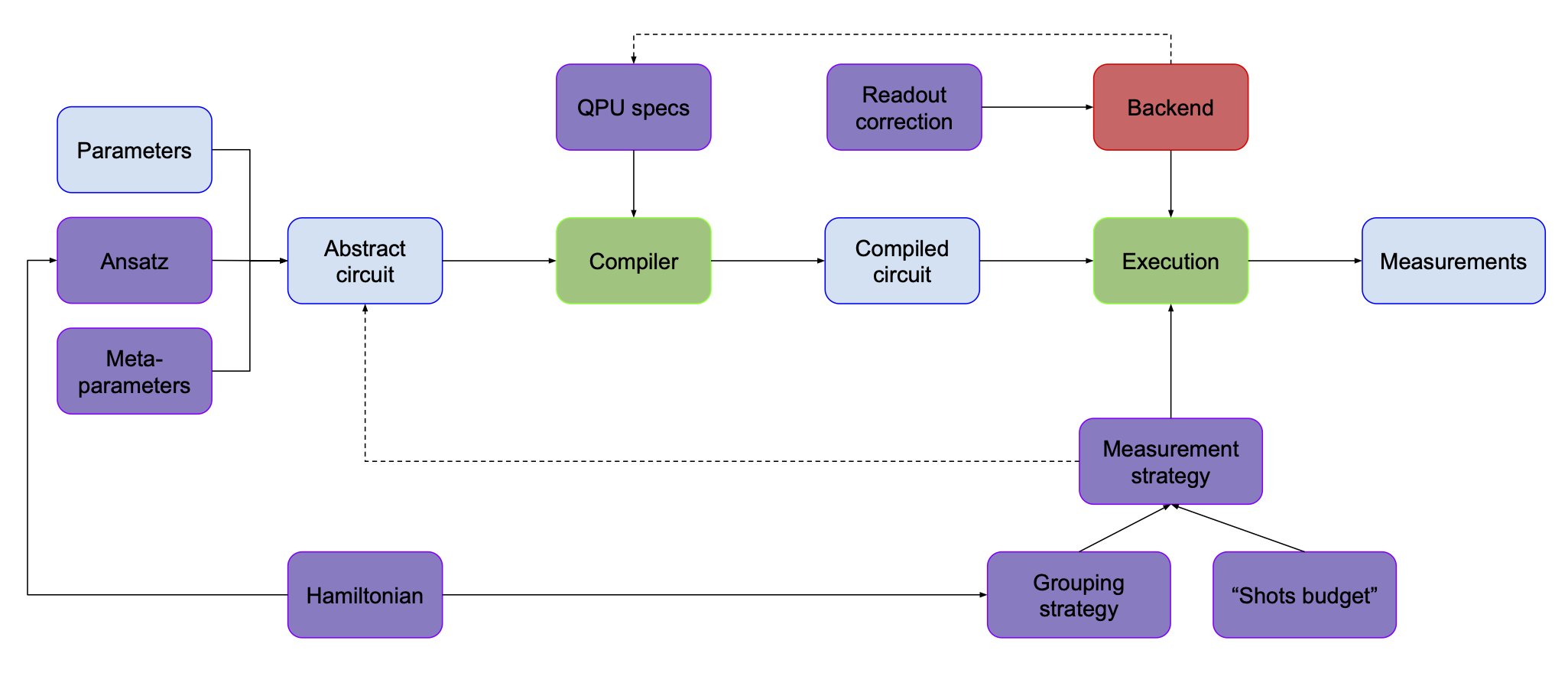}
\caption{Adding measurement to our previous diagram}
\end{figure}

As you can see, I've also added a Hamiltonian box. It might influence
how our ansatz looks like (e.g.~in QAOA) or whether we need to rotate
the state before measuring (in VQE), but it also guides the grouping
strategy.

So we're getting the measurements at the end, but are they something
that we really care about? Aren't they just a means to an end? Sure they
are! In the case of VQE we want to calculate the expectation value of
some operator---for each term you take the measurements, we check what
eigenvector they represent and calculate the corresponding eigenvalue
(for the details please check out
\hyperlink{section-02}{chapter 1 about VQE}). For QAOA we don't really need to do all that, we can just
take the bitstring we got, plug it into the cost function and get the
cost value. As you can see, the exact thing that happens here might
vary, so let's just call this part ``postprocessing'' and agree that it
translates all the measurements we got into ``energy'' or ``cost
value.''

\begin{figure}
\centering
\includegraphics{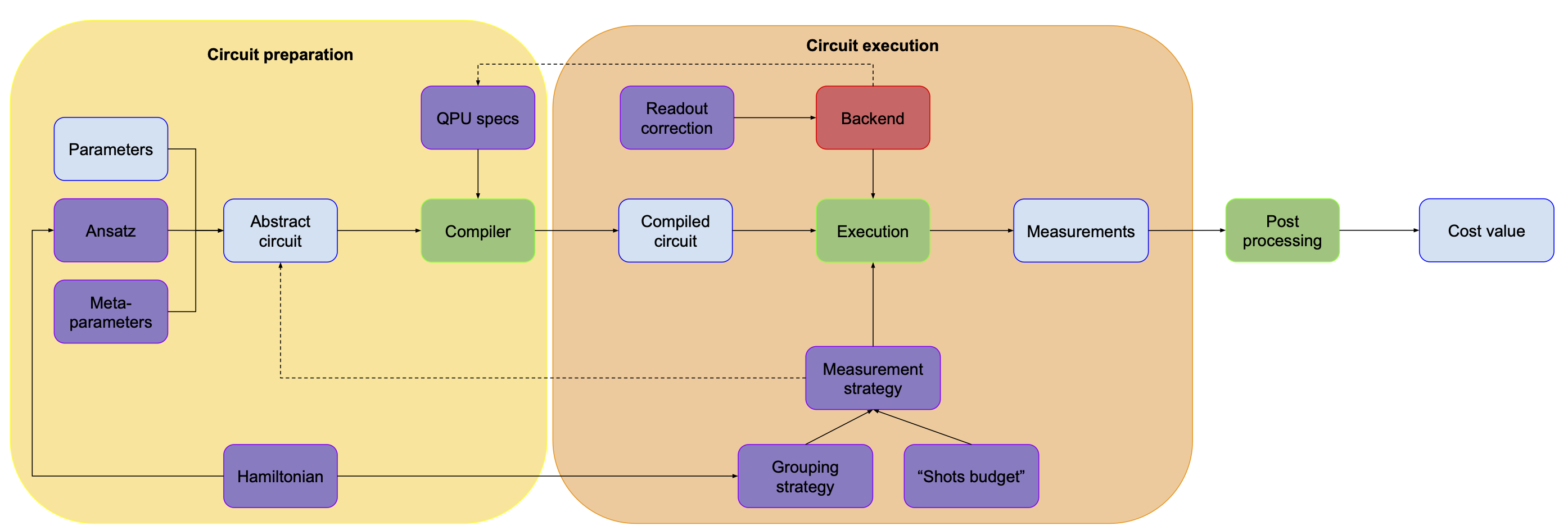}
\caption{Adding postprocessing to our previous diagram}
\end{figure}

The picture so far can be divided into two parts. One is everything we
do to generate a circuit that we'll run on our hardware. The second one
is everything we do with this circuit---how we run it, what we do with
the measurements etc. It's by no means a formal distinction, but I found
it helpful to think about it in such a way.

That's already quite a lot, but we're just starting! You see, all that
we have seen so far is about running a circuit for a single set of
parameters. But how do we know which parameters we need to choose?

Glad that you've asked!

Welcome to the land of the

\hypertarget{optimization-loop}{%
\subsection{Optimization Loop}\label{optimization-loop}}

Basically, the optimization loop looks like this:

\begin{figure}[H]
\centering
\includegraphics{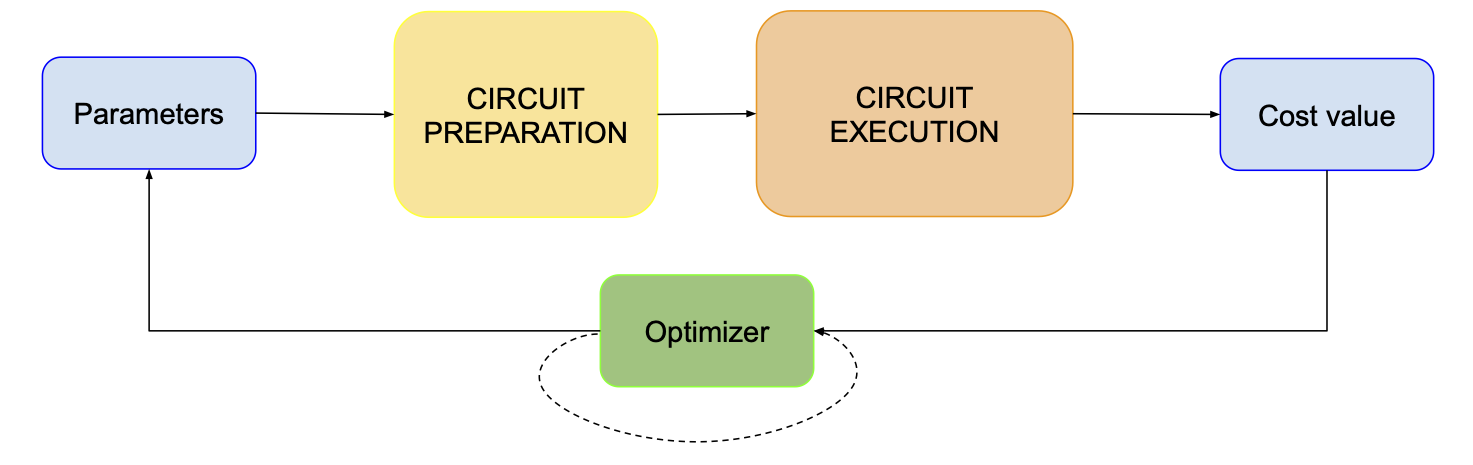}
\caption{Basic diagram of an optimizer}
\end{figure}

We take some parameters, prepare a circuit based on that, execute and
measure that, calculate the energy value and pass it to the optimizer
alongside the initial parameters. Given that optimizer can actually need
past values to work properly, there's also this dashed loop around it.

The new block here is ``Optimizer,'' but actually we haven't discussed
``Parameters'' properly yet, so let's start from that.

Parameters\ldots{} we know the optimizer updates parameters based on the
value of the cost function, but where does the first set of parameters
come from? They come from some ``parameter initialization'' procedure.
In order to initialize them, we need to actually know a couple of
things:
\begin{itemize}
    \tightlist
    \item How many parameters do we have?
    \item Are they constrained in any
way?
    \item What method do we want to use for the initialization?
\end{itemize}

Answers to the first two questions depend specifically on the ansatz
that we're using and the problem we want to solve. For example in the
case of QAOA our number of parameters depends solely on the number of
layers and not on the problem---we just have a single pair of
($\beta$, $\gamma$) per layer. Also, since our ansatz has periodic
behavior with respect to these angles, we might want to constrain them
to ($0$, $2\pi$). There's still the last question to answer---how to choose good initial angles? The most basic method is to just choose them at random, but there are many other methods to get better initial values and it turns out to influence the performance of the algorithm a lot---definitely more than I suspected.

So our picture looks like this now:

\begin{figure}[H]
\centering
\includegraphics{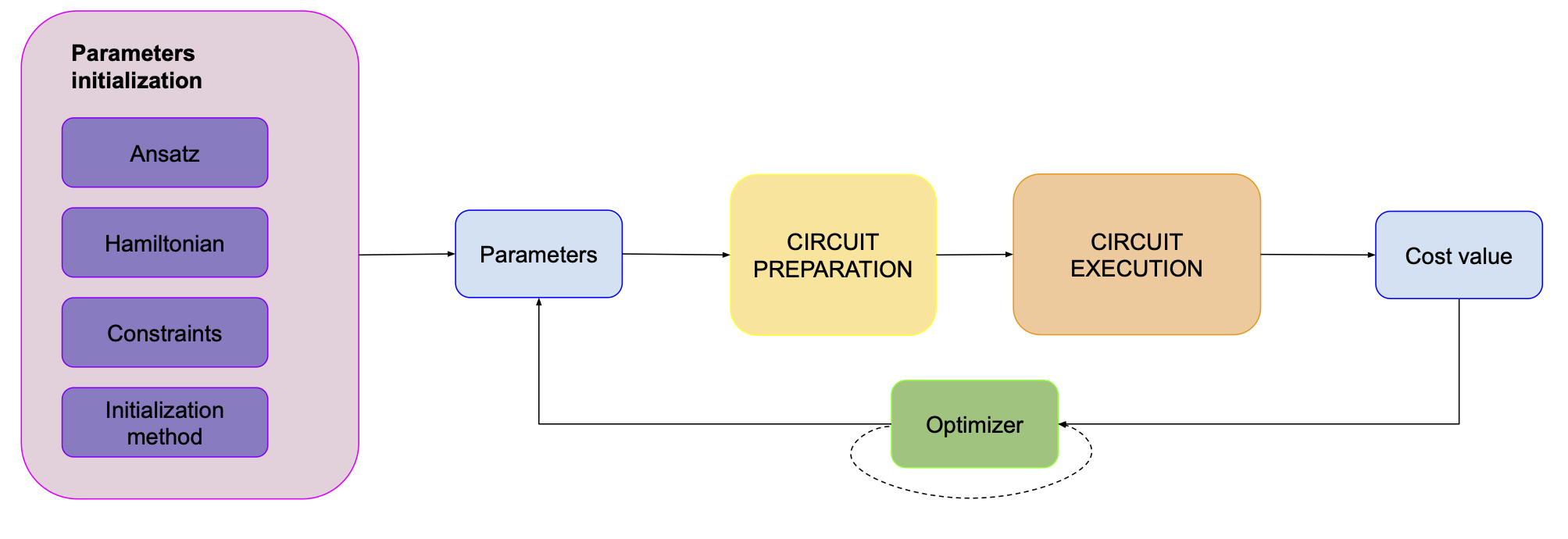}
\caption{An optimizer with initialization}
\end{figure}

We know the initial parameters, we run our circuit, and we get some energy value. We pass it to the optimizer and it spits out new parameters. But how does this happen?

\hypertarget{optimizers---gradients}{%
\subsubsection{Optimizers---gradients}\label{optimizers---gradients}}

In the process of writing, I have realized that it's hard to talk about this topic without some basic concepts from the theory of optimization. Let's say we have some function (we usually call it the cost function), which depends on N parameters. Whenever we put these parameters into our function, it returns a single number (in some cases it might be a vector, but that's uncommon in this context). If we calculated the values of the function for all possible sets of parameters, we would get a full optimization landscape. It's very convenient to think about it in terms of N-dimensional space, where each set of parameters represents a point (or vector) in such space. Another useful concept is a gradient, which tells us what's the direction and rate of the fastest increase of the function for any point.

If this sounds abstract, you can think about a map. Our function in this case will take two parameters---latitude and longitude---and return a single value: height. Global (\emph{nomen omen}) maximum would be Mount Everest and the global minimum Mariana Trench. A steep slope means the gradient is big in that direction, a plateau means it's small (or even zero). You can generalize such a map to an N-dimensional space and most intuitions still hold.

In the case of the optimization problem, we usually want to find the global minimum and there are many methods to do that. For the sake of simplicity let's start with a basic one called ``gradient descent.'' It works as follows:

\begin{enumerate}
\def\labelenumi{\arabic{enumi}.}
\tightlist
\item Select a point.
\item Calculate the value of the cost function.
\item  Calculate a gradient.
\item Make a step in the direction indicated by the gradient---since we want to get to the minimum, we're looking for the negative values of the gradient.
\item You're at a new point, go back to 2.
\item  Finish if no further improvement is possible.
\end{enumerate}

Now we're ready to get back to our VQAs. If we translate this algorithm into the world of VQAs it looks something like this:

\begin{enumerate}
    \tightlist
    \item Select initial parameters
    \item Run your circuit and get the energy value.
    \item Calculate gradients.
    \item Update parameters
    \item Go back to 2.
    \item Finish if no further improvement is possible.
\end{enumerate}

But how do you calculate gradients? Sometimes we know the exact mathematical formula, but in the case of QC that's usually not the case. One of the most basic methods to approximate it is called ``finite differences.'' Let's say we have a function of one variable $f(x)$.We can check what's the value of the gradient by calculating its values at
$f(x+\epsilon)$ and $f(x-\epsilon)$, where $\epsilon$ is just a very small number. The gradient is

\begin{equation}g(x) = \frac{1}{2\epsilon} (f(x+\epsilon) - f(x-\epsilon)) \end{equation}

If we have a function with more variables, we need to check it for various combinations of $ x_i \pm \epsilon$, but the general idea stays the same.

One of the problems with calculating gradients this way is that now we have to evaluate our cost function 3 times every iteration: $f(x)$,
$f(x+\epsilon)$ and $f(x-\epsilon)$.
If we have more variables, we will need even more evaluations. This might mean that most of your calculations on a quantum computer actually go towards calculating gradients.

Fortunately, there are some other ways of calculating gradients that work better than this. However, some of them might require modifying your circuit. Also, there are some optimizers that don't even use gradients, but describing all that would be way beyond the scope of these notes.

This part was quite lengthy, but it illustrates a couple of points pretty well:

\begin{itemize}
    \item Notice that there's nothing ``quantum'' about the optimization process we used here. In general, most (if not all) of the optimizers used in VQAs don't really care whether they optimize parameters for quantum computers, neural networks, or a mathematical function. They just care about the input and output.
    \item There's actually one thing that is specific for QC (though it doesn't necessarily apply to quantum only). We have limited precision while calculating the value of the cost function (and its gradients) on a quantum computer, associated with the finite number of measurements.
    \item It's easy to miss some big performance issues when you just look at the big picture. For example, if we understood all the pieces of the VQA except for the optimizer we're using, we could still end up with very bad performance due to a naive method of calculating gradient.
    \item VQAs have some distinct qualities which make some optimizers work well and some do not. For example, every calculation of the cost function is computationally expensive, since you need to use a QPU to get it. Another example is the precision issue that I've mentioned above.
\end{itemize}

\hypertarget{optimizers---other-aspects}{%
\subsubsection{Optimizers---other
aspects}\label{optimizers---other-aspects}}

We have initial parameters, we know how to calculate gradients (or we decide to go with a gradient-free method) so what else do we need to know about the optimizer part?
 
The first thing is that an optimizer might have multiple hyperparameters, like how big the steps should be in a gradient method or how many iterations we want to have. These hyperparameters might change throughout the optimization process (if you're unfamiliar with the concept of hyperparameters, \href{https://machinelearningmastery.com/difference-between-a-parameter-and-a-hyperparameter/}{this article} might be useful).

Another thing, quite distinct from the optimizer's parameters, is its constraints. In some cases, we might need to put some constraints on the parameters (e.g.~they all need to sum up to 1), which influences how the optimizer behaves. An example could be a problem where we know that our quantum state should have a fixed number of particles and our quantum circuit should not represent states with other numbers, even though it's possible with the ansatz we use.

Lastly, an optimizer might actually modify some of the elements inside the "Circuit preparation/circuit execution'' box, most notably the measurement strategy. The easiest example that comes to my mind is the one that I've already mentioned, i.e.~changing the number of shots during the training. We can argue that in the first iterations when we're exploring the optimization landscape, we might not need a lot of precision and we can save some QPU time by doing a smaller number of shots. However, once we start converging on the minimum, a small difference in parameters might have huge consequences, so we would rather get higher precision.

So this is what our picture looks like now:

\begin{figure}[H]
\centering
\includegraphics{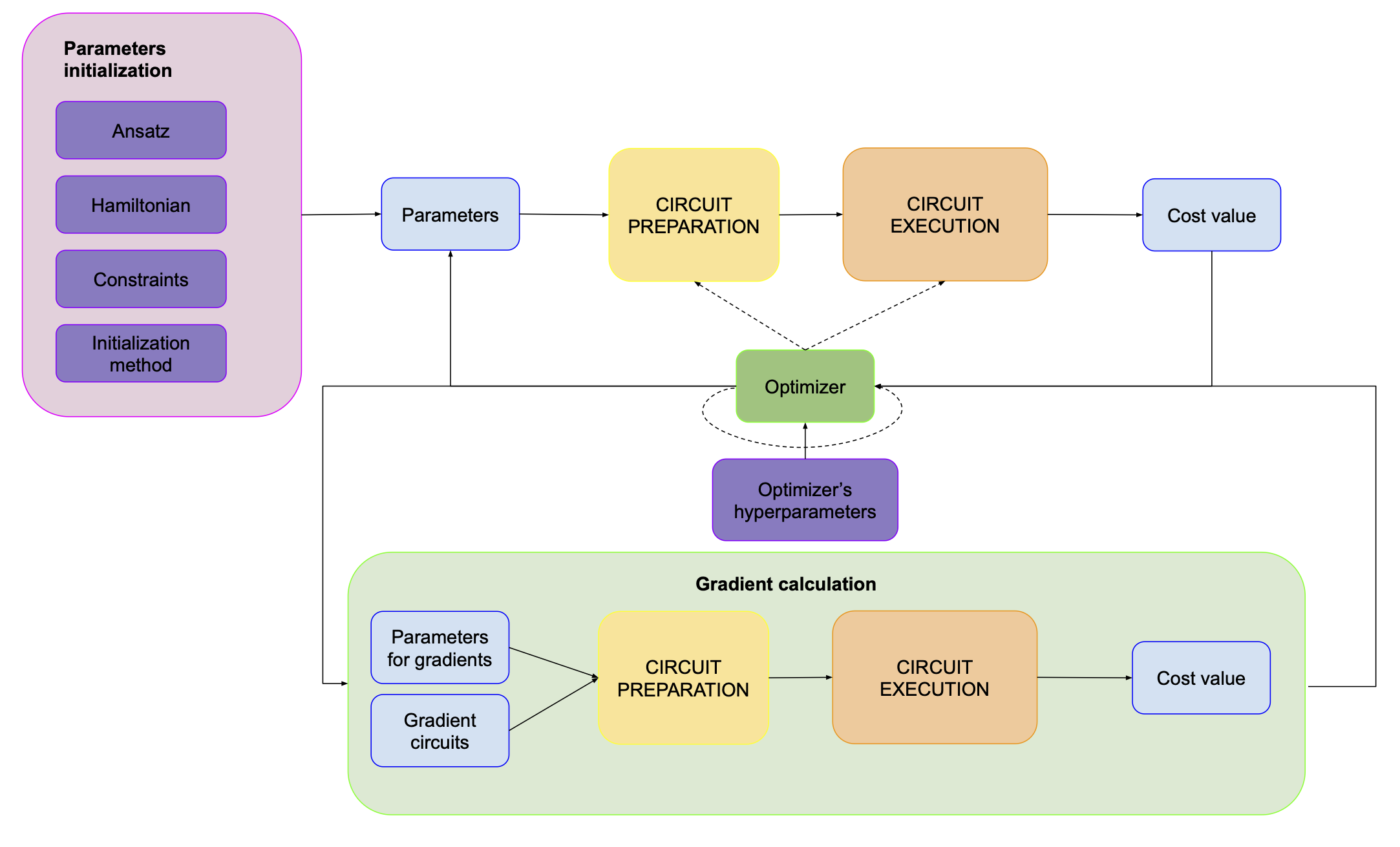}
\caption{Diagram of an optimizer with gradient calculation}
\end{figure}

At the bottom, you can see the gradient calculation loop. For the sake of not over complicating this plot we'll hide it inside the optimizer:

\begin{figure}[H]
\centering
\includegraphics{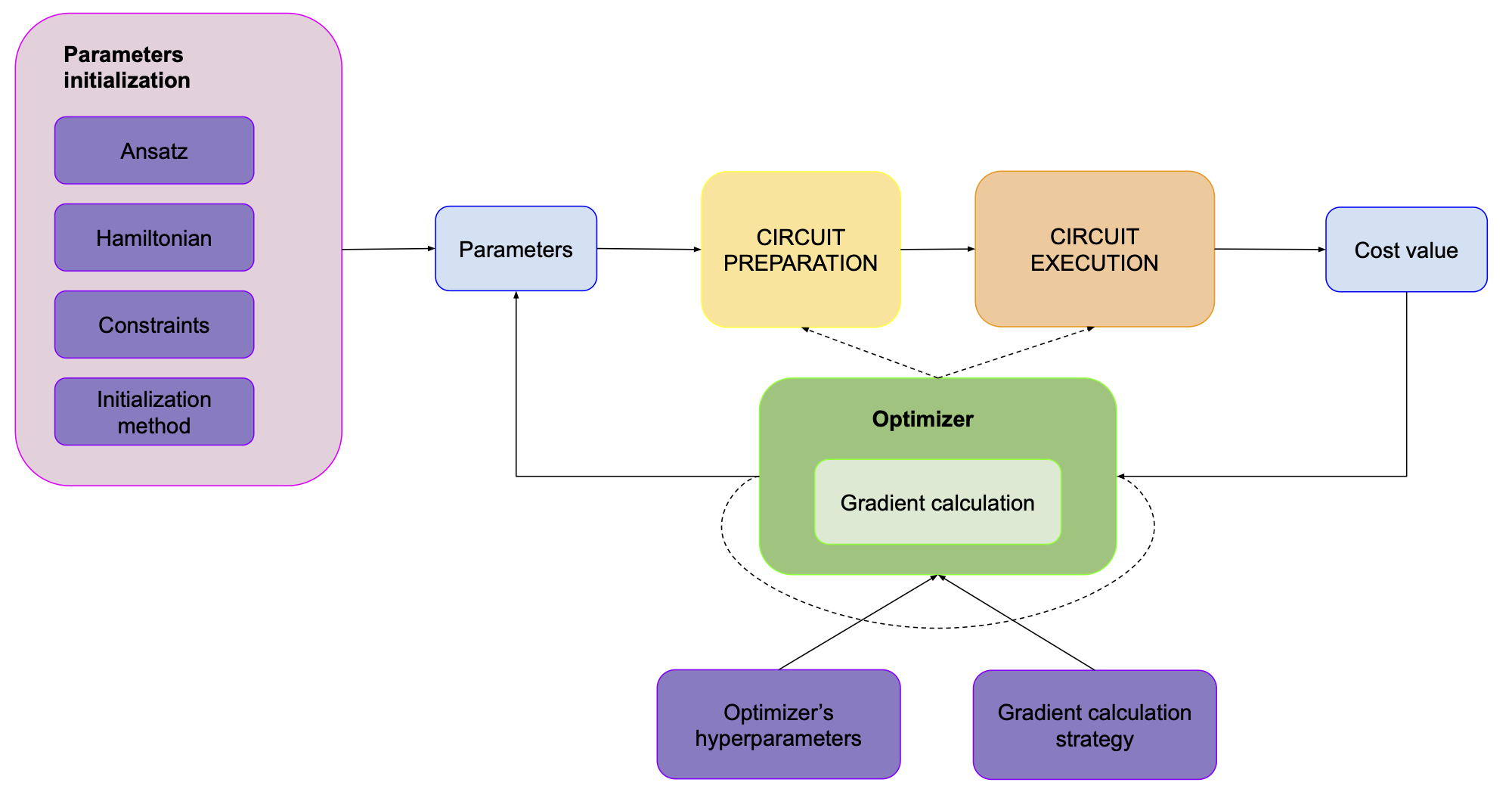}
\caption{Diagram of an optimizer with gradient calculation, simplified}
\end{figure}

This picture is quite accurate for standard versions of VQAs that you usually encounter. However, I'd like to cover two more cases here, which often show up in the literature

\hypertarget{outside-loops}{%
\subsection{Outside loops}\label{outside-loops}}

\hypertarget{layer-by-layer-training}{%
\subsubsection{Layer-by-layer training}\label{layer-by-layer-training}}

Many ansatzes (e.g.~QAOA) can have multiple layers. One of the problems with many layers is that the more layers you have, the more parameters there are to optimize which means it's harder to find the optimal values. One of the methods that's often used to help with that is layer-by-layer training. We first find the optimal parameters for the first layer of our ansatz. We keep them and then add another layer. We optimize the parameters for both layers and then add another.

In such a case our picture would look something like this:

\begin{figure}[H]
\centering
\includegraphics{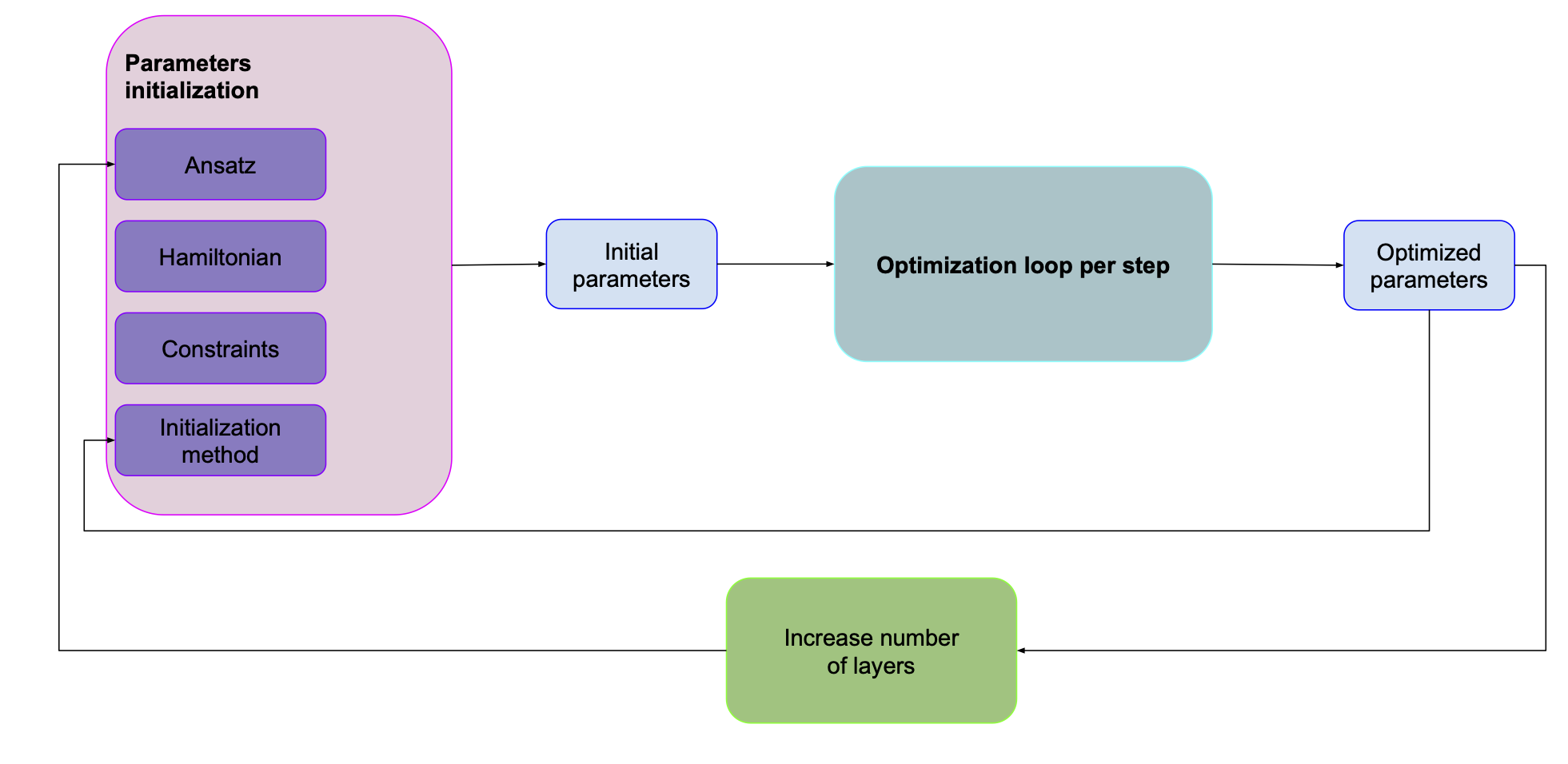}
\caption{Diagram of an optimizer with layer-by-layer training}
\end{figure}

It's pretty similar to the previous one, except that our parameter initialization strategy depends on the values we found for a smaller number of layers. Obviously, we might also want to change some other parameters like measurement strategy or optimizer hyperparameters.

\hypertarget{adaptive-circuits}{%
\subsubsection{Adaptive circuits}\label{adaptive-circuits}}

Recently (2020) there have been many propositions of algorithms that throughout the optimization modify the structure of the circuit itself, not only the values of its parameters. As an example, I'd like to use the \href{https://arxiv.org/abs/2010.00629}{PECT algorithm}.

The main idea of this algorithm is that instead of optimizing all the parameters at once, we remove some of the parameters (and the corresponding gates) and optimize only a fraction of the parameters. We have two optimization loops---one selects which parameters we should select for optimization, and the other finds the best values of these parameters.

Here we don't change the structure of the circuit itself, rather just which gates we treat as ``parameterizable.'' However, there are other proposals as well, where we change the structure of the circuit itself. Layer-by-layer can be treated as an example of such. Another one would be using genetic algorithms to find the optimal ansatz for our problem. Since the structure of the circuit ``adapts'' during the optimization, such algorithms are sometimes called "adaptive algorithms'' (see e.g.~\href{https://arxiv.org/abs/1812.11173}{Adapt-VQE}).

\hypertarget{summary}{%
\subsection{Summary}\label{summary}}

Look at where you are:

\begin{figure}[H]
\centering
\includegraphics{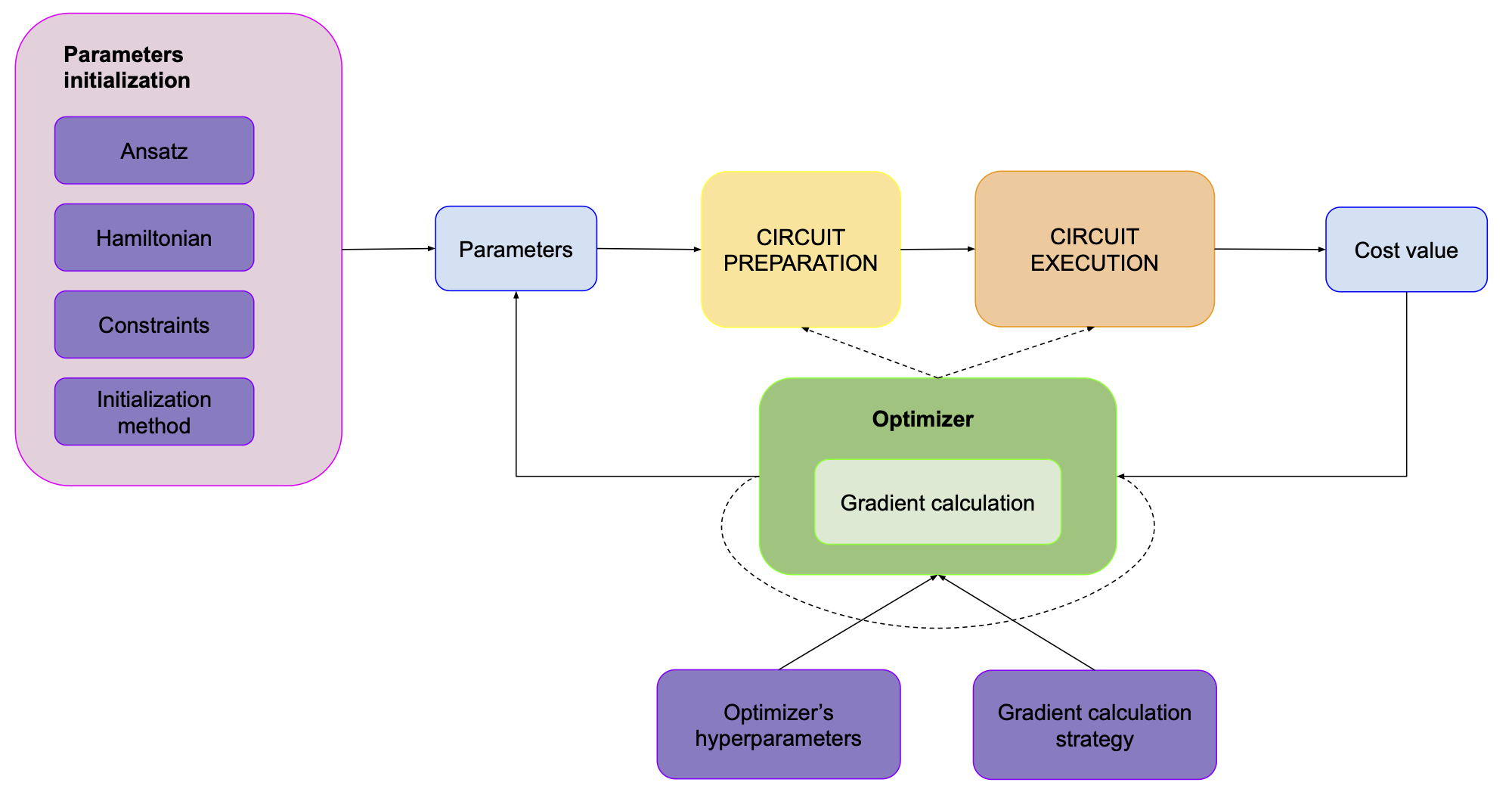}
\caption{Optimizer with gradient again}
\label{fig:enter-label}
\end{figure}

Look at where you started:

\begin{figure}[H]
\centering
\includegraphics{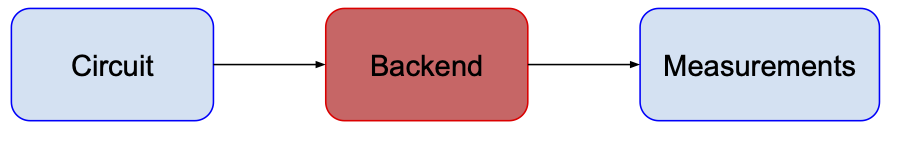}
\caption{Basic diagram}
\end{figure}

I tried to make this chapter comprehensive, but it is actually not getting into too many details. Basically any single element that we discussed can be subject for a separate scientific paper---and many are. If you want to get deeper into specific issue with the VQAs and what are some methods for solving them, I've described them in \hyperlink{section-05}{VQA---challenges and state of research}, the next chapter of these notes.

I hope you enjoyed reading and it helped you better understand quantum computing in the NISQ era.

\newpage

\hypertarget{section-5}{\section{VQA --- Challenges and state of research}}

In previous parts we learned how these algorithms work in theory, now it's time to learn more about some challenges, their extensions and improvements, and practical considerations about them.

In this section, we'll focus on the general issues with running VQAs on quantum computers, regardless of their specific structure. We'll go into more detail about challenges unique for (or at least more characteristic to) VQE and QAOA in the next two sections.

While writing this part I heavily relied on two excellent review papers: \href{https://arxiv.org/abs/2101.08448}{from Alan Aspuru-Guzik group of University of Toronto} and \href{https://arxiv.org/abs/2012.09265}{a big collaboration of various institutions}. You can find a lot more details there!

Let's begin!

\hypertarget{hardware-related-problems}{%
\subsection{Hardware-related problems}\label{hardware-related-problems}}

No matter what algorithm you want to run on the NISQ devices, you need to deal with the following issues:

\begin{itemize}
\tightlist
\item Noise
\item Connectivity
\item Size of the device
\item Limited gate-set
\end{itemize}

\hypertarget{noise-types}{%
\subsubsection{Noise types}\label{noise-types}}

Understanding how noise and error impact quantum computation is an entire subfield known as quantum characterization, verification and validation (QCVV). We'll stick to just a cursory description of a few important concepts.

On a physical device, two important concepts involving error are:

\begin{itemize}
\tightlist
\item Decoherence---unwanted interaction between the qubits and their environment that causes the quantum state of the qubits to lose its purity.
\item Control error---physical operations (e.g.~laser pulses) that are not exactly as intended, causing the actual quantum gates to differ from the target quantum gates.
\end{itemize}

In practice, these are quite intertwined and difficult to separate. Generally, they lead to three types of error in a quantum computation:

\begin{itemize}
\tightlist
\item Coherent error---during a quantum circuit the quantum state is shifted to a different quantum state
\item Stochastic error---during a quantum circuit the quantum state becomes an average of different quantum states
\item Measurement error---you might have a perfect quantum computer with zero noise and then at the end of the circuit, your measurement procedure might assign an incorrect state when you measure it
\end{itemize}

Coherent and stochastic error differ in how they cause overall error in a quantum circuit to accrue. Coherent error accrues less favorably in that the quantum state can continue to be shifted in the wrong direction. With stochastic error there can be some canceling out of the errors so that on average they leave you close to the intended state.

If you'd like to learn more, see section 8.3 of Nielsen and Chuang. For those looking for a deeper rabbit hole, here are some more papers: \href{https://arxiv.org/abs/1510.05653}{Kueng et al.}, \href{https://arxiv.org/abs/2104.01119}{Zhang et al.}, \href{https://arxiv.org/abs/1710.02270}{Bravyi} or \href{https://arxiv.org/abs/2001.09980}{Geller and Sun}.

\begin{figure}
    \centering
    \includegraphics{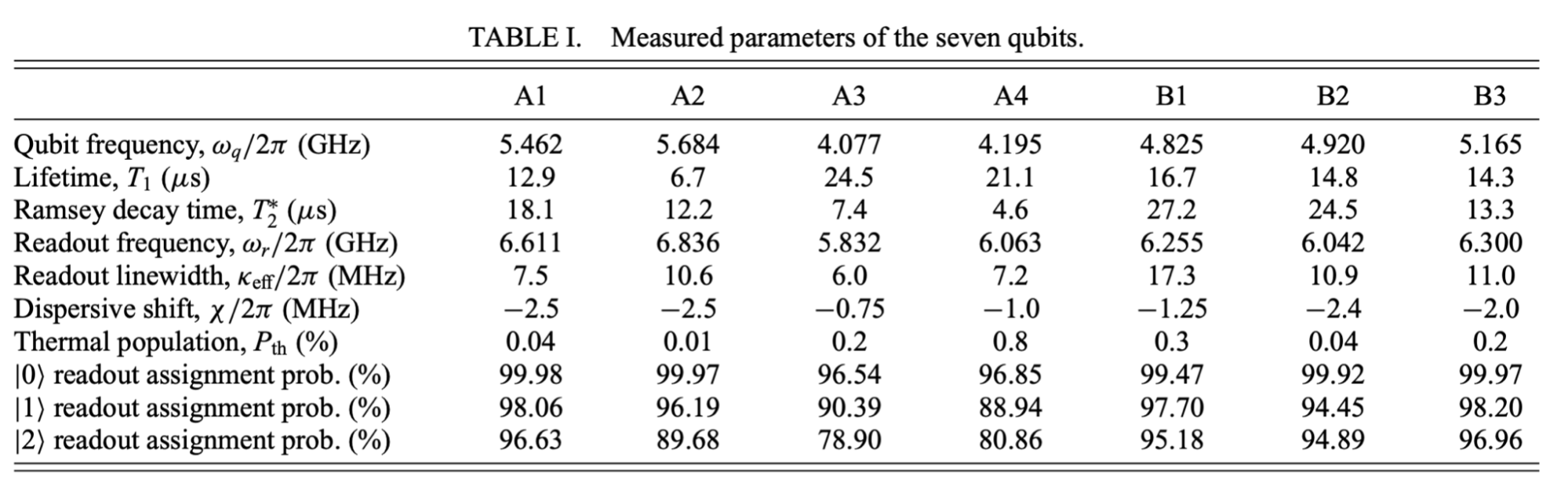}
    \caption{A table displaying the specs of the device}
    \label{fig:table_specs}
\end{figure}

\begin{figure}
    \centering
    \includegraphics[width=10cm]{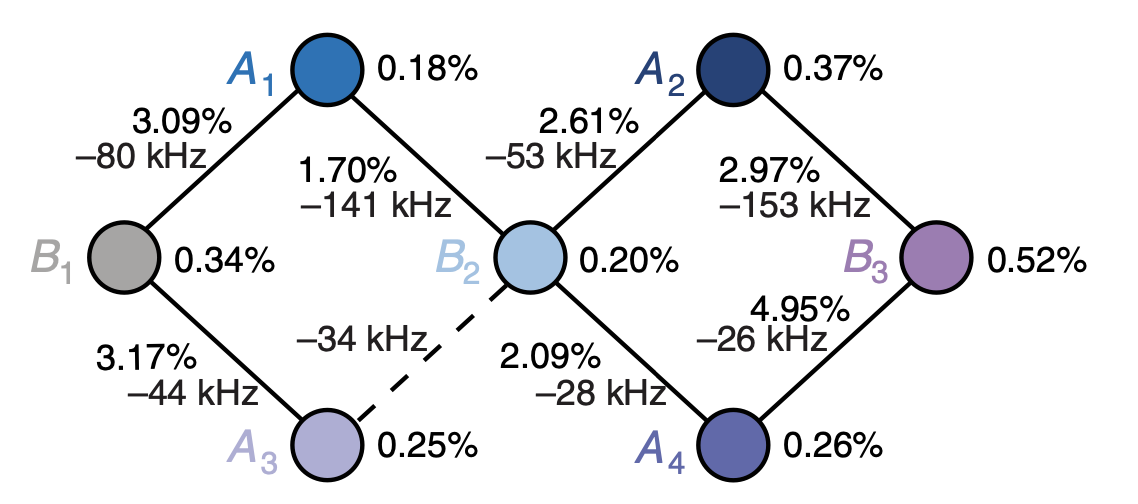}
    \caption{A diagram depicting the layout of the 7 qubits of the device}
    \label{fig:layout_specs}
\end{figure}

To give you some ideas about how it looks like on real devices we have today, let's take a look at the specs of the device from ETH Zurich used in \href{https://journals.aps.org/prxquantum/abstract/10.1103/PRXQuantum.1.020304}{this paper}. You can see the specs in Figs \ref{fig:table_specs} and \ref{fig:layout_specs}.


As we can see, all the qubits have very different characteristics.
There's a lot of data here, so let's focus on the following values:

\begin{itemize}
\tightlist
\item
  T1 and T2---they roughly define the lifetime of a qubit. T1 defines
  how long does it take for the qubit to go from $ | 1 \rangle $ to
  $ | 0 \rangle $ and T2 measures how quickly qubit loses its phase
  (see
  \href{https://ocw.mit.edu/courses/mathematics/18-435j-quantum-computation-fall-2003/lecture-notes/qc_lec19.pdf}{section
  1.1 here}).
\item
  Readout assignment probability---this basically defines the
  measurement error. As you can see it's different for states
  $ | 0 \rangle $ and $ | 1 \rangle $ (you can disregard state
  $ | 2 \rangle $).
\item
  Gate errors---the percentage values next to nodes (green) indicate
  one-qubit errors and next to edges the two-qubit errors (blue). This
  basically tells you what's the probability that a given gate won't do
  what it's supposed to.
\item
  Gate speed---how long it takes to execute one gate.
\end{itemize}

Ok, what does it all mean? Let's use the most optimistic values for
simplicity---lifetime of the qubit equal to 27.2$\mu s$, gate errors
0.18\% and 1.7\%, and gate speed 50ns (see Fig 1c from the paper).

This means that we can apply at most

\begin{equation}
    \dfrac{24.5 \mu \text{s}}{50 \text{ns}} \approx 500
\end{equation}

gates to still be able to get
reasonable results. What does it look like if we take gate errors into
account? The probability that the result you get for a single qubit is
correct is equal to

\begin{equation}
    (1 - \text{gate error})^{(\text{number of gates})}
\end{equation}

So if you
want to be 90\% sure that you got the right result, you can run 50
gates. If 80\% is enough that's about 123 gates. What about 2-qubit
gates? Well, you can use only 6 of them for the 90\% case and 13 for the
80\% case.

To make things even worse, once you make a measurement, some of the
measurements will be wrong anyway. I couldn't find the number in this
paper, but for Google's Sycamore chip, readout error was 3.8\% (see
\href{https://www.nature.com/articles/s41586-019-1666-5}{Fig 2.}).

Btw. I highly recommend the paper that I took the data from, it shows
you what's the state-of-the-art implementation of QAOA on a real device.
And here's \href{https://arxiv.org/abs/2004.04197}{another good example
from Google}.

There's also statistical uncertainty (sometimes also called ``sampling
noise'')---this is the error that we introduce into our estimates by the
fact that we cannot directly measure the wavefunction, we need to sample
from it. So having 100 measurements gives you results with more
uncertainty than having 10,000 measurements. We'll probably talk more
about how to deal with that in the next article.

A good example of how the noise affects result are the plots of the
optimization landscapes of 1 layer QAOA which came from Google's
Sycamore chip:

\begin{figure}[H]
\centering
\includegraphics[width=8cm]{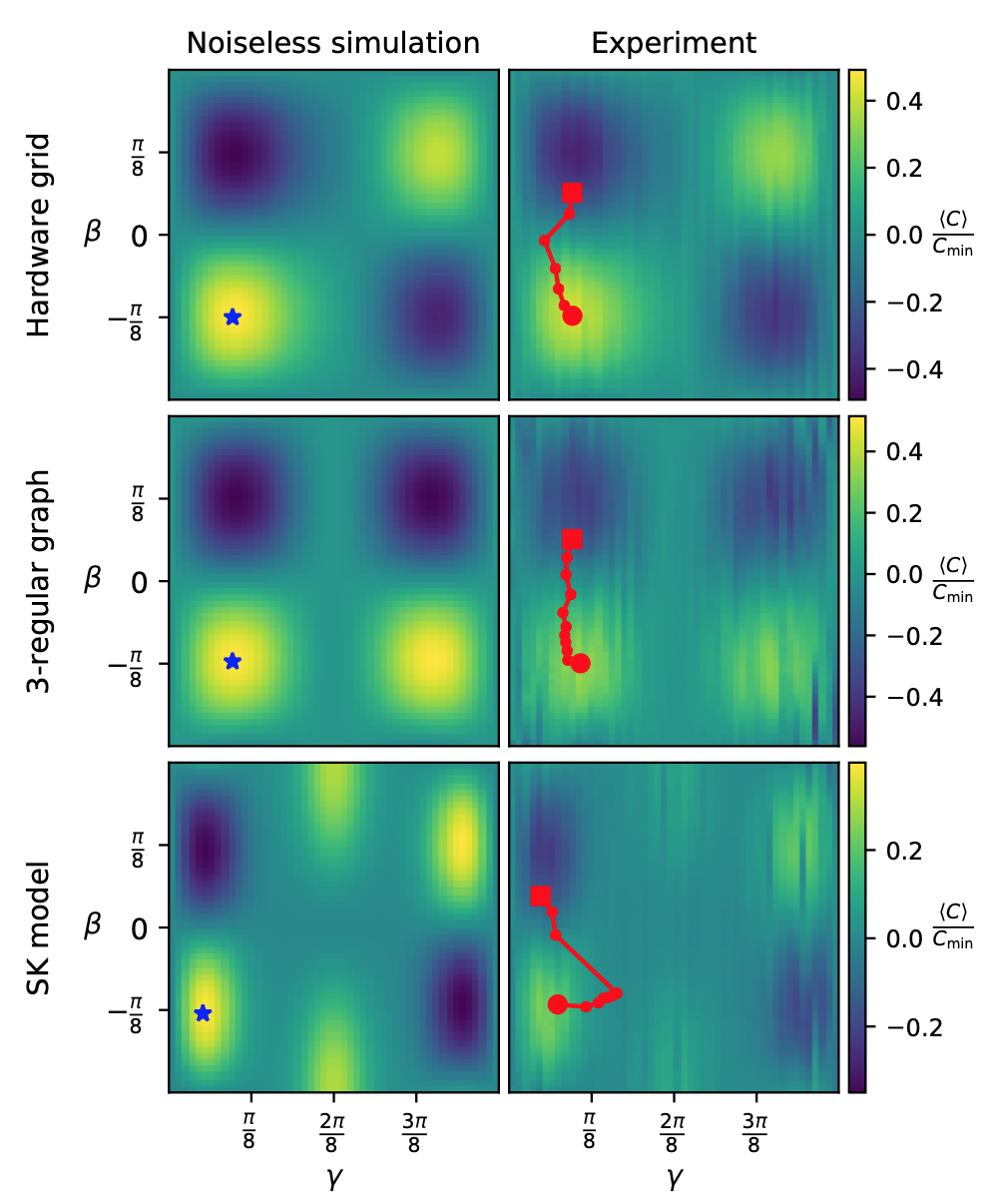}
\caption{Energy landscape of 1-layer QAOA---notice how the experimental landscape looks different (more ``noisy'') compared with the theoretical one. This picture is taken from \href{https://arxiv.org/abs/2004.04197}{Fig 3 in this
paper}.}
\end{figure}

One way to deal with these issues (at least some of them) is to use Quantum Error Correction (QEC). QEC is a set of techniques that allow you to correct some errors that can happen in your circuit. If you`ve ever heard about the distinction between ``physical'' and ``logical'' qubits, then you might know that by ``logical'' people usually mean ``perfect, error-corrected qubits.'' Once we have such qubits, we will
be in the realm of Fault-Tolerant Quantum Computing (FTQC) The problem is that in order to implement one error-corrected qubit you have to use a LOT of physical qubits (which also depends on the level and type of errors), as we encode one logical qubit in an entangled state shared by
multiple physical qubits. We can design this state in such a way that the quantum information is protected against different types of noise. But since in this article we're talking about variational algorithms and the NISQ era, we won't actually go into this, if you're interested check out chapter 10 from Nielsen and Chuang.

Another, quite straightforward, way to deal with some of these issues is to simply run short circuits. The fewer gates you have, the lower the chance something will go wrong. Obviously, improving the hardware also helps (or, as it turns out, improving control software responsible for
gates, as shown \href{https://arxiv.org/abs/2010.08057}{here}). But let's get into some more algorithmic techniques.

\hypertarget{error-mitigation}{%
\subsubsection{Error mitigation}\label{error-mitigation}}

Instead of correcting the errors and getting perfect results, we can try to \emph{mitigate} the errors. There are many interesting techniques (see \href{https://arxiv.org/abs/2011.01382}{here} for a good review), here I'll describe one of them to give you a general idea of how it can
be achieved. It's important to mention that these methods allow us to better estimate the expectation values of some operators. This means that we'll be talking about getting certain real numbers with a specific precision, not about getting the correct set of 0s and 1s.

One of the methods is called ``Zero Noise Extrapolation''
(\href{https://journals.aps.org/prl/abstract/10.1103/PhysRevLett.119.180509}{source}). The main idea is that if we just run our circuit, it will be affected by a certain base level of noise. Our goal would be to know the result for the ``zero noise'' case, but we can't just fix the hardware so that it doesn't have any noise. However, we can artificially increase (scale) the level of noise. So we'll run our circuit a couple of times, with different levels of noise, and then we'll extrapolate it to see what happens for the ``zero noise'' case.

\begin{figure}
\centering
\includegraphics[width=10cm]{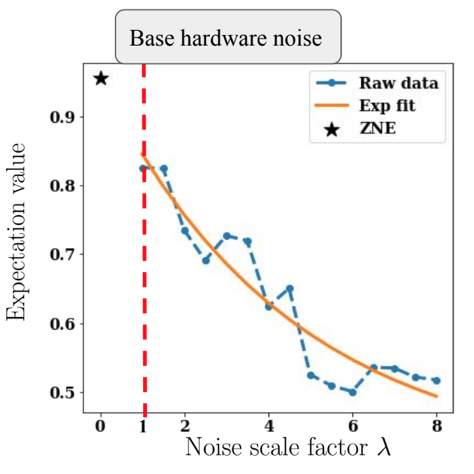}
\caption{An example of how zero-noise extrapolation is done. The star showcases the extrapolated expectation value.}
\label{fig:ZNE}
\end{figure}

Fig. \ref{fig:ZNE} shows the idea.

How do you artificially increase the noise? Well, there are a couple of methods to do that, one of them involves simply duplicating gates---we replace gate G by (G, G', G), where by G' I mean the inverse of G.

There are other error mitigation techniques, but in general, they involve running several modified versions of the original circuit and performing some post-processing to get the value we're interested in.

If you'd like to use error mitigation techniques, there's an open-source Python library called \href{https://github.com/unitaryfund/mitiq}{mitiq} developed by folks from \href{https://unitary.fund}{Unitary Fund}. You can learn more about ZNE and other methods, as well as about Mitiq from
this talk by Ryan LaRose (which is also the source of the plot you see above).

There's one more important caveat here---what I described is mitigation by ``classical postprocessing.'' There's another way to achieve error mitigation by performing certain actions during the execution of the circuit which counteracts the noise.

\hypertarget{compilation}{ \subsubsection{Compilation}\label{compilation}}

Well\ldots{} If noise itself wasn't bad enough, there are other issues that make its existence even worse. It's the device connectivity (i.e., which qubits are directly connected, so you can use a two-qubit gate between them) and gate set (i.e.~which gates you can directly execute on the hardware).\footnote{For explanation of why these are problematic please see the earlier \hyperlink{section-04}{section about how VQAs work}.}

How do you deal with these? Two basic methods are compilation and ansatz design, and in this section, we'll talk about the first of these. To be honest---I know very little about compilation, so this section definitely doesn't do justice to the topic, though I think it's important to include it for the sake of completeness.

What is compilation? It's transforming one quantum circuit into another one that does exactly the same thing.

Usually, we do the compilation for the following reasons (in no particular order): 

\begin{itemize}
    \tightlist
    \item Our circuit doesn't map well on the hardware architecture (dealing with connectivity) 
    \item Our circuit uses gates that are not directly implemented on the device (dealing with the native gate set)
    \item We want to minimize the number of gates (reducing the impact of noise). Usually, we care mostly about 2-qubit gates, as they are much more noisy than 1-qubit gates.
\end{itemize}

There are different approaches to compilation, here I present three that
I'm aware of: 
\begin{itemize}
    \item Rule-based---we define a set of rules (e.g.~how to decompose a gate into a set of other gates) that we then simply apply to the circuit.
    \item We treat circuits as some mathematical structure and perform some mathmagic to simplify it, for example using ZX calculus. Don't ask me for any more details, I've personally chosen to just think about it as mathmagic (though to be honest I've heard \href{https://arxiv.org/abs/2012.13966}{this review} is a decent intro to ZX calculus, I simply never got to reading it. There's also this \href{https://github.com/Quantomatic/pyzx}{open source package}.).
    \item Machine Learning---well, you can also just throw some machine learning at the problem, as they did in \href{https://arxiv.org/abs/2007.14608}{Paler et al.}
\end{itemize}

If anyone is interested in the topic
\href{https://si2.epfl.ch/~demichel/research/quantum.html}{this article}
might be helpful. Another important theoretical result to mention is the
Solovay-Kitaev theorem---it basically says that we can approximately
compile any unitary operation into a limited set of gates quite
efficiently (with some caveats, of course, more on
\href{https://en.wikipedia.org/wiki/Solovay-Kitaev_theorem}{wiki}).

That's it for now, if I learn more about the topic I'll revisit and
improve this part. Let's now get into the topic much closer to my heart.
Namely\ldots{}

\hypertarget{optimization}{%
\subsection{Optimization}\label{optimization}}

Ok, we've covered some general, hardware-related issues. Now it's time
to talk about the more algorithmic part. One of the central components
of VQAs is the optimization loop and there's indeed a lot of research on
this topic. So let's see what are some of the main challenges involved:

\hypertarget{barren-plateaus}{%
\subsubsection{Barren plateaus}\label{barren-plateaus}}

Imagine you want to find the lowest point in some area. How would you do
that? Well, you can just look around, follow the slope and you'll
eventually get somewhere. Doesn't sound like the best possible
strategy---if you're in the mountains you'll soon find some valley,
though not necessarily the lowest/deepest one. But at least you're
getting somewhere. Do you know what's a nightmare scenario in such a
case?

A huge desert.

You're not able to see anything but flat sand all the way to the horizon
and the landscape constantly changes as the wind reshapes the dunes.

What's the only chance you have of actually finding what you're looking
for? Start from a point that's so close to the valley you're looking for that you just can't miss it. What's another name for a landscape like this?

A barren plateau.

In 2018 \href{https://www.nature.com/articles/s41467-018-07090-4}{Jarrod McClean et al.} pointed out that there are two big problems with training variational quantum circuits.

The first is that you can't just run a single circuit and learn what the value of the gradient is (we've discussed it in the previous
\hyperlink{section-02}{VQA section}). You need to repeat a certain procedure several times and the more times you do it, the better accuracy you get. This is not a big problem if your gradient is huge; let's say it has a value of 9001. You just repeat the procedure several times and you know the ballpark. But what if your gradient has a value of 0.00001? Well, you have to run many many more circuits (for those interested---in the best case it scales as $O(1/ \epsilon)$, where $\epsilon$ is desired accuracy.)\footnote{Numbers here are totally fake, I'm just making a point.}

The second is, that for random quantum circuits the gradient is really small in most places except for some small area where it's not. And the chance that it's arbitrarily close to zero for a random point grows exponentially with the number of qubits and number of parameters (see \href{https://pennylane.ai/qml/demos/tutorial_local_cost_functions.html}{this PennyLane tutorial} for more context).

Ok, now let's put it together in plain English. In most places, the gradient is close to 0 and from the practical perspective, there's a limit to the precision up to which you can estimate it. So basically unless you start close to good parameters, you have no clue how to tweak your parameters to get anything reasonable. And the more qubits or gates you have, the worse it gets.

\begin{figure}
\centering
\includegraphics{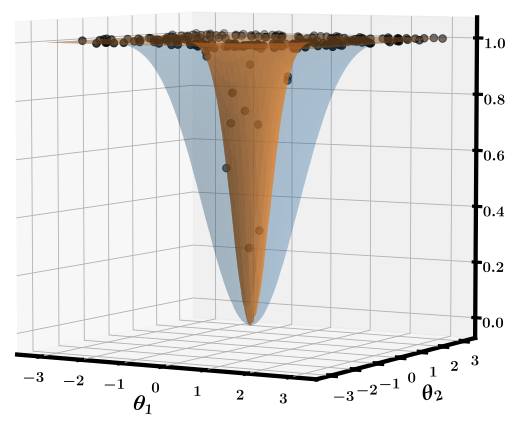}
\caption{Cost function with 4 variables (blue) vs. 24 (orange). \href{https://arxiv.org/abs/2001.00550}{Source: Cerezo et al.}}
\label{fig:barren_plateau}
\end{figure}

I also like Fig. \ref{fig:barren_plateau} which visualizes the problem. It shows how the landscape becomes harder to optimize when we increase number of variables in the cost function.

As a side note, I find this problem shows quite well one of the problems with the state of quantum computing research. As long as you're running small simulations, this effect is too small to be noticed. So some previous small experiments might be fundamentally flawed because if you tried to scale them up, you would hit the barren plateau problem. On the other hand, there might be other problems that we'll notice only when we'll get to hundreds of qubits. These are exciting times to be working on QC!

\hypertarget{dealing-with-barren-plateaus}{%
\subsubsection{Dealing with barren
plateaus}\label{dealing-with-barren-plateaus}}

Not surprisingly, these findings resulted in quite some stir in the community. People started coming up with various methods to solve this issue and here we'll go through a couple of them. Let's look at the two common approaches for dealing with this problem.

The first way is to simply initialize your parameters close enough to the minimum so that your optimizer can find its way to the minimum. This might sound like a no-brainer---sure we want to have a good way of initializing the parameters, right? Actually, it's not that obvious, as in classical machine learning, it usually is not that big of a problem and you can get away with initializing parameters randomly. So how do you choose a reasonable parameter initialization? There are several methods to do that, one of them is the so-called \href{https://arxiv.org/abs/2006.14904}{``layer by layer'' (LBL) training}.\footnote{Another example: \href{https://quantum-journal.org/papers/q-2019-12-09-214/}{E. Grant et al.}}

Many ansatzes we use consist of layers. Basically, all the layers are identical and (at least in theory), the more layers you add, the more powerful your circuit is (QAOA is a good example). So here the idea is that we start from training our algorithm for one layer. Since barren plateaus depend on both the number of qubits and the depth of the circuit, it should allow us to avoid them. Then, once the 1st layer is optimized, we treat it as fixed and train parameters for the second layer. Once its training is finished, we proceed with the next one until we have all the layers we wanted. Now we have some initial guesses of the parameters and we proceed to train more than one layer at a time, e.g.~25\% of them.

So at first, we try to avoid barren plateaus by training only a handful of parameters at the time. And then, we try to avoid them by starting from a point that is already a pretty good guess.

The second idea is about restricting the parameter space---if you can design your ansatz specifically for a given problem, it might be less expressible (we'll talk about expressibility in a moment), but as long as it can find the solution to your problem, it doesn't really matter.

Just to put things in perspective, since the initial paper published in March 2018, I've counted 18 papers with the words ``barren plateaus'' in the title and the phrase was cited 263 times according to Google Scholar. If this problem sounds interesting to you, some good papers to read would be those about what are the sources plateaus
(\href{https://arxiv.org/abs/2010.15968}{entanglement} and
\href{https://arxiv.org/abs/2007.14384}{noise}), or some
\href{https://www.nature.com/articles/s41467-021-21728-w}{mathematical ways} to avoid them.

\hypertarget{choice-of-optimization-methods}{%
\subsubsection{Choice of optimization
methods}\label{choice-of-optimization-methods}}

A question that I sometimes hear is ``are there optimizers that are specific for quantum computing?'' In principle no, as you can treat the calculation of the cost function in the variational loop as a black box, and therefore you can use any optimization method for that. However, there are certain challenges associated with our particular black box that might make some optimizers a much better fit than the others.

So the first challenge is that in general, we consider the time spent running algorithms on a quantum chip much more expensive (money-wise) than on a classical one. Therefore, since we use it exclusively for calculating the value of the cost function, evaluating the cost function is the most expensive part of the algorithm. Thus, it is something we want to do as little as we can get away with. (Isn't it somewhat ironic, that when we run calculations of a quantum computer, we want to use it as little as possible?) The practical meaning of this is that we want to use optimizers that can work well while making a relatively small number of evaluations.

The second challenge is the probabilistic nature of the cost function
evaluation. In order to deal with it, you need to repeat your circuit
more times to have better accuracy. Or using an optimizer, which works
well with some level of noise/uncertainty.

The third challenge is the existence of noise. Noise might have
different effects on the landscape of the cost function. One is
``flattening of the landscape,'' another might be the existence of some
artifacts. Both these effects are visible in the plot we've QAOA
landscapes we've seen before.

Last, but not least, it's extremely hard to study the behavior of the
optimizers. They often rely on hyperparameters (e.g.: step size in
gradient descent), their behavior might be different for different
problems. This is a problem not only for QC, this is the same for
classical optimization problems or regular Machine Learning. Also, the
cost of implementing and checking the performance of various optimizers
is really high, so researchers usually either decide to use one that
worked for them in the past or one that is commonly used or perhaps
check a couple of them and pick the one that looks reasonable. Yet
another issue is that sometimes these methods are studied without the
noise (where they work well), but they don't perform that well in the
presence of noise.

My plan was to follow this section with a selection of some widely used
optimizers, and explanations why some of them are widely used. But after
digging into the literature, I wasn't able to come up with anything
satisfactory. It basically looks that right now we have little to no
idea why certain optimizers work while others don't. So if you're
interested in this topic, I recommend the following papers:
\href{https://arxiv.org/abs/2005.11011}{Sung et al.} and
\href{https://arxiv.org/abs/2004.03004}{Lavrijsen et al.}. Excellent
work which shows how hard it is to analyze these and how much we still
need to learn.

\hypertarget{ansatz-design}{%
\subsection{Ansatz design}\label{ansatz-design}}

We've already mentioned ansatzes in the previous section, but let's now
make them our main focus. Ansatz design is an active field of research,
some of my favorite papers are those by
\href{https://arxiv.org/abs/1905.10876}{Sukin Sim and other folks from
Zapata},
\href{https://quantum-journal.org/papers/q-2021-03-29-422/}{Lena Funcke
et al.} and \href{https://arxiv.org/abs/2105.01114}{J. Lee et al.}.
Since I like to be practical, let's look at this from a practical
perspective:

\begin{itemize}
\tightlist
\item
  How can you tell whether an ansatz is a good one?
\item
  How to design a good ansatz?
\end{itemize}

\hypertarget{is-my-ansatz-any-good}{%
\subsubsection{Is my ansatz any good?}\label{is-my-ansatz-any-good}}

In the NISQ world, when you have an ansatz, you want to ask yourself two
questions:

\begin{itemize}
\tightlist
\item
  How powerful is my ansatz?
\item
  How much does it cost to use it?
\end{itemize}

What does powerful mean in this context? One component is its
expressibility---i.e.~whether we can create an arbitrary state with it.
The best illustration of this concept I know is the picture below
(\href{https://arxiv.org/abs/1905.10876}{source Sim et al.}). It shows
how changing the circuit changes the amount of the space that we can
cover.

\begin{figure}
\centering
\includegraphics{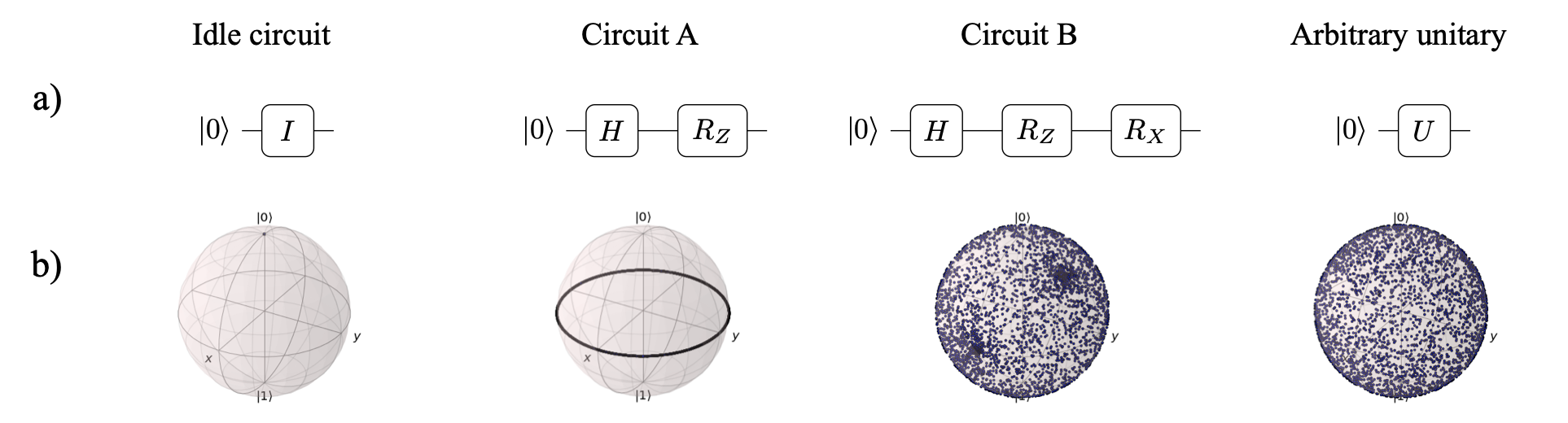}
\caption{Levels of expressibility of various circuits}
\end{figure}

Another useful metric is the entangling capability of an ansatz. We want
the ansatz to be able to produce highly entangled states, as the more
entanglement there is, the more ``quantum'' it is and potentially more
useful.

These metrics are the best we have, but unfortunately, they are not
really very good. Expressibility might not be a good metric for large
circuits as more expressive circuits might be also more prone to barren
plateaus or have more parameters and be harder to optimize. Entangling
capability---well, creating circuits that are hard to simulate just for
its own sake doesn't make much sense.

Therefore, we can use these two metrics as something that can help us
rule out bad ansatzes, but not necessarily find the best ones. Some
further reading: \href{https://arxiv.org/abs/2102.01659}{Haug et al.}
and \href{https://arxiv.org/abs/2001.00550}{Cerezo et al.}

When it comes to cost, typical metrics to measure the cost are:

\begin{itemize}
    \tightlist
    \item Circuit depth---the more shallow the circuit, the better chance that noise won't destroy our quantum state.
    \item Circuit connectivity---since on many NISQ devices we cannot directly connect arbitrary qubits, it might be beneficial to have an ansatz which only requires connectivity between nearest neighbors.\footnote{This depends on the hardware type, as it's a much bigger issue for superconducting qubits rather than ion traps.}
    \item Number of parameters---fewer parameters mean an easier job for optimization.\footnote{Though there are some counterarguments, see
  \href{https://arxiv.org/abs/2010.00157}{Kim et al} or
  \href{https://arxiv.org/abs/2105.01114}{Lee et al.}}
    \item Number of
two-qubit gates---usually two-qubit gates are much more noisy than
one-qubit gates, so this number is often used to compare different
circuits.
    \item Gate types---it's a little bit more indirect, but if we have
an ansatz which uses gates that are not available on our hardware, we
will need to decompose gates, which increases circuit depth.
\end{itemize}

\hypertarget{how-to-design-a-good-ansatz}{%
\subsubsection{How to design a good
ansatz}\label{how-to-design-a-good-ansatz}}

Now we want to have an ansatz for which expressibility and entangling
capabilities are high, while we keep all the associated costs to a
minimum, right?

Easier said than done :) How do you come up with the idea for the ansatz
in the first place? At the moment there are two main schools:
problem-motivated and hardware-motivated ansatz design.

To design a problem-motivated ansatz, you do things like: 
\begin{itemize}
    \tightlist
    \item Thinking hard about the problem you want to solve.
    \item Analyzing whether there is any structure in the problem you can exploit
    \item Reading existing literature on ``classical'' (meaning not QC) methods for solving this problem to get inspired.
\end{itemize}

To design a hardware-motivated ansatz, you do things like: 
\begin{itemize}
    \tightlist
    \item Memorizing technical specs of the device you have
    \item Figuring out how to make maximally expressive ansatz given the hardware constraints
    \item Reading existing literature on this type of hardware to find some tricks you can exploit or traps you can fall into
\end{itemize}

As always, both methods have some pros and cons that I've summarized
below, and combining both of them probably is the best solution.

\textbf{Problem-motivated}

Pros:

\begin{itemize}
\tightlist
\item Can exploit problem structure
\item Can be used with different devices
\item Doesn't require intricate knowledge about the hardware
\end{itemize}

Cons:
\begin{itemize}
    \tightlist
    \item Might not be implementable on specific hardware
    \item Might perform poor on real hardware while working perfectly in theory
    \item Probably will work only for a narrow class of problems
\end{itemize}

\textbf{Hardware-motivated}

Pros: 
\begin{itemize}
    \tightlist
    \item Allows to squeeze the most out of the device
    \item Doesn't require intricate knowledge about the problem domain
    \item Can be used for solving multiple problems
\end{itemize}

Cons: 
\begin{itemize}
    \tightlist
    \item Might not take advantage of the problem's structure
    \item  Is probably useful only for a very specific device
\end{itemize}

One consequence of coming from the problem-focused approach is that you
don't necessarily want to have the maximally expressible ansatz. While
such an ansatz would by definition cover your solution, it will also
cover all the solutions that you might a priori know are useless. Often
we have some knowledge about the problem and we know that a certain
family of states does not contain the ground state, so we can make our
ansatz simpler (and hence lower the cost) by designing it in a way that
excludes such states.

\hypertarget{how-to-get-a-good-ansatz-without-designing-it}{%
\subsubsection{How to get a good ansatz without designing
it?}\label{how-to-get-a-good-ansatz-without-designing-it}}

There's yet another method for finding a good ansatz---let the algorithm
figure it out. To do that, we use what are called ``adaptive''
algorithms. The main idea is that we modify not only the parameters of
the circuit, but also the structure of the circuit itself during the
optimization.

One example of such a method and the first such algorithm proposed is a
method called ADAPT-VQE. Here's a general description of how it works:\footnote{For a more detailed description please see
\href{https://www.nature.com/articles/s41467-019-10988-2}{Grimsley et
al.}}

\begin{enumerate}
\def\labelenumi{\arabic{enumi}.}
\tightlist
\item
  Define an ``operator pool''---this basically contains the bricks that
  you'll be building your algorithm from.
\item
  Create some initial reference state as the first iteration of your
  ansatz.
\item
  Calculate the gradient of the expectation value for each operator in
  the pool using your ansatz.
\item
  Add the operator with the biggest gradient to the ansatz (along with a
  new variational parameter).
\item
  Run ``regular VQE'' with this ansatz to optimize all the parameters of
  the ansatz.
\item
  Go to step 3.
\end{enumerate}

Another example of an adaptive method is
\href{https://arxiv.org/abs/2010.00629}{PECT}, which I have mentioned in
the previous section,
\hyperlink{section-04}{VQAs---how do they work?}.

Here are some comments about these methods:

\begin{itemize}
    \item These algorithms are
usually more difficult to implement on the classical side, as they
involve more than just taking a parametrized circuit and adjusting the
parameters, but also changing its structure.

    \item It's not obvious whether
such an approach actually yields better results or produces them faster.
There are two effects that are at play---on one hand, the algorithm
should need more time than a conventional one because it needs to find
the correct structure of the ansatz and optimize its parameters at the
same time (in point 5 of ADAPT-VQE we're running a full, regular VQE!).
On the other hand, doing both at the same time makes the optimization
process actually simpler as it doesn't introduce (or gets rid of)
parameters that are useless. In principle it should find better
solutions, however, it might take longer---it's hard to say before
actually running it for a specific problem.

    \item These algorithms seem
especially promising for NISQ devices, as they produce ansatzes that are
shallower than those designed by hand (and hence there's less room for
noise), but also they naturally find circuits that take into account all
the quirks of a specific device. For example, in ADAPT-VQE you can
create an operator pool in such a way that it leads to
hardware-efficient ansatzes (see
\href{https://journals.aps.org/prxquantum/abstract/10.1103/PRXQuantum.2.020310}{Tang et al.}).
\end{itemize}

If you want to learn more about ADAPT-VQE and the problem of ansatz
design in general, Sophia Economou (one of the authors of ADAPT-VQE)
gave an \href{https://youtu.be/pDI6uFW2bu4?si=YJgZPMt2CunU9Blr}{excellent 25-minute talk} about these during one of Quantum
Research Seminar Toronto (QRST).

As a side note, this approach reminds me of a NEAT
(\href{http://nn.cs.utexas.edu/downloads/papers/stanley.ec02.pdf}{Neuroevolution
of Augmenting Topologies}) algorithm used for classical neural networks.
You can find \href{https://www.youtube.com/watch?v=qv6UVOQ0F44}{an
excellent video} showing how a NEAT algorithm learns to play Mario.

\hypertarget{interpolating-algorithms}{%
\subsection{Interpolating algorithms}\label{interpolating-algorithms}}

The approach which makes me particularly excited is something I'll call
``interpolating algorithms.''\footnote{There's no proper name for this
  class of algorithms in the literature yet.} What do we mean by
``interpolating''? The fact that the performance of these algorithms can
interpolate between that of near-term and far-term algorithms. You can
adjust some hyperparameters of the algorithm depending on what hardware
you're running on, so no matter what stage of development of quantum
hardware we're currently at (from now to perfect qubits), you can find
hyperparameters that will allow you to actually run the algorithm within
the limitations of the hardware and squeeze most out of it.

Imagine you have circuit A that allows you to estimate the ground state
of some Hamiltonian. If you want to estimate it to precision
$\epsilon$, you would need to run circuit A
$N_A=\frac{1}{\epsilon^2}$ times. You also have circuit B, that allows
you to get the same precision, but you only need to run it
$N_B = \log(\frac{1}{\epsilon})$ times. However, circuit B is
$\frac{1}{\epsilon}$ times longer and hence not very practical for the
devices we have today. But what if we had some way to interpolate
between these two approaches so that we would have a parameter that
makes the circuit longer, but also decreases the number of samples
needed?

Let's see what it could look like in practice. Let's say
$\epsilon=10^{-3}$. This means we need to run circuit A 1,000,000
times and each execution of circuit A takes 1ms. So the total runtime
will be 1,000s. For circuit B we need to run it $\log(1000)$ times, which
is just 3 times (sic!), though each execution takes 1 second. Hence we
would get our result in just 3 seconds. Unfortunately, as we have seen
earlier, our circuit probably won't be able to run for as long as 1
second anytime soon. However, this class of algorithms gives us a way to
construct such a circuit that it requires less measurements but its
execution time still fits our hardware.

While this explanation is \textbf{oversimplified}, going into more
details is, again, way beyond the scope of this article. If you'd like
to learn more, three examples of similar approach are
\href{https://arxiv.org/abs/1802.00171}{``alpha-VQE''},
\href{https://arxiv.org/abs/2012.03348}{``Power Law Amplitude
Estimation''} and
\href{https://www.zapatacomputing.com/publications/juice/}{``Bayesian
Inference with Engineered Likelihood Functions for Robust Amplitude
Estimation''} by my colleagues from Zapata. (I strongly recommend
watching \href{https://www.youtube.com/watch?v=RifDO1zBYjI}{this
3.5-minute video} explaining the gist of it).

\hypertarget{other-issues}{%
\subsection{Other issues}\label{other-issues}}

This article has been quite dense, so here I wanted to just point to
some other final issues that are prevalent in contemporary research,
without spending too much time on any of them:

\hypertarget{lack-of-common-benchmarks-and-standardization}{%
\subsubsection{Lack of common benchmarks and
standardization}\label{lack-of-common-benchmarks-and-standardization}}

As you could see from the section about optimizers, it's really hard to
benchmark certain solutions. It's not only extremely costly to compare
multiple methods, but it's also really hard to design such an experiment
in a way that makes the comparison fair and broadly useful.

On top of that, as can be expected for a discipline at such an early
stage, we lack standardization. This is healthy, as it allows for more
experimentation and exploration, but it makes it much harder to compare
results, as you basically never compare apples to apples.

\hypertarget{we-have-no-idea-what-were-doing}{%
\subsubsection{We have no idea what we're
doing}\label{we-have-no-idea-what-were-doing}}

The truth is that QC is a totally new paradigm of computation that we
don't comprehend. I`ve had the opportunity to talk with some excellent
researchers in the field and while their level of understanding and
intuition about these matters is far beyond my reach, they're quite open
about the fact that we all just started scratching the surface. There
are some fundamental questions that no one knows the answers to, which
is both a challenge and a source of excitement.

\hypertarget{scaling-for-bigger-devices}{%
\subsubsection{Scaling for bigger
devices}\label{scaling-for-bigger-devices}}

In most of our research, we're limited in what we can analyze by the
size of the devices we're able to simulate with computers (rarely beyond
30 qubits). Bigger devices are extremely scarce---there is literally a
handful of them in the world---so we have very little understanding of
how these methods will scale beyond 50 qubits. And while some results
seem independent of size or we have some theory that explains how they
will behave, for many we don't. And right now there's no other way of
knowing other than building bigger devices and trying them out.

\subsection{Closing notes}
Thank you for going through this whole chapter! If this was of interest for you, you’ll definitely like the next chapter of these notes, where I focused on VQE. 

\newpage

\hypertarget{section-6}{\section{VQE --- Challenges and state of research}}

In the previous section, we talked about challenges associated with Variational Quantum Algorithms in general. In this section, we'll focus more specifically on Variational Quantum Eigensolver (VQE). All of the things we've covered previously are relevant for VQE as well, but here we'll focus on challenges and progress that are more specific for VQE.

If you don't remember exactly how VQE works, you can refresh your memory with the first section of the document---\hyperlink{section-02}{VQE explained}.

One more thing---while I was working on this article a new preprint came out on arxiv by \href{https://arxiv.org/abs/2111.05176}{Tilly et al}. It is a really great piece of work and it made my life much easier. Therefore I consider this article a light-weight review of the current state of VQE, but if you'd prefer an actual deep scientific dive, that's the place to go!

Without further ado, let's start with\ldots{}

\hypertarget{ansatz-design}{%
\subsection{Ansatz design}\label{ansatz-design}}

Ansatz design is a big part of ongoing research for VQE. However, I won't be going into more details in this article, for a couple of reasons:

\begin{itemize}
\tightlist
\item
  I covered ansatz design in
  \hyperlink{section-05}{the
  previous part about VQA challenges}.
\item
  Reviewing various ansatzes would require some introduction to quantum
  chemistry, which is beyond the scope of this piece.
\end{itemize}

If anyone is interested, you can find a great review in
\href{https://arxiv.org/abs/2103.08505}{Fedorov et al.} or in more
details in \href{https://arxiv.org/abs/2109.15176}{Anand et al.} And if
you need some introduction to quantum chemistry, the first part of
\href{https://arxiv.org/abs/1812.09976}{this review by Cao et al} is a
decent starter.

\hypertarget{hamiltonian-construction}{%
\subsection{Hamiltonian construction}\label{hamiltonian-construction}}

This is a topic that requires way more quantum chemistry background than
I have. I found section 3 and 4 of
\href{https://arxiv.org/abs/2111.05176}{Tilly et al.} covers this topic
way better than I would, so if you're looking for a good reference,
that's the place. It contains explanations of concepts such as the
Jordan-Wigner or Bravyi-Kitaev transformations, which allow us to
translate Hamiltonians from the realm of quantum chemistry to quantum
computing and which are definitely some terms I've come across a lot but
did not understand for quite a long time.

It's also worth checking out
\href{https://warrenalphonso.github.io/qc/hubbard}{this tutorial} to the
Fermi-Hubbard model by Warren Alphonso, it also explains some of these
concepts very well.

Ok, after these two disappointingly short sections, let's get into
concepts which do not require background in quantum chemistry!

\hypertarget{measurement-problem}{%
\subsection{Measurement problem}\label{measurement-problem}}

VQE is basically an estimation problem---given a parameterized quantum
circuit you want to estimate the expectation value of an operator
(Hamiltonian) with certain precision. Two big issues are how to design
the circuit and how to get the right parameters, but we've talked about
them a lot in
\hyperlink{section-05}{the
previous section on VQA challenges}. Now let's talk about ``certain
precision''---we'll work on a concrete example of a $CO_2$ molecule.
To be clear, this problem is associated with every ``cost function
evaluation'' in VQE that we do while optimizing parameters, not just the
final result.

A standard accuracy that we desire in quantum chemistry is called
``chemical accuracy'' and is equal to 1kcal/mol,\footnote{Why does anyone
  use this particular number? As it gives you order-of-magnitude correct
  answers to reactions rates, based on Arrhenius equation at room
  temperature.} which is roughly equal to 1.6 mHa\footnote{mHa is
  \href{https://en.wikipedia.org/wiki/Hartree}{a unit of energy} used in
  quantum chemistry.}). But for this exampe we want to be extra safe---in the end, these quantum devices are far from perfect, so let's aim for
an accuracy of $ \epsilon = 0.5 mHa$. Fortunately, the relation
between the number of measurements we need to do and the target accuracy
is fairly simple: $ M = \frac{K}{\epsilon^2}$. $K$ is a constant
that depends on a molecule of interest, the estimation strategy
used\footnote{For example it might change depending on whether you group
  measurements or not.} and certain other assumptions about the problem.
In \href{https://arxiv.org/abs/2012.04001}{Gonthier et al} they
calculated that for $CO_2$ it's $K=8000$. After substituting values
in the equation above it turns out we need 32 billions measurements to
properly estimate the expectation value of interest. This number seems
high, but to give you a better sense of how much it actually is, this is
equivalent to (given some reasonable assumptions described in
\href{https://arxiv.org/abs/2012.04001}{Gonthier et al} ) roughly 39
days of calculations.

Let's stop here for a moment to appreciate that.

We need to run a quantum computer, for 39 days, non stop, to get a
\textbf{single} estimation of energy.\footnote{Nowadays you need to
  recalibrate your device every couple of hours in order to make sure
  its performance doesn't degregate,
  \href{https://twitter.com/mstechly/status/1466883105072066561?s=20}{source}.}

And you know how many energy estimations you need to perform to actually
find good parameters for VQE if we run an optimization loop? I tried it
for a simple $H_2$ molecule with much more optimistic assumptions and
got 87, but it can easily increase to thousands.

Therefore, one of the biggest challenges for VQE is how to reduce this
number of measurements and fortunately, there are a couple of methods to
do this. I described some of them in
\hyperlink{section-05}{the
previous section on VQA challenges}, like ``interpolating
algorithms,'' and below you can find the description of how grouping and
measurements allocation works. They tackle different components in the
``overall VQA structure'' (see
\hyperlink{section-04}{the previous section ``VQAs---how do they work?''}, so some of these methods can be used in
conjunction.

\hypertarget{grouping}{%
\subsection{Grouping}\label{grouping}}

VQE is all about finding the lowest eigenvalue of a given Hamiltonian,
which is expressed as a sum of Pauli terms. For each term we need to run
a separate circuit and perform multiple measurements. Then we combine
the results of all the runs and we calculate the energy of a given state
based on that.

However, there's a big problem with this naive way of using VQE. Let's
say our Hamiltonian looks like this: $ X_0 + Y_1 + Z_2$. It consists
of three terms and in the most basic implementation we would just run a
separate circuit for each term. However, each term affects a different
qubit, and hence, you can run just one circuit and measure all of them
at the same time (we say that these terms are \textbf{co-measurable}).

Let's consider a slightly more complicated example:

\begin{equation} Z_0 \cdot X_1 + Y_1 \cdot X_2 + X_2 \cdot X_3 + X_0 + Z_3 \end{equation}

There are a couple of ways to gather these terms in co-measurable
groups, one being:

Group 1: $ Z_0 \cdot X_1 + Z_3 $ \\
Group 2: $ Y_1 \cdot X_2 + X_0 $ \\
Group 3: $ X_2 \cdot X_3 $

And another being:

Group 1: $ Z_0 \cdot X_1 + X_2 \cdot X_3 $ \\
Group 2: $ Y_1 \cdot X_2 + X_0 + Z_3 $

Obviously the second allows us to perform a smaller number of
evaluations to get the same result.

It turns out that performing grouping efficiently is already a pretty
hard problem---there has been a lot of research in recent years to find
better algorithms to do this. You can find a summary of these methods in
\href{https://www.nature.com/articles/s41534-020-00341-7/tables/1}{Table
1 of Huggins et al}. Some key takeaways from this table for our
discussion:

\begin{itemize}
\tightlist
\item
  Number of groups depends polynomially on the number of
  qubits.\footnote{It's actually in the number of spin orbitals, which
    corresponds to the number of qubits in the most commonly used
    Jordan-Wigner transform.} While ``polynomial scaling'' usually
  means ``good'' in algorithms, $N^4$ might be actually pretty brutal
  in practice.
\item
  There are different tradeoffs that we need to take into account when
  choosing grouping methods. Some work best with certain QPU topologies,
  some others increase the number of gates, etc.
\item
  Smaller number of groups does not necessarily imply a smaller total
  number of overall measurements! More groups might actually allow us to
  get higher precision by achieving better energy estimation per group
  with the same number of measurements.\footnote{For more details see
    \href{https://www.nature.com/articles/s41534-020-00341-7/tables/1}{Huggins
    et al} and section 5.3 from Tilly et al., or section V of
    \href{https://journals.aps.org/prxquantum/pdf/10.1103/PRXQuantum.2.040320}{Yen
    et al.}}
\end{itemize}

Section 5 of \href{https://arxiv.org/abs/2111.05176}{Tilly et al.}
provides an introduction into some of the grouping methods, so if this
topic is of interest for you, that's where you should go.

\hypertarget{measurements-allocation}{%
\subsection{Measurements allocation}\label{measurements-allocation}}

Let's say we have our groups now and we want to actually run the
circuits and get some measurements. Should we just go ahead and do that?
Well, not necessarily. Let's say we have a Hamiltonian that looks like
this:

\begin{equation} H = H_1 + H_2 + H_3 = 5 Z_0 + 3 Z_1 + 2 Z_0 \cdot Z_1 \end{equation}

And a circuit which creates a state:

\begin{equation} | \psi \rangle = \cos{\left(\frac{\pi}{6}\right)} |00 \rangle +  \sin{\left(\frac{\pi}{6}\right)} |10 \rangle \end{equation}

By calculating expectation values by hand, we can see that

\begin{equation}
\begin{aligned}
    \langle H_1 \rangle &= 5 \langle Z_0 \rangle = 5 \cdot 0.5 = 2.5 \\
    \langle H_2 \rangle &= 3  \langle Z_1 \rangle = 3 \cdot 1 = 3 \\
    \langle H_3 \rangle &= 2 \langle Z_0 \cdot Z_1 \rangle = 2 \cdot 0.5 = 1 \\
    \langle H \rangle &= 6.5 
\end{aligned}
\end{equation}

Looks easy, right?

But now let's see what happens if we try to measure these on a
QPU,\footnote{Well, to be honest with you, I have used a simulator ;)}
with 100 shots for each term.\footnote{We often say ``shots'' instead of
  ``measurements'' in QC.}

\begin{equation}
\begin{aligned}
    \langle H_1 \rangle &= 5 \cdot 0.66 = 3.3 \\
    \langle H_2 \rangle &= 3 \cdot 1.0 = 3 \\
    \langle H_3 \rangle &= 2 \cdot 0.58 = 1.16 \\
    \langle H \rangle &= 7.46
\end{aligned}
\end{equation}

Ok, let's try once again, now with 1000 shots:

\begin{equation}
\begin{aligned} 
 \langle H_1 \rangle &= 5 \cdot 0.488 = 2.44 \\
 \langle H_2 \rangle &= 3 \cdot 1.0 = 3 \\
 \langle H_3 \rangle &= 2 \cdot 0.504 = 1.008 \\
 \langle H \rangle &= 6.448
\end{aligned}
\end{equation}

As we can see, the estimates we got are not exact and they depend on the
number of shots. So we can increase the accuracy by increasing the
number of shots, problem solved, right? Well, not really, as when we
increase the number of shots, we also increase the time of running
calculations.

Can we be smarter about it? Absolutely\ldots{}

One can use a fixed total number of shots (budget) and allocate it among
various groups. So let's say we have 300 shots in our budget, how should
we allocate it? The first approach is to do a uniform allocation,
i.e.~100 shots for each operator. This is what we did in the first
example. But we can do better than this. We know that the operators have
weights 5, 3 and 2. So it would make sense to put most shots from our
budget towards the first operator and least shots towards the last one.
Let's say we do that proportionally---all weights add up to 10, so we
use $\frac{5}{10}$ shots for $H_1$ (150), $\frac{3}{10} $ for
$H_2$ (90) and $\frac{2}{10}$ for $H_3$ (60). What did we get?

\begin{equation}
\begin{aligned}
 \langle H_1 \rangle &= 5 \cdot 0.53 = 2.65 \\
 \langle H_2 \rangle &= 3 \cdot 1.0 = 3 \\
 \langle H_3 \rangle &= 2 \cdot 0.66 = 1.32 \\
 \langle H \rangle &= 6.97
\end{aligned}
\end{equation}

This result is better from what we got from just using 100 shots per
operator. But this could be a random fluke, so in order to evaluate
that, we should calculate the standard deviation (\textbf{std}) of our
final energies. So after repeating my experiment 10000 times, it turns
out that for uniform shot allocation, std is 0.469 and for proportional
shot allocation it's 0.420.

You might wonder---proportional allocation is definitely better than
uniform, but what is the optimal allocation strategy? Well, as always,
the answer is not simple. It will generally depend on the exact grouping
method used, and getting a perfect answer is generally not possible:
indeed, the number of shots necessary depends on the standard deviation
of each individual term, which itself depends on the current
wavefunction on the quantum computer. In some cases it is possible to
mathematically solve for the optimal allocation by making some
assumptions about these standard deviations, for more details see
\href{https://iopscience.iop.org/article/10.1088/1367-2630/aab919}{Rubin
et al.} section 5.1.
\href{https://arxiv.org/pdf/2004.06252.pdf}{Arrasmith et al.} is also a
good reference for this problem.

What we did so far was allocation of shots for a single evaluation of
energy. But what if we wanted to have a shot budget for the whole
optimization process? Perhaps at the beginning of the optimization we
don't need as much precision as towards the end? One approach to this
has been proposed in \href{https://arxiv.org/abs/1912.06007}{Cade et
al.} (but also described very well in
\href{https://arxiv.org/abs/2111.13454}{Bonet-Monroig et al.} ), which
authors call ``3-stage-approach.'' It works by defining a shot budget
for the whole optimization process and later dividing the optimization
in 3 stages. As an example, they use 100, 1000 and 10000 samples in
stages 1, 2 and 3 accordingly. However, different phases have assigned
different numbers of energy evaluations. The ratio they proposed is
10:3:1, so for the first phase they use $\frac{10}{10+3+1}$ of all
energy evaluations, in the second $\frac{3}{14}$ and just
$\frac{1}{14}$ in the last.

\begin{figure}
\centering
\includegraphics{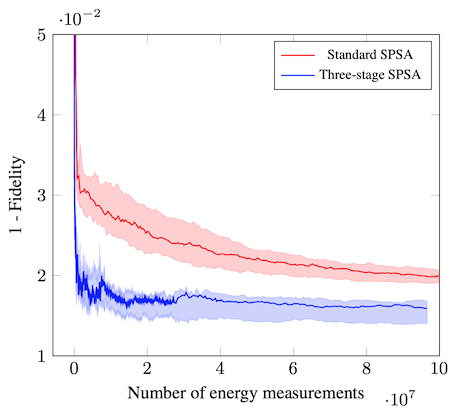}
\caption{Experimental results of the 3-stage approach. This picture is Fig. 6 from
\href{https://arxiv.org/abs/1912.06007}{Cade et al.}}
\label{fig:three_stage}
\end{figure}

This allows for much more efficient usage of the available resources, as
you can see in Fig. \ref{fig:three_stage}.

This is just one way to approach this problem, you can find others in
\href{https://arxiv.org/abs/2108.10434}{Gu et al.} or
\href{https://arxiv.org/abs/2004.06252}{Arrasmith et al.}

\hypertarget{vqe-beyond-ground-states}{%
\subsection{VQE beyond ground states}\label{vqe-beyond-ground-states}}

VQE is usually used for finding the ground state of a given Hamiltonian.
However, it could also be used to solve other problems, such as:
\begin{itemize}
    \tightlist
    \item Finding excited states of the system
    \item Calculating vibrational spectrum of a molecule
\end{itemize}

\hypertarget{finding-excited-states-of-a-system}{%
\subsubsection{Finding excited states of a
system}\label{finding-excited-states-of-a-system}}

In many cases we're interested in finding the excited states of a given
system, not just the ground state. For example, in order to get
information about the color of a material we need to know what the
energy difference between the ground stare and first excited state of
the molecule is. One approach to do this has been described in
\href{https://arxiv.org/abs/1603.05681}{McClean et al.} and
\href{https://journals.aps.org/prx/pdf/10.1103/PhysRevX.8.011021}{Colless
et al.} It basically boils down to the following steps:

\begin{enumerate}
\def\labelenumi{\arabic{enumi}.}
\tightlist
\item
  Run a regular VQE in order to find the ground state of the system.
\item
  Once the algorithm has converged, modify the Hamiltonian in such a way
  that it contains information about excited states.
\item
  Measure new Hamiltonian.
\item
  Perform some classical postprocessing of the results.
\item
  Voila!
\end{enumerate}

There are many other variations of VQE, if this is of interest to you I
recommend section 9 of \href{https://arxiv.org/abs/2111.05176}{Tilly et
al.}

\hypertarget{calculating-vibrational-modes}{%
\subsubsection{Calculating vibrational
modes}\label{calculating-vibrational-modes}}

Another interesting property that we might want to learn is associated
with the vibrations existing in the molecule. What is it about? Well,
atoms in the molecules are not rigid, they are in constant motion in
respect to each other and this movement, the vibrations, can affect the
properties of the molecule in non-trivial ways. This is important for
topics such as astrochemistry or modeling fuel combustion, but also
particularly useful for a class of molecules that have certain
structures.

In principle, we can use the old good VQE to solve it---we just need
the to use a different Hamiltonian, which describes those vibrational
modes. However, since this problem is different from the electronic
structure problem that we usually solve, there are somewhat different
considerations, e.g.~we use different mappings to construct the
Hamiltonian or we need to pay more attention to the excited states.

If you would like to learn more about this topic, I talked about it with
Nicolas Sawaya, research scientists working at Intel in
\href{/episodes.html\#NISP2}{my podcast} and here you can find
\href{https://arxiv.org/abs/2009.05066}{his paper} where he describes
these ideas in more details.

\hypertarget{outside-of-quantum-chemistry}{%
\subsubsection{Outside of quantum
chemistry}\label{outside-of-quantum-chemistry}}

In principle you can use VQE for any situation where you have a matrix
and you want to learn what its lowest eigenvalue is.

One example can be using VQE for solving combinatorial optimization
problems. Usually, we think about using QAOA for such problems, but QAOA
also takes a Hamiltonian as a problem, so we could use VQE to find its
ground state as well. If you'd like to better understand the difference,
I have covered the difference between these two methods in the previous
\hyperlink{section-03}{``QAOA explained'' section}. An example of using VQE for
combinatorial optimization problems can be found e.g.: in
\href{https://arxiv.org/abs/2102.05566}{Liu et al.}

Apart from that, I have not encountered any particularly exciting or promising use cases for VQE outside of chemistry, so I'm not going to talk about it too much. However, if you know one, please leave me a comment under this article and I will be happy to include it :)

\subsection{Closing notes}
As it turns out, getting to the forefront of VQE actually requires more quantum chemistry background than I expected, hence section ended up being shorter than I expected.

If you'd like to get deeper into the topic, follow the references I left throughout the text and make sure to spend some time with \href{https://arxiv.org/abs/2111.05176}{Tilly et al}!

\newpage

\hypertarget{section-7}{\section{QAOA --- Challenges and state of research}}

In this section, we will delve into the details of QAOA---variational quantum algorithm for solving combinatorial optimization problems.

Update: about a month after writing this a review paper on the topic came out by \href{https://arxiv.org/abs/2306.09198}{Blekos et al.} I don't refer to it here, but it covers many of the variants of QAOA and is much more comprehensive than this blogpost could ever be! I recommend reading it to anyone interested!

\hypertarget{qaoa-vs-qaoa}{%
\subsection{QAOA vs QAOA}\label{qaoa-vs-qaoa}}

Before we even get into a detailed discussion about the algorithm
itself, let's first talk about naming. The original QAOA algorithm
presented by \href{https://arxiv.org/abs/1411.4028}{Farhi et al.} is
called Quantum Approximate Optimization Algorithm. However, in 2018
\href{https://www.mdpi.com/1999-4893/12/2/34}{Stuart Hadfield et al.}
presented a more general algorithm which is called Quantum Alternating
Operator Ansatz. Which also spells ``QAOA.'' You might have a couple of
questions now:

\begin{itemize}
\tightlist
\item
  Why would they do that?
\item
  What's the difference between these two?
\item
  If I hear ``QAOA'' what exactly do people mean by it?
\end{itemize}

Let me quote Hadfield's paper:

\begin{quote}
We reworked the original acronym so that ``QAOA'' continues to apply to
both prior work and future work to be done in this more general
framework. More generally, the reworked acronym refers to a set of
states representable in a certain form, and so can be used without
confusion in contexts other than approximate optimization, e.g., exact
optimization and sampling.
\end{quote}

This means that ``the original QAOA'' is just a subset of ``a more
general QAOA.'' One of the differences is that in the original QAOA the
Hamiltonians which we use are defined in a specific way, while the more
general framework allows for much more flexibility. Plus the more
general framework is not designed specifically for ``approximate
optimization'' but also for other uses.

So if you hear ``QAOA'' which one people mean? Well, I'd argue that most
of the time people mean the ``more general QAOA,'' even if they don't
necessarily realize that or what's the actual difference. I think that
most of the time if someone means ``original QAOA'' specifically they
refer to it as ``vanilla QAOA,'' ``basic QAOA,'' ``Farhi's QAOA'' or
something similar.

\hypertarget{problem-related-issues}{%
\subsection{Problem-related issues}\label{problem-related-issues}}

There are three topics I wanted to talk about, which are more about the
problems we're trying to solve rather than the QAOA algorithm itself.
Dealing with these challenges is important for any algorithm which tries
to solve optimization problems, quantum or classical. Some are more
apparent in one class of algorithms, some in others, but over the years
I learned that they all can heavily influence the results of your
optimization routine, so I think it's worth discussing them here.

These are:

\begin{itemize}
\tightlist
\item Problem formulation
\item Constraints
\item Problem encoding
\end{itemize}

\hypertarget{problem-formulation}{%
\subsubsection{Problem formulation}\label{problem-formulation}}

In general, when you try to solve optimization problems, there is a huge
payoff for understanding your problem better and coming up with a better
formulation. By formulation I mean understanding the real-world problem
and expressing it in a way that an algorithm can solve and maybe adding
some problem-specific improvements to the algorithm. This is in
opposition to coming up with a better optimization method for a given
class of problems, fine tuning it, etc.

Let's take a look at the following problem---you are in charge of IT
infrastructure at a campus. You have a team of 5 people and you have to
service equipment in various buildings at the campus. Every day you get
calls from people around the campus about the issues that your team
needs to solve. For each request you:

\begin{itemize}
\tightlist
\item
  identify where the issue happened
\item
  estimate how long it will take to solve it.
\end{itemize}

The problem is how to assign your team in an optimal way?

One approach would be to model this as Capacitated Vehicle Routing
Problem (CVRP, it's a more complicated variant of Traveling Salesman
Problem).

In such formulation you can say that:

\begin{itemize}
\tightlist
\item
  The capacity of each person is 6 hours per day
\item
  Each person shouldn't spend more than 1 hour driving around the
  campus, which translates into a constraint that the route itself needs
  to be under 1 hour.
\item
  You are fine with not solving all the requests every day (i.e.~not
  visiting all the cities)
\end{itemize}

You also decide that your cost function is the number of issues you can
solve per day.

Do you think it's a good formulation?

Well, it depends. Below you can find a couple of questions that one
needs to ask in order to model this specific problem better:

\begin{itemize}
\tightlist
\item
  Are all the people on the team equally skilled? If one of them is new
  to the job and another one is a veteran, you should probably start
  expressing the time estimates differently, e.g.~depending on one's
  experience or in abstract ``points'' rather than time. In the latter
  case, the expert person would have higher capacity.
\item
  How accurate are your time estimates for solving those problems? If
  they're not very accurate, the mathematical model might have a huge
  mismatch with reality. If they're not very accurate, can you model how
  inaccurate they are, perhaps using a normal distribution?
\item
  How much time does it take to travel between different buildings on
  campus? If there are 3 buildings which are all 5 minutes apart,
  perhaps it's worth modeling them as one node---that would decrease
  the complexity of the graph. If getting from one building to another
  takes less than 10 minutes, perhaps it's negligible and you would be
  better off modeling it as a scheduling problem and not CVRP?
\item
  Do issues have different importance?
\item
  Given that your cost function depends on the number of problem solved
  per day, it might lead to a situation where one 6-hour long problem
  never gets scheduled, because there are always multiple 1-hour
  problems popping everywhere. Is that acceptable?
\item
  Can some problems be solved remotely? If so, how would you model that?
\item
  What if a particular problem takes more than a day to solve?
\end{itemize}

I hope that's enough to make this point, perhaps the most important from
the whole blogpost:

You can have the absolutely best algorithm for solving combinatorial
optimization problems, but in the end, if your problem model doesn't
match reality well, who cares?

\hypertarget{constraints}{%
\subsubsection{Constraints}\label{constraints}}

Imagine you have a store and you sell two products: X and Y. You want to
maximize the profits from selling them. Profit equals revenue minus
cost:

\begin{equation} P = R - C \end{equation}

Revenue is equal to the amount of X sold times its price and same for Y.

\begin{equation} R = p_x x + p_y y \end{equation}

The costs are trickier, as the more products you have, the pricier it is
to store it:

\begin{equation} C = c_x x^2 + c_y y^2 \end{equation}

Your goal is to order the optimal amount of X and Y so that you can
maximize your profit.

Well---that's a simple quadratic function. You can solve it
analytically or just use a simple gradient descent optimizer and find it
in no time. But let's add some constraints to make it more realistic.
Let's say you have a limited capacity of your storage, so:

\begin{equation} 0 < x + y < C_1 \end{equation}

Also, they shouldn't be negative: $ x \ge 0 $ and $ y \ge 0 $.

On top of that, your supplier tells you that process for creating X and
Y is correlated and you can't just buy only X or only Y, but there's
also some upper cap on their production in relation to each other, so
you need to add the following constraint:

\begin{equation} C_2 \le x * y \le C3 \end{equation}

Oh, and on top of that, X is sold in discrete quantities ($x$ is an
integer), but Y is a liquid so you can buy any quantities ($y$ is a
real number).

Well, now our problem is much more interesting and it's not immediately
obvious how to solve it. As you might guess, people came up with some
techniques to deal with such constraints. Unfortunately not all of these
methods are applicable to all types of cost functions, all types of
constraints or all types of optimizers. As it's quite a broad topic,
here I'll focus only on two methods that you might encounter in
QAOA-related literature problems.

First of them works by using something called a ``penalty function.'' It
is a way of modifying your cost function, by introducing some artificial
terms which would ``penalize'' if the variables don't meet the
constraint. Let's use our favorite MaxCut. As you might recall from
previous parts, the cost function of MaxCut is:

\begin{equation} C = \sum_{(i,j) \in E } w_{ij} (1-x_i x_j),\end{equation}

where $E$ is a set of all the edges in the graph.

But what if we had a constraint that one group needs to consist of
exactly 5 elements? How could we express it mathematically? Since
$x_i$ is either 0 or 1, we can arbitrarily say that we want the group
represented by ones to have exactly 5 elements. In such a case we know
that $ \sum x_i = 5 $. We can also rewrite it as:
$ 5 - \sum x_ii = 0$. How do we incorporate this in our cost function?
Well, here's the trick. We transform it into the following term:

\begin{equation} P = P_1 (5 - \sum x_i)^2 \end{equation}

and then just add it to the cost function:

\begin{equation} C_{new} = C + P \end{equation}

Since we want to minimize the value of $C_{new}$, any solution $x$
which doesn't meet the constraint will make the value of $C_{new}$
higher, so it will be a worse solution. Since $P_1$ is an arbitrary
constant which doesn't have any real-world meaning, we can select it to
be arbitrarily high. E.g. if our values C are usually in the range of
0-10, we can set $P_1=100$, so that any minor violation of the
constraint would result in an extremely high penalty to the
cost.\footnote{In practice, tuning the values of the penalties constants
  might be a significant problem in itself. Too small and it doesn't
  serve its function. Too high and it might lead the optimizer to behave
  in unexpected ways. Not to mention interplay between multiple
  penalties!} Basically what we did is we transformed a constrained
problem (cost function C and our constraint) into an unconstrained one
(just cost\_function $C_{new}$ without any constraints).

This method has some pros and cons---it's relatively simple and it
works for basically any problem, as you can always modify your cost
function. However, it doesn't work well with all types of optimization
methods and is a little bit more tricky to use with inequality
constraints (i.e.~$ \ge, \le$). You can find examples of this being
incorporated into the cost function in everyone's favorite
\href{https://www.frontiersin.org/articles/10.3389/fphy.2014.00005/full}{``Ising
formulations of many NP problems'' by Andrew Lucas}, see e.g.: section
7.1 . Another example of how to do that is in section 3.1 of
\href{https://arxiv.org/abs/1804.09130}{this paper} (again by Stuart
Hadfield).\footnote{It also shows how to map real functions to Ising
  Hamiltonians---something I didn't know is possible and find it pretty
  cool!}

There is another way to avoid getting solutions you don't want, and it
works particularly well with problems where solutions have some inherent
structure, as is the case for many combinatorial optimization problems.
Let's take the Traveling Salesman Problem (TSP) as an
example\footnote{If you are not familiar with it, you can find an
  explanation
  \href{https://en.wikipedia.org/wiki/Travelling_salesman_problem}{on
  wiki}.}. For 5 cities the solution can look like this: {[}5, 2, 3, 1,
4{]}, which specifies in what order we visit cities. An incorrect
solution might look like this: {[}5, 2, 3, 1, 2{]}---as you can see
we're visiting city 2 twice in this case, but we're missing city 4,
which violates the constraints.

How can we ensure our solution is always correct? There are two
components to the solution:

\begin{itemize}
\tightlist
\item
  Start from a valid solution (i.e.~one doesn't violate constraints)
\item
  Create new solutions by performing only operations which transform a
  valid solution into another valid solution.
\end{itemize}

In the case of TSP such operations would be swapping elements in the
list. So we could transform {[}5, 2, 3, 1, 4{]} into {[}4, 2, 3, 1,
5{]}, which is also a valid solution, but with this method we would
never get anything like {[}5, 2, 3, 1, 2{]}.

Easy to do with lists, but how does one do that with quantum states? I
talk more about it in
\protect\hyperlink{quantum-alternating-operator-ansatz}{the section
about alternative QAOA ansatz}!

\hypertarget{encoding}{%
\subsubsection{Encoding}\label{encoding}}

Another important topic is how do you encode your problem to fit a
particular algorithm. Let's think about how to describe the TSP problem
in a mathematical form and different ways to encode solutions to it.

The most natural (at least for me) way is to define a distance matrix. It is a matrix which tells us what's a difference between each pair of cities. For three cities it can look like this:

\begin{center}
    \begin{tabular}{l|lll}
      & A & B  & C  \\ \hline
    A & 0 & 5  & 7  \\
    B & 5 & 0  & 10 \\
    C & 7 & 10 & 0 
    \end{tabular}
\end{center}

We know that the distance from A to B is 5km, from B to C is 10km and
from A to C is 7km. How would we encode the solution? Simply as: {[}A,
B, C{]}, {[}B, C, A{]} etc.

We can easily switch to integer labels (A $\to$ 1,
B $\to$ 2, C $\to$ 3), which will make some math down the
road easier (like indexing rows and columns of the distance matrix).

Ok, but what if our machine has a limited number of integers it can work
with for expressing the solution. In the case of a quantum computer it's
actually just 0 or 1. So how would we encode the solution to our 3-city
problem?

Well, there are multiple ways to do this. The simpler one (perhaps) is
called ``one-hot encoding'' or ``unary encoding.'' In this encoding, we
encode each city like this:

\begin{itemize}
    \tightlist
    \item A $\to$ 100
    \item B $\to$ 010
    \item C $\to$ 001
\end{itemize}

Now if we want to represent a route: A $\to$ C $\to$
B, we can write it as ``100 001 010.'' Each bitstring will consist of
three blocks, each represent one moment in time. And then each block
will consist of three bits, each representing a city. So in this case
having ``001'' in the middle means, that we visit city ``C'' in the
second time step. This allows us to easily encode constraints, such as
``you can be only in one city at the time,'' as you just need to check
if any given block contains only a single 1.

However, one thing people often say when looking at the unary encoding
is that it's very space inefficient. Indeed, it will take you $n^2$
bits to encode a solution for the problem with $n$ cities. So perhaps
we could use binary encoding. In this case we will have:

\begin{itemize}
    \tightlist
    \item A $\to$ 00
    \item B $\to$ 01
    \item C $\to$ 10
\end{itemize}

Now, we use only 6 bits to encode the solution and it actually scales as
$n \cdot \log(n)$, which is pretty good!

There are, however, a couple of issues here. First, the bitstring ``11''
is undefined. You can write a route containing it, but it doesn't mean
anything. One way to deal with it is to add a constraint that will
prohibit its occurrences, but it adds more complexity. Second, encoding
the ``basic constraints'' of the problem, like ``each city needs to be
visited only once'' is more involved than in the previous case.

There's plenty of trade-offs when it comes to choosing encoding, and you
can come up with many different variants (or even combining multiple
encoding at the same time). I won't get into any details, but if you are
interested, I recommend reading
\href{https://www.nature.com/articles/s41534-020-0278-0}{this paper by
Nicholas Sawaya et al.}

We didn't even discuss how those encodings impact the construction of
the cost Hamiltonian for our problem, as it would be rather tedious. I
just wanted to signal this as another important decision which you need
to think about when you try to solve optimization problems, especially
with quantum computers.

\hypertarget{parameter-concentration}{%
\subsection{Parameter concentration}\label{parameter-concentration}}

Ok, now we're done with those ``general optimization topics,'' let's get
into some QAOA-specific topics. Let's start with a very interesting
phenomenon first described by
\href{https://arxiv.org/abs/1812.04170}{Brandao et al}. Consider a class
of problems: MaxCut on 3-regular graphs.\footnote{3-regular means that
  each node connects to exactly three other nodes. So each node has
  exactly 3 neighbours.} Let's take one problem instance (i.e.~one
specific graph) and find good parameters for it. And now let's take 100
other graphs of the same size and see how well these parameters would
do?

It turns out that good parameters for one instance work very well on 100
other similar instances of the same size. But that's not everything---they also work very well on graphs of different sizes, but of the same
type. So for example a graph with 10 nodes and the other one with 20
nodes. Also, good parameters work equally well and bad parameters work
equally badly. This inspired us to do a little bit of investigation with
the visualization package we developed at Zapata,
\href{https://github.com/zapatacomputing/orqviz}{orqviz}. As you can see
in the plots below, for several graphs from the same family which differ
in size, the cost landscapes look very similar around the same
parameters. I'm skimming over a lot of details here, see
\href{https://arxiv.org/abs/2111.04695}{the paper} for details.

\begin{figure}
\centering
\includegraphics{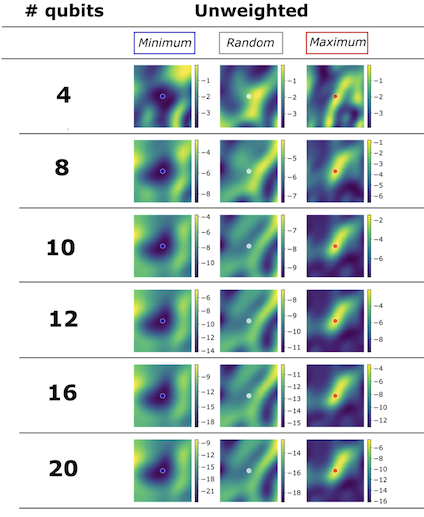}
\caption{Notice the similarity in cost landscapes between various graph sizes. This plot was generated by orqviz and it's taken from Fig. 2 of \href{https://arxiv.org/abs/2111.04695}{the orqviz paper}.}
\end{figure}

It's pretty crazy if you think about it---we can take one graph, find
good parameters for QAOA and just use them for another graph and it just
works. For now this has been investigated only for simple types of
problems and I don't think this will hold for much bigger real-world
problems, but it's still an extremely interesting phenomenon to
investigate. I think
\href{https://link.aps.org/accepted/10.1103/PhysRevA.103.042612}{this
paper by Jonathan Wurtz} gives some insight on what might be going on,
especially for 3-regular graphs.

\hypertarget{modified-cost-functions}{%
\subsection{Modified cost functions}\label{modified-cost-functions}}

In many VQAs we are interested in calculating the expectation value of a
particular operator. This is a very reasonable thing to do if the
expectation value is what we're looking after. While using QAOA the
expectation value corresponds to the value of the cost function.
However, we don't really care about the value of the cost function.
Sure, it would be nice to get to the minimum, but its numerical value is
usually not what we're after. What we are after is the solution of our
problem which gives us this low cost function value.

Before we go further let's revisit a basic question: how do we use
quantum computers to calculate the expectation value?

\begin{enumerate}
\def\labelenumi{\arabic{enumi}.}
\tightlist
\item
  We run a particular circuit and obtain a bitstring.
\item
  We use this bitstring to calculate the energy value based on some
  Hamiltonian.
\item
  And since quantum computers are probabilistic, we need to run it N
  times to get good statistics.
\end{enumerate}

Now let's imagine that we have an ansatz and two sets of parameters.
Here are two tables, one shows the energy of each bistring and the other
one the measurements we get when we run the circuits with those two sets
of params:

\begin{table}[H]
    \begin{center}
    \begin{tabular}{l|l}
          & Energy  \\ \hline
        bitstring\_1 & 10  \\
        bitstring\_2 & 1 \\
        bitstring\_3 & 0 \\
        bitstring\_4 & 1000 
    \end{tabular}
    \end{center}
    \caption{Energy of each bitstring}
    \label{tab:bitstring_energy}
\end{table}

\begin{table}[H]
    \begin{center}
    \begin{tabular}{l|lllll}
      & bitstring\_1 & bitstring\_2 & bitstring\_3 & bitstring\_4 & expectation value \\ \hline
    params\_1 & 900 & 100 & 0 & 0 & 9.1  \\
    params\_2 & 0 & 0 & 900 & 100 & 900 \\
    \end{tabular}
    \end{center}
    \caption{Bitstring distributions when measured with 1000 shots, and the expectation values of the distributions. Expectation value is calculated by taking a weighted average of the bitstring energies.}
\end{table}

Which set of params is better?

Well, if we just look at the expectation value, the first one is much
better. But if we also look at the quality of the bitstrings we obtained---that's a different story. Second set of parameters gives us
bitstring\_3. And bitstring\_3 has the lowest energy value from all the
bitstrings we've seen for this problem---perhaps it's even the best
possible solution to our problem? And, as we mentioned, in the
optimization problems we are more interested in obtaining the best
possible solution to our problem, rather than obtaining the best
expectation value.

The phenomenon we see here is a result of superposition. In the case of
Ising Hamiltonians representing combinatorial optimization problems,
there is a quantum state which gives you a particular bitstring
(representing your solution) with 100\% probability. But since we have
superposition, we can have states which represent more than one solution
at once (as seen above). This results in this somewhat unintuitive
situation, where we have a state with higher expectation value which
provides a much better solution than another state with lower
expectation value. Can we do something about it? Sure we can!

What if we simply discarded 50\% percent of measurements with the
highest cost value? This is how our table would look like:

\begin{table}[H]
    \begin{center}
    \begin{tabular}{l|lllll}
      & bitstring\_1 & bitstring\_2 & bitstring\_3 & bitstring\_4 & expectation value \\ \hline
    params\_1 & 400 & 100   & 0     & 0     & 8.2  \\
    params\_2 & 0   & 0     & 500   & 0     & 0 \\
    \end{tabular}
    \end{center}
    \caption{Bitstring distributions after discarding the 500 highest-energy (worst) samples, and the expectation values of the new distributions.}
\end{table}

This technique is called CVaR and has been proposed by
\href{https://quantum-journal.org/papers/q-2020-04-20-256/}{Barkoutsos
et al.}. It can also be applied to other VQAs, I recommend checking out
the paper as it's a really good read. Results in
\href{https://journals.aps.org/prresearch/pdf/10.1103/PhysRevResearch.4.023225}{this
follow-up work} suggests that CVaR is a must-have tool in anyone's
QAOA's toolbox. It's worth mentioning that apart from CVaR, there is
another method which works similarly, called ``Gibbs Objective
Function.'' In this case the cost function is defined as:

\begin{equation} f = - \log(e^{-\eta E})\end{equation}

By using such a cost function instead of a standard one, we increase the
weight of the low-energy samples. The
\href{https://journals.aps.org/prresearch/abstract/10.1103/PhysRevResearch.2.023074}{original
paper by Li et al.} provides a good intuition about how and why it
works, so if you're interested, I recommend reading section III of it.

\hypertarget{other-ansatzes}{%
\subsection{Other ansatzes}\label{other-ansatzes}}

One of the fundamental building blocks of VQAs is ansatz. We discussed
its importance in detail in
\hyperlink{what-the-hell-is-an-ansatz}{the
section about VQA challenges} and what are some considerations
in using one versus another.

In
\hyperlink{section-03}{the
first QAOA section} we have described the most basic
version of the ansatz for QAOA, let's do a quick recap here.

QAOA ansatz is generated by two Hamiltonians, summarized by table \ref{tab:two-qaoa-hamiltonians}.

\begin{table}
    \begin{center}
    \renewcommand{\arraystretch}{1.25} 
    \begin{tabular}{l|l}
    $H_B$ & $H_C$ \\ \hline
    Associated with the angle beta & Associated with the angle gamma \\
    Independent of the cost function & Is defined by the cost function \\
    Called ``Mixing Hamiltonian'' & Called ``Cost Hamiltonian'' \\
    \end{tabular}
    \end{center}
    \caption{Comparison of the two QAOA Hamiltonians}
    \label{tab:two-qaoa-hamiltonians}
\end{table}

QAOA creates the state shown in equation \ref{eq:qaoa-state}.

There can be a lot of variation in how you construct the circuit within
this framework, here I will go through two of them: Warm-start QAOA
ansatz and Hadfield's QAOA.

These are by no means the best possible ansatzes, but I think they show
well how one can approach constructing new ansatzes.

\hypertarget{warm-start-qaoa}{%
\subsubsection{Warm-start QAOA}\label{warm-start-qaoa}}

In optimization we have a concept of ``warm-starting'' the optimization
process. It happens when we use information from previous optimization
runs to start with some reasonable initial parameters. In QAOA we can
use some classical optimization process to get some reasonable
solutions. Then we can inject this information into our ansatz, which
hopefully will work better now. How does it work in practice?

Let's say we have a problem, for which the solution is a binary string,
and a classical solver which can solve a ``relaxed version'' of this
problem. ``Relaxation'' here means that instead of limiting ourselves to
integers, our solution can contain any real number between 0 and 1. This
is reasonable, as often continuous versions of a problem are simpler to
solve than discrete ones. We denote our classical solution as $|c|^*$
and based on this we construct a vector $\theta$ for which every i-th
element is defined as: $\theta_i = 2 \arcsin{(\sqrt{c_i^*})}$. Now for
the initial state we use state, where each qubit is set to
$|i \rangle = RY(\theta_i) |0 \rangle$. In this state the probability
to measure given qubit in state $|1 \rangle$ is equal to $c_i^*$. On
top of that, we also modify the mixing Hamiltonian to the following
form:

\begin{equation} H_M^{ws} = \sum_{i=0}^{n-1} H_{M,i}^{ws} \end{equation}

where

\begin{equation} H_{M,i}^{ws} = -\sin{\theta_i} X - \cos{\theta_i} Z \end{equation}

and $X, Z$ are Pauli operators. This is easily implementable using
$R_Y$ and $R_Z$ gates.

There is a lot more to this, so if you're interested I recommend
\href{https://arxiv.org/abs/2009.10095}{Egger et al} which shows how
this basic idea can be extended and elaborate much more on why these
choices have been made. Also, it's worth noting that there are other
approaches to warm-starting QAOA, such as
\href{https://dl.acm.org/doi/10.1145/3549554}{Tate et al} (and
\href{https://arxiv.org/abs/2112.11354}{the follow-up work}).

\hypertarget{quantum-alternating-operator-ansatz}{%
\subsubsection{Quantum Alternating Operator
Ansatz}\label{quantum-alternating-operator-ansatz}}

This section is about the framework that was proposed by Stuard Hadfield
(\href{https://academiccommons.columbia.edu/doi/10.7916/D8X650C9}{his
thesis} and \href{https://www.mdpi.com/1999-4893/12/2/34}{Hadfield et
al.}). I won't go into too many technical details about it, as it's
described very well in there, with examples for specific problems and
reasoning behind it. Here I would like to just give you a general
understanding of how this family of ansatzes work. As initially
proposed, it was intended to be used with constrained problems (please
read \protect\hyperlink{constraints}{the section about constraints}
first if you haven't already).

The main idea is as follows:

\begin{enumerate}
\def\labelenumi{\arabic{enumi}.}
\tightlist
\item
  Start from a quantum state which represents a valid solution.
\item
  Use mixing Hamiltonian which allows you to evolve your state only
  within feasible subspace. Which means it transforms quantum states
  representing valid solutions into other quantum states representing
  other valid solutions.
\end{enumerate}

Therefore a circuit here is built from three elements:

\begin{enumerate}
\def\labelenumi{\arabic{enumi}.}
\tightlist
\item
  Initial state---we need to start in a state which represents a
  feasible solution to the problem. Not the optimal one (this would
  pretty much kill the purpose), but just some which doesn't violate any
  constraints. This can also be a superposition of multiple feasible
  states, which usually gives better results.
\item
  Mixing Hamiltonian---this Hamiltonian is responsible for changing the
  provided state. So it can be implemented in such a way, that we only
  allow for the operations that transform feasible states into feasible
  states. For example in the case of TSP this means having a Hamiltonian
  which represents list permutation. It also gives a more flexible
  framework for constructing these operators, as they no longer need to
  necessarily be exponentials of a single Hamiltonian, as in regular
  QAOA.
\item
  Cost Hamiltonian---this is our old good cost Hamiltonian, no changes
  here :)
\end{enumerate}

As I've learned in some of my own projects, implementing an ansatz from
this class can give you a significant boost in the quality of the
results. However, it comes at a price---those circuits are usually
deeper than the alternatives. Let's discuss such tradeoffs in the next
section!

\hypertarget{implementation-tradeoffs}{%
\subsection{Implementation tradeoffs}\label{implementation-tradeoffs}}

As we discussed, solving a problem using QAOA requires making several
decisions---how to formulate the problem, how to encode it into an
Ising Hamiltonian, what ansatz to choose, which cost function to use,
etc.

Each of these comes at a certain cost---let's look at the choice of the
ansatz. When we consider different options it seems that Hadfield's
ansatz looks like a decent idea. However, it incurs some extra costs---first, we need to create the initial state. Depending on the particular
case, it might be either easy and cheap (in terms of number of required
gates) or very non-trivial, especially if we want to create a
superposition of many feasible initial states. Another hurdle is
implementing the mixing Hamiltonian. The standard ``all-X'' mixing
Hamiltonian can be implemented by a single layer of RZ gates. On the
other hand, some of the Hamiltonians proposed in the paper are pretty
complicated and implementing them might require creating a circuit with
hundreds of gates. Fortunately, the cost Hamiltonian will be the same
most of the time.

Let's say that for the particular problem that we want to solve, the
circuit using Hadfield's ansatz is 10x longer than the basic one. This
means that it will run 10x longer. Can this be even worth it? Well, it
might turn out that our ansatz gives us the correct answer with 10\%
probability, while using vanilla ansatz it would be 0.1\%. So even
though we need 10x more time to run our circuit, we need to run it only
100 times, to get 10 good samples, while with the basic version, we
would need to run it 10000 times. Therefore, using a more expensive
ansatz would still give us an order of magnitude improvement in the
runtime.

Obviously, this is an overly simplified toy example---in the real case
the analysis would be much, much more difficult to do---we would need
to model how different circuits are susceptible to noise, try to
estimate how fast the optimization of the parameters converges in
various cases etc. Taking all of those factors and various combinations
into account is impossible and requires extensive experimentation to
come up with the set of methods that will be appropriate for solving the
problem at hand.

However, certain calculations can be done relatively quickly---for
example, if we have some theory to back up a claim that a particular
method would require getting 1000 more samples than another one, or if
one circuit is 1000x longer than the other one, we can probably rule
them out, cause it's unlikely, they will yield enough benefits, given
that they already have 3 orders of magnitude longer runtime. Therefore,
it's always good to look critically at the methods we plan to use---even if it's not possible to get the precise numbers, we might still get
some good insight about the tradeoffs.

\hypertarget{parameter-initialization}{%
\subsection{Parameter initialization}\label{parameter-initialization}}

Imagine two people are looking for treasure on an island. One is a 9
years old girl with a broken leg, walking with crutches. The other one
is a 30-years old commando who spends two months a year on survival
trips to the Amazon jungle. They both have the same map to the treasure
and both start the treasure hunt from different places on the map.

Who is more likely to win?

Well, I wouldn't bet too much on a girl.

However, what if I told you that the girl was lucky enough to start 10
meters away from the treasure chest, while the guy starts on the
opposite side of the island, behind 2 rivers and an active volcano?

This changes the picture.

It's similar to optimization---you will spend much less time looking
for optimal parameters if the initial parameters are close to optimal.
And starting from a bad place will make things so much harder for you.
Since VQAs are all about finding optimal parameters for a given problem,
it's no surprise that people spend a lot of time trying to find a way
for good parameter initialization.

Here I will cover a couple of methods for parameter initialization---layer-by-layer optimization (LBL), INTERP and schedules. One thing
before we start---even though I call them ``parameter initializations''
strategies, they are sometimes tightly related to the optimization
process.

The simplest one is LBL---here we start from a small number of layers
which is easy to optimize due to the small number of parameters. Once we
have found optimal params, we keep them, add a new layer, initialize it
with random parms and repeat the process. And again, and again until we
reach our target number of layers. There are different variants of this
approach, but you can see that it's pretty resource-intensive, as we
have to basically run about N optimization loops, where N is the target
number of layers.

This has been improved by introducing a method called INTERP. It's based
on the observation that gammas and betas seem to increase monotonically
(see fig.~2 from
\href{https://journals.aps.org/prx/abstract/10.1103/PhysRevX.10.021067}{Zhou
et al.}). Therefore, we don't have to initialize the new layer randomly---if we have good parameters for layer N-1, we can predict what would
be good parameters for N, tune all the params a little bit and repeat
the process. This can be done using a straightforward interpolation, but
the authors also suggested a more sophisticated method called
``FOURIER''---I recommend reading the original paper, it's extremely
interesting. There is also \href{https://arxiv.org/abs/2209.01159}{a
recent work by Sack et al}, which provides some more theoretical
background and intuition why and how INTERP works.

Methods like this will work well for a relatively small number of
layers, but what if we talk about the future where we could have
thousands of layers? It turns out that in this limit we might be able to
get decent results by using something called ``schedules.'' In
\href{https://arxiv.org/abs/2108.13056}{this paper} the authors decided
that they'll choose values for betas and gammas defined by:

\begin{equation}
\begin{aligned}
    \gamma(f) &= \Delta f \\
    \beta(f) &= (1- \Delta) f
\end{aligned}
\end{equation}

Here $f$
has values from 0 to 1 and for a case with a total of $p$ layers, we
substitute $f$ with $f_j = \frac{j}{p+1}$ for a $j$-th layer.
$\Delta$ is a free parameter we need to pick ourselves. Therefore
schedule is basically a curve from which we take our parameters. This
work by \href{https://quantum-journal.org/papers/q-2021-07-01-491/}{Sack
and Serbyn} also discusses similar ideas.

I like how these methods fit together---we've seen in the INTERP paper that the parameters change monotonically, so why not just make a simple assumption that the change is linear and see where it leads us? Since finding optimal parameters is an extremely computationally expensive problem, it would be great if using schedules like this (or more involved) would spare us optimizing parameters or provide us with really good initial params. But to see how this will play out in practice, I think we need to start running more of these on real devices.

\hypertarget{closing-notes}{%
\subsection{Closing notes}\label{closing-notes}}

There are a lot of other developments in the QAOA space.
\href{https://journals.aps.org/prl/abstract/10.1103/PhysRevLett.125.260505}{RQAOA},
\href{https://arxiv.org/abs/2109.11455}{Multi-angle QAOA} and
\href{https://arxiv.org/abs/2005.10258}{ADAPT-QAOA} to name a few.
However, if I tried to cover all these topics, this would soon become a review paper, so I had to stop somewhere :)

I hope you found this chapter helpful in understanding some more advanced concepts around QAOA and now it will be easier for you to explore these concepts by yourself.

\newpage

\hypertarget{section-8}{\section{Conclusion}}

Well, that's it! I hope reading this was enjoyable experience and it made you much more knowledgable on the topic of VQAs!

There is a big number of people who helped me write all my blogposts and this document. Each of them took some time from their life to review a draft, explain some concepts or edit this PDF. I'm deeply grateful for each contribution!

Here they are, in roughly chronological order: Rafał Ociepa, Ntwali Bashige, Peter Johnson, AJ Roeth, Sukin Sim, Tanisha Bassan, Tomasz Sabała, Davit Khachatryan, Borja Peropadre, Alejandro Perdomo-Ortiz, Yudong Cao, Rishi Sreedhar, Jhonathan Romero-Fontalvo, Alba Cervera-Lierta, Alex Juda, Stefano Mangini, Mark Cunningham, Pierre-Luc Dallaire-Demers, Ahmed Darwish, Sophia Economou, Nicolas Sawaya, Jérôme Gonthier, Boniface Yogendran, Matt Kowalsky, Stuart Hadfield, Elias Ancer, Laura Gao.

Once again---if you would like to see what I'm up to these days, please visit my blog \href{https://mustythoughts.com/}{Musty Thoughts}. And if you found this document useful or have any feedback, reach out to me at \href{mailto:michal@mustythoughts.com}{michal@mustythoughts.com}. 

Have a nice day!

Michał



 


\end{document}